\documentclass[review]{elsarticle}
 
\usepackage{lineno,hyperref}
 
\modulolinenumbers[5]

\usepackage[T1]{fontenc}
\usepackage{amsmath}
\usepackage{amssymb}
\usepackage{graphicx}
\usepackage{esint}

\journal{Spatial Statistics}

\biboptions{authoryear}

\begin{document}

\begin{frontmatter}

\title{Beyond correlation in spatial statistics modeling}

\author[myaddress]{Jhan Rodr\'iguez \corref{corrAuthor}}
\cortext[corrAuthor]{Corresponding author}
\ead{Jhan.Rodriguez@iws.uni-stuttgart.de}

\author[myaddress]{Andr\'as B\'ardossy}

\address[myaddress]{Institute for Modelling Hydraulic and Environmental Systems, Universit\"at Stuttgart}

\begin{abstract}
We introduce a model for spatial statistics which can account explicitly for interactions among more than two field components at a time. The theoretical aspects of the model are dealt with: cumulant and moment generating functions, spatial consistency and parameter estimation. On the basis of a detailed synthetic example, we show the kind of inference about the (partially observed) spatial field that can be very wrong, if one validates his model by checking only one and two dimensional marginal fit, and covariance function fit. We suggest statistics that can be used
additionally for model validation, which help assess interdependence among groups of variables. The implications of considering multivariate interactions for intense daily precipitation forecasting over a small catchment in southeastern Germany (that of the Saalach river) are investigated.
\end{abstract}

\begin{keyword}
Multivariate Interdependence\sep Cumulant Generating Function\sep Non-Gaussian Fields
\end{keyword}

\end{frontmatter}

\linenumbers

\section{\label{sec:Introduction}Introduction}

In the context of precipitation downscaling, \citet{bardossy_multiscale_2012}
found that observed clustering patterns of very high values at multiple
locations on the target scale could not be reproduced by simulated
data, even after site-wise (i.e. marginal) bias correction and correlation
bias correction of the simulations. That is, even though the marginal
distributions and the inter-site correlations of the data simulated
for the validation period were identical to those of observed data
in the same period, clustering patterns of observed data were still
not properly recovered, particularly for clusters with very high values: 
the data simulated were unable to recover the distribution of the
sum of blocks of four sites or ``pixels''. This may lead to substantial underestimation
of flood return periods, since the occurrence of unexpected clusters
of intense rainfall over a catchment can result in great floods, not
expected in centuries. 

The paper of \citet{bardossy_multiscale_2012} adds evidence to that
of \citet{bardossy_copula_2009} about the need to build spatial models
which can consider, explicitly, simultaneous interactions among more
than two of the components of the modeled field. The present paper
proposes one such model, and elaborates on its theory. It is conceived
as an initial step in a research direction which has gone mostly unnoticed.
It intends to help start a wider discussion on the topic of multivariate
interactions for spatial statistics models.

In our exposition, we focus for simplicity on the class of second
order stationary processes, possibly after subtracting a trend field.
We also assume a two dimensional isotropic field, and that the studied
spatial process takes values only on a finite number of locations
over a grid. However, these simplifications are by no means restrictive
of the methodology presented in this paper; they just help to make
exposition easier. 

Consider spatially labeled locations $\left\{ \mathbf{s_{1}},\mathbf{s_{2}},\ldots,\mathbf{s_{J}}\right\} $,with
$\mathbf{s_{j}}\in\mathbb{R}^{2}$, and let 
\[
\left(\mathbf{Z\left(s_{1}\right)},\mathbf{Z\left(s_{2}\right)},\ldots,\mathbf{Z\left(s_{J}\right)}\right)
\]
 represent a random quantity taking values at the given locations,
so that a field of variable $\mathbf{Z}$ is obtained. Association
between every two components of this field can be modeled in terms
of covariance function, $C$, 
\begin{equation}
cov\left(\mathbf{Z\left(s_{i}\right),}\mathbf{Z\left(s_{j}\right)}\right)=C\left(\left\Vert \mathbf{s_{i}}-\mathbf{s_{j}}\right\Vert \right)\label{eq:cov_funct01}
\end{equation}
where $\left\Vert \mathbf{s_{i}}-\mathbf{s_{j}}\right\Vert $ is the
euclidean distance between $\mathbf{s_{i}}$ and $\mathbf{s_{j}}$.
This covariance function must ensure positive-definiteness of the
resulting covariance matrix. For example, two popular covariance functions
are: 
\begin{description}
\item [{Powered-exponential:}] given by equation 
\begin{equation}
C\left(d\right)=\sigma_{0}^{2}.I\left(d=0\right)+\sigma_{1}^{2}\exp\left(-\left(d/\theta_{1}\right)^{\theta_{2}}\right)\label{eq:Power-exp_cov}
\end{equation}
where $I\left(*\right)$ stands for the indicator function.
\item [{Mat\'ern's:}] given by equation 
\begin{equation}
C\left(d\right)=\sigma_{0}^{2}.I\left(d=0\right)+\sigma_{1}^{2}\left[2^{\theta_{2}-1}\Gamma\left(\theta_{2}\right)\right]^{-1}\left[d/\theta_{1}\right]^{\theta_{2}}K_{\theta_{2}}\left(d/\theta_{1}\right)\label{eq:Matern_cov}
\end{equation}
where $\Gamma\left(*\right)$ stands for the Gamma function and $K_{\theta_{2}}\left(d/\theta_{1}\right)$
for the modified Bessel function of the second kind of order $\theta_{2}$
(see, for example \citet{abramowitz_handbook_1972}).
\end{description}
Parameters $\left(\theta_{1},\theta_{2},\sigma_{0}^{2},\sigma_{1}^{2}\right)$
are the covariance function parameters. Hence, only a reduced number
of parameters must be estimated in order to find the covariance between
every two components $\mathbf{Z\left(s_{i}\right)}$ and $\mathbf{Z\left(s_{j}\right)}$,
given locations $\mathbf{s_{i}}$ and $\mathbf{s_{j}}$. 

The Normal model is a common model in Spatial Statistics for components
corresponding to every finite set of locations, 
\[
\left(\mathbf{Z\left(s_{1}\right),}\ldots,\mathbf{Z\left(s_{J}\right)}\right)\sim N_{J}\left(\mu,\Sigma\right)
\]
where the covariance matrix is given by $\Sigma_{ij}=C\left(\left\Vert \mathbf{s_{i}}-\mathbf{s_{j}}\right\Vert \right)$.
Under the Gaussian model, the whole distribution is defined by a vector
of means $\mu\in\mathbb{R}^{J}$ and parameters $\left(\theta_{1},\theta_{2},\sigma_{0}^{2},\sigma_{1}^{2}\right)$,
which determine matrix $\Sigma$. It is often the case that the mean
vector is represented as a function $\xi$ of the geographic coordinates
of $\mathbf{s_{j}}$, or of an additional variable (\textquotedbl{}external
drift\textquotedbl{}) related to such location ,
\begin{equation}
\mu_{j}=\xi\left(\mathbf{s_{j}}\right)\label{eq:external_drift}
\end{equation}

For a new location $\mathbf{s_{k}}\notin\left\{ \mathbf{s_{1},s_{2},s_{3},\ldots,s_{J}}\right\} $,
the joint distribution of 
\[
\left(\mathbf{Z\left(s_{1}\right),}\ldots,\mathbf{Z\left(s_{J}\right)},\mathbf{Z\left(s_{k}\right)}\right)
\]
 can be readily found under the Normal model: one adds component $\mu_{k}=\xi\left(\mathbf{s_{k}}\right)$
to the means vector, and extends the covariance matrix by $\Sigma_{ik}=C\left(\left\Vert \mathbf{s_{i}}-\mathbf{s_{k}}\right\Vert \right)$,
for each $\mathbf{s_{i}}\in\left\{ \mathbf{s_{1},s_{2},s_{3},\ldots,s_{J}}\right\} $.
The model is thus completely specified. This is one of the reasons
why the Normal model is very convenient conceptually, and is often
used in practice, if necessary after applying a suitable transformation
to data (see section \ref{sec:A-glimpse-at}).

The family of elliptical distributions can be seen as the wider family
to which both the multivariate Normal and the multivariate Student
distributions belong. The classical definition, according to \citet{cambanis_theory_1981},
is as follows:
\paragraph{Definition} 
Let $\mathbf{X}$ be a J-dimensional random vector, $\mu\in\mathbb{R}^{J}$
and $\Sigma$ a $J\times J$, non-negative definite matrix. Let $\phi_{\mathbf{X}-\mu}\left(\mathbf{t}\right):\mathbb{R}^{J}\rightarrow\left[0,+\infty\right)$
be the characteristic function of $\mathbf{X}-\mu$. If $\phi_{\mathbf{X}-\mu}\left(\mathbf{t}\right)=\Psi\left(\mathbf{t}\Sigma\mathbf{t^{'}}\right)$
for some function $\Psi\left(s\right):\left[0,+\infty\right)\rightarrow\left[0,+\infty\right)$,
then we say that $\mathbf{X}$ has an elliptically contoured distribution
with parameters $\mu$ and $\Sigma$.
 
In case $\mathbf{X}-\mu$ is Normally distributed with means vector
$\mathbf{0}$ and covariance matrix $\Sigma$, one has of course $\Psi\left(s\right):=\exp\left(-\frac{1}{2}s\right)$.

An elliptically distributed vector $\mathbf{X}-\mu$ can always be
represented as
\begin{equation}
\mathbf{X}-\mu=R\times\mathbf{U}\times\Sigma^{\frac{1}{2}}\label{eq:representation1}
\end{equation}
where $\Sigma^{\frac{1}{2}}$ is a $J\times J$ matrix such that $\Sigma^{\frac{1}{2}}\times\mbox{\ensuremath{\left(\Sigma^{\frac{1}{2}}\right)^{T}}}=\Gamma$,
for example its Cholesky decomposition factor; $R\geq0$ is a non-negative
random variable; and $\mathbf{U}\in\mathbb{R}^{J}$ is a random vector
uniformly distributed on the boundary of the unit hypersphere (see
\citet{cambanis_theory_1981}). Variable $R$ receives the name of
``generating variable'', and together with $\Sigma$ determines
the specific characteristics of $\mathbf{X}$, most importantly, its
tail behavior. The generating variable is what really marks the difference
among the several elliptical distributions one might construct. 
\paragraph{Example 1}
\label{ejamplo_Gaussian}In case $\mathbf{X}-\mu$ is Normally distributed
with means vector $\mathbf{0}$, then generating variable $R$ is
distributed as a $\chi$ distribution with $J$ degrees of freedom.
That is, $R^{2}\sim\chi_{J}^{2}$, a chi-squared distribution with
$J$ degrees of freedom. 
 
Another well-known case is that of the multivariate student distribution
with $\nu$ degrees of freedom, for which $R^{2}\sim J\times F_{J,\nu}$,
and $F_{J,\nu}$ represents the Fisher distribution with $J$ and
$\nu$ degrees of freedom.

Despite being a generalization to the Normal model, which pervades
the Spatial Statistics literature, elliptical models are not part
of current practice in the area. For example, among other excellent
books on the subject, no mention is made about elliptical distributions
at \citet{le_statistical_2006,cressie2011statistics,cressie_statistics_1991,diggle2007model,banerjee2003hierarchical}.
This may have to do with the inconvenient the model presents for interpolation
or ``kriging'': For the multivariate Normal and student models,
it has already been seen that the distribution of the generating variable
depends on the dimension of the vector $\mathbf{X}\in\mathbb{R}^{J}$,
which means that function $\Psi$ must also change. Since our model
is defined in terms either of $\Psi$, as in definition 1, or in terms
of generating variable $R$, as in representation (\ref{eq:representation1}),
it is not clear in general into what these parameters will turn when
extending the model to $k$ ``ungauged'' sites, whereby $\mathbf{X}\in\mathbb{R}^{J+k}$.
This issue is addressed in this paper.

Our intention in dealing with elliptical distributions is to consider
interdependence among variables that cannot be quantified in terms
of correlations or covariances alone, which concepts form the core
of dependence modeling in current spatial statistical practice. The
topic of \textquotedbl{}beyond correlation interdependence\textquotedbl{}
has been addressed in itself by \citet{CumulantsRodriguezBardossy}.
We intend here to give an implementation to the ideas presented at
\citet{CumulantsRodriguezBardossy} in the context of Spatial Statistics. 

At \citet{CumulantsRodriguezBardossy}, a distinction is drawn between
interaction ``parameters'', and interaction ``manifestations''.
The latter are subject-matter specific statistics connected with sub-vectors
of the analyzed random vector, $\mathbf{X}$, and dependent on the
type of association among the components of such sub-vectors; they
have a relevant interpretation for the researcher. Interaction ``parameters''
can be seen as convenient building blocks of a (low dimensional) model
or dependence structure that can somehow reproduce the target interaction
manifestations. It is argued that the joint cumulants of $\mathbf{X}$
are legitimate extensions to correlation coefficients to more than
two variables, and as the building blocks referred to above. The cumulant
generating function is then accordingly referred to as a ``dependence
structure''.

In the present paper, we show how we can build a low dimensional (regarding
the number of parameters to fit) spatial model on the basis of joint
cumulants, i.e. on the basis of a cumulant generating function. This
model can be considered a natural extension to the Normal model. A
Normal model is built on the order one and two joint cumulants only,
namely a means vector $\mu$ containing the order one cumulants, and
an array of covariances $cov\left(X_{j},X_{i}\right)$ containing
the order two joint cumulants. In the extension here presented, higher
order joint cumulants can be considered without increasing prohibitively
the number of parameters to fit. 

The rest of the paper is structured as follows. Section \ref{sec:The-proposed-model}
is theoretical; it introduces our model via its cumulant generating
function; it is intended to make clear why the model actually considers
interactions among groups of variables with a minimum of parameters.
Section \ref{sec:Model-Implementation} is a transition toward practical
applicability; it shows the probability density of the model, and
how to ensure spatial consistency; a basic parameter estimation procedure
is presented. Section \ref{sec:Simulation,-interpolation-and} shows
how to obtain unconditional realizations of the field for arbitrary
dimensions, how to find conditional distributions given partial observations
of the field, and how to simulate conditional fields of an arbitrary
dimension. Section \ref{sec:A-simulation-based-illustration} can
be considered the core of this paper; by means of a synthetic example,
it explores the kind of inference that can go wrong when using a model
predicated on a combination of bi-variate connections only; it also
suggests some statistics that can help to identify interdependence
features of data beyond correlation. Section \ref{sec:A-glimpse-at}
analyzes the implications of multivariate interdependence for flood
risk assessment in the Saalach river catchment, in southeastern Germany;
a space-time model, whose structure is provided by a latent Gaussian
field, is fitted to daily precipitation from 2004 to 2009; this latent
Gaussian structure is replaced by a quasi-Gaussian structure which
possesses interactions beyond correlations, and the forecasts of the
two versions are compared; in addition, the conditional rainfall field
of June 1st 2013 over the catchment is analyzed in the light of both
models. Section \ref{sec:Discussion-and-work} contains a short discussion
and intended future work.

\section{\label{sec:The-proposed-model}The proposed model}

In the following, we assume the existence of sufficiently many product
moments of $\mathbf{X}$; sufficient so as to provide a practically
useful approximation to the processes modeled. Then it is more convenient,
for our purposes, to conceptualize elliptical distributions in terms
of their moment generating function. 

It might be protested that moments (and hence cumulants) of sufficiently
high orders might not exist for the ``true'' probability distribution
of the process under analysis. We would answer that such distributions
can always be sufficiently (i.e. for practical purposes) \emph{approximated}
by a distribution with existing moments of all orders. See, for example
\citet{1987}, where the authors introduce a semi-parametric model,
similar to an Edgeworth expansion. This model possesses moments of
all orders. Yet, under minimal conditions it can approximate \emph{any}
continuous distribution on $\mathbb{R}^{J}$, provided sufficiently
many factors are added to the sum defining the model. Additionally,
\citet{del_brio_gramcharlier_2009,mauleon2000testing,perote_multivariate_2004}
present variants of the model of \citet{1987}, and show how they
can be effectively applied to modeling heavy tailed data, both univariate
and multivariate. 

We say, then, that random vector $\mathbf{X}\in\mathbb{R}^{J}$ is
``elliptically distributed'' if and only if its moment generating
function can be written as 
\begin{equation}
M_{\mathbf{X}-\mathbf{\mu}}\left(\mathbf{t}\right)=E\left(e^{\left\langle \mathbf{t},\mathbf{X}-\mu\right\rangle }\right)=\Upsilon\left(\mathbf{t^{T}}\text{\ensuremath{\Sigma}}\mathbf{t}\right)\label{eq:Original_mgf_elliptical}
\end{equation}
for some function $\Upsilon:\mathbb{R}\rightarrow\mathbb{R}$, and
some $\mathbf{\mu}\in\mathbb{R}^{J}$. For the sake of simplicity,
we assume for now that $\mu=\mathbf{0}$. 

Consider a moment generating function of the form
\begin{equation}
M_{\mathbf{X}}\left(\mathbf{t}\right)=\exp\left(\delta\left(\frac{1}{2}\mathbf{t^{T}}\Sigma\mathbf{t}\right)\right)\label{eq:archetype_mgf}
\end{equation}
for some function $\delta:\mathbb{R}\rightarrow\mathbb{R}$. Then
the cumulant generating function (c.g.f.) of $\mathbf{X}$ is given
by
\begin{equation}
K_{\mathbf{X}}\left(\mathbf{t}\right):=\log\left(M_{\mathbf{X}}\left(\mathbf{t}\right)\right)=\delta\left(\frac{1}{2}\mathbf{t^{T}}\Sigma\mathbf{t}\right)\label{eq:cumulants_eq}
\end{equation}

This function $\delta\left(y\right)$ can be formally expanded in
its Taylor Series around zero,
\begin{multline}
\delta\left(y\right)=c_{0}+\frac{c_{1}}{1!}y+\frac{c_{2}}{2!}y^{2}+\frac{c_{3}}{3!}y^{3}+\frac{c_{4}}{4!}y^{4}+\ldots\\
=c_{0}+\frac{c_{1}}{1!}\left(\frac{1}{2}\mathbf{t^{T}}\Sigma\mathbf{t}\right)+\frac{c_{2}}{2!}\left(\frac{1}{2}\mathbf{t^{T}}\Sigma\mathbf{t}\right)^{2}+\frac{c_{3}}{3!}\left(\frac{1}{2}\mathbf{t^{T}}\Sigma\mathbf{t}\right)^{3}+\ldots\\
=\delta\left(\left(\frac{1}{2}\mathbf{t^{T}}\Sigma\mathbf{t}\right)\right)\label{eq:Taylor_aux}
\end{multline}
where $c_{r}=\frac{d^{r}}{dy^{r}}\delta\left(y\right)\mid_{y=0}$. 

A little thought shows that the assumption $\mathbf{\mu}=\mathbf{0}$
implies that $c_{0}=0$. Thus, by virtue of (\ref{eq:cumulants_eq})
and (\ref{eq:Taylor_aux}) combined, we have that the c.g.f can be
written as
\begin{equation}
K_{\mathbf{X}}\left(\mathbf{t}\right)=c_{1}\frac{1}{2}\mathbf{t^{T}\Sigma t}+\frac{1}{2!}c_{2}\left[\frac{1}{2}\mathbf{t^{T}\Sigma t}\right]^{2}+\frac{1}{3!}c_{3}\left[\frac{1}{2}\mathbf{t^{T}\Sigma t}\right]^{3}+\ldots\label{eq:archetypal_cgf}
\end{equation}

This c.g.f. was studied by \citet{Steyn19931}, in his attempt to
introduce more flexibility into the elliptical distributions family.
Our proposed model for spatial statistics is given by expansion (\ref{eq:archetypal_cgf}),
up to an (application specific) expansion order $K\in\left\{ 1,2,3,4,\ldots\right\} $.
That is, our proposed model is given by a covariance/correlation matrix,
$\Sigma_{J\times J}$, together with a set of coefficients $c_{1},c_{2},\ldots,c_{K}$.
Coefficients corresponding to a higher order, $c_{r>K}$, are left
undetermined but will be automatically fitted in the presence of data,
by means of the implementation of the model given at section (\ref{sec:Model-Implementation}).
Such an implementation circumvents the inconvenience of a model introduced
in terms of a c.g.f., by dealing with the equivalent density function
instead. 

From the definition of our model (\ref{eq:archetypal_cgf}), some
remarks are immediately in place and are given below.

\subsubsection*{The introduced model as extension to the Normal model}

Firstly, the c.g.f. (\ref{eq:archetypal_cgf}) boils down to that
of the Normal distribution by setting $c_{r}:=0$, for $r>1$. Hence
the proposed model can be seen as a natural extension to the Normal
model which, under $\mu=\mathbf{0}$, is entirely determined by its
covariance coefficients 
\begin{equation}
\frac{\partial^{2}K_{\mathbf{X}}\left(\mathbf{t}\right)}{\partial t_{2}\partial t_{1}}\mid_{\mathbf{t}=\mathbf{0}}=c_{1}\Sigma_{ij}=cov\left(X_{i},X_{j}\right)\label{eq:cov01}
\end{equation}

From (\ref{eq:cov01}), the need to assume either $c_{1}$ fixed,
or $\Sigma$ a correlation matrix, becomes evident: otherwise it will
be impossible to identify them separately. In this research, we define
$\Sigma$ to be a covariance matrix, whereas $c_{1}=1$, unless otherwise
stated.

\subsubsection*{Joint cumulants and product moments}

Secondly, the joint cumulants of a random vector having a c.g.f as
in (\ref{eq:archetypal_cgf}) are readily found by differentiating
$K_{\text{\ensuremath{\mathbf{X}}}}\left(\mathbf{t}\right)$ with
respect to the indexes of the joint cumulant, and evaluating the result
at $\mathbf{t}=\mathbf{0}$. \citet{CumulantsRodriguezBardossy} show
why it is reasonable to call joint cumulants multivariate \textquotedbl{}interaction
parameters\textquotedbl{}.

All joint cumulants of odd order, $\kappa^{j_{1},\ldots,j_{k}}$ ($k$
odd), are zero for our dependence model. For $k$ an even integer,
joint cumulants are given by:

\begin{eqnarray}
\kappa^{j_{1},j_{2}} & = & \frac{c_{1}}{2}\left\{ \Sigma_{j_{1}j_{2}}+\Sigma_{j_{2}j_{1}}\right\} \nonumber \\
\kappa^{j_{1},j_{2},j_{3},j_{4}} & = & \frac{c_{2}}{2!}\frac{1}{2^{2}}\left\{ \Sigma_{j_{1}j_{2}}\Sigma_{j_{3}j_{4}}+\Sigma_{j_{1}j_{3}}\Sigma_{j_{2}j_{4}}+\Sigma_{j_{1}j_{4}}\Sigma_{j_{2}j_{3}}\right\} \nonumber \\
 & \vdots\nonumber \\
\kappa^{j_{1},\ldots,j_{k}} & = & \frac{c_{\frac{k}{2}}}{\frac{k}{2}!}\frac{1}{2^{\frac{k}{2}}}\left\{ \sum_{j_{1},\ldots,j_{k}=1}^{J}\Gamma_{j_{1}j_{2}}\ldots\Gamma_{j_{k-1}j_{k}}\right\} \label{eq:cumulants_archetypal}
\end{eqnarray}
and so on, as shown in  \ref{sec:Derivation-of-joint}. In
this manner, interaction among sets of four, six or more variables
can be conveniently summarized.

It will be convenient to define ``covariance interdependence factor''
$\varrho\left(j_{1},\ldots,j_{k}\right)$ as the sum of the products
of the covariance coefficients at (\ref{eq:cumulants_archetypal}).
Specifically,
\begin{eqnarray*}
\varrho\left(j_{1},j_{2}\right) & = & \Sigma_{j_{1}j_{2}}\\
\varrho\left(j_{1},\ldots,j_{4}\right) & = & \Sigma_{j_{1}j_{2}}\Sigma_{j_{3}j_{4}}+\Sigma_{j_{1}j_{3}}\Sigma_{j_{2}j_{4}}+\Sigma_{j_{1}j_{4}}\Sigma_{j_{2}j_{3}}\\
\varrho\left(j_{1},\ldots,j_{6}\right) & = & \Sigma_{j_{1}j_{2}}\Sigma_{j_{3}j_{4}}\Sigma_{j_{5}j_{6}}+\Sigma_{j_{1}j_{3}}\Sigma_{j_{2}j_{4}}\Sigma_{j_{5}j_{6}}+\ldots+\Sigma_{j_{1}j_{6}}\Sigma_{j_{2}j_{4}}\Sigma_{j_{5}j_{3}}
\end{eqnarray*}
and so on. This is a ``potential'' interdependence factor, since
its effect on higher order interdependence parameters (i.e. joint
cumulants of even order $k$ greater than 2), is only present if its
corresponding coefficient $c_{k/2}$ is non-zero. So, every joint
cumulant at (\ref{eq:cumulants_archetypal}) can be written as
\begin{equation}
\kappa^{j_{1},\ldots,j_{k}}=\frac{c_{\frac{k}{2}}}{\frac{k}{2}!}\frac{1}{2^{\frac{k}{2}}}\times\varrho\left(j_{1},\ldots,j_{k}\right)
\end{equation}

The interdependence parameter (i.e. joint cumulant) of order $k>2$
of our model can then be conceptually split into two components: On
the one hand, a ``covariance interdependence component'', $\varrho\left(j_{1},\ldots,j_{k}\right)$,
which can be estimated \emph{low-dimensionally} via covariance function
fitting, as usual in Geostatistics. On the other hand, an interdependence
``enhancing'' parameter $c_{k/2}$, whose departure from zero determines
the departure from zero of the $k$-th order joint cumulant. As illustrated
in \citet{CumulantsRodriguezBardossy}, these joint cumulants can
be connected with relevant interaction manifestations (such as the
differential entropy of the distribution, or the distribution of the
sums of the components of the field). As a consequence, one can can
try fitting the research-specific interaction manifestation, which
is not explainable in terms of correlations, by fitting parameters
$c_{2},c_{3},\ldots$

An expansion for the moment generating function (m.g.f.) for $\mathbf{X}$
will be now introduced. By setting shorthand notation 
\[
y:=\frac{1}{2}\mathbf{t^{T}\Sigma t}
\]
the dependence structure (\ref{eq:archetypal_cgf}) can be written
\begin{equation}
K_{\mathbf{X}}\left(\mathbf{t}\right)=\frac{c_{1}}{1!}y+\frac{c_{2}}{2!}y^{2}+\frac{c_{3}}{3!}y^{3}+\ldots\label{eq:moment_gen_fun_1}
\end{equation}

On the other hand, the definition of our dependence structure, given
originally by (\ref{eq:archetype_mgf}) implies that we can write,
using the same shorthand notation as above,
\begin{multline}
\exp\left(K_{\mathbf{X}}\left(\mathbf{t}\right)\right):=M_{\mathbf{X}}\left(\mathbf{t}\right)=\\
\exp\left(\delta\left(y\right)\right)=1+\frac{m_{1}}{1!}y+\frac{m_{2}}{2!}y^{2}+\frac{m_{3}}{3!}y^{3}+\ldots
\end{multline}
for certain coefficients $m_{1},m_{2},m_{3},\ldots$, at least for
$y$ in a neighborhood of zero (that is, for $\mathbf{t}$ in a sufficiently
small neighborhood of $\mathbf{0}$). Summarizing, we have that

\begin{equation}
\log\left(1+\frac{m_{1}}{1!}y+\frac{m_{2}}{2!}y^{2}+\frac{m_{3}}{3!}y^{3}+\ldots\right)=\frac{c_{1}}{1!}y+\frac{c_{2}}{2!}y^{2}+\frac{c_{3}}{3!}y^{3}+\ldots
\end{equation}
and then we can obtain, as in the case of the one-dimensional cumulants
in terms of the one-dimensional moments (see, e.g. \citet{kendall_advanced_1969,smith1995recursive}
), coefficients $m_{1},m_{2},m_{3},\ldots$ in terms of $c_{1},c_{2},c_{3},\ldots$,
by 

\begin{eqnarray}
c_{1} & = & m_{1}\nonumber \\
c_{2} & = & m_{2}-m_{1}^{2}\nonumber \\
c_{3} & = & m_{3}-3m_{2}m_{1}+2m_{1}^{3}\nonumber \\
c_{4} & = & m_{4}-4m_{3}m_{1}-3m_{2}^{2}+12m_{2}m_{1}^{2}-6m_{1}^{4}\label{eq:cums_in_terms_moms-1}
\end{eqnarray}
which after some algebraic manipulation, returns,
\begin{eqnarray}
m_{1} & = & c_{1}\nonumber \\
m_{2} & = & c_{2}+c_{1}^{2}\nonumber \\
m_{3} & = & c_{3}+3c_{2}c_{1}+c_{1}^{3}\nonumber \\
m_{4} & = & c_{4}+4c_{3}c_{1}+3c_{2}^{2}+6c_{2}c_{1}^{2}+c_{1}^{4}\label{eq:cums_in_terms_moms-2}
\end{eqnarray}

So, we have shown that the moment generating function at (\ref{eq:archetype_mgf})
can be written as
\begin{equation}
M_{\mathbf{X}}\left(\mathbf{t}\right)=1+\frac{m_{1}}{1!}\left(\frac{1}{2}\mathbf{t^{T}\Sigma t}\right)+\frac{m_{2}}{2!}\left(\frac{1}{2}\mathbf{t^{T}\Sigma t}\right)^{2}+\ldots
\end{equation}
which is similar to the expansion of $K_{\mathbf{X}}\left(\mathbf{t}\right)$,
except for the leading term 1 and coefficients $m_{r}$, $r=1,2,\ldots$.
We express product moments analogously as joint cumulants by
\begin{equation}
\mu^{j_{1},\ldots,j_{k}}:=E\left(X_{j_{1}}\ldots X_{j_{k}}\right)
\end{equation}
where $j_{r}\in\left\{ 1,\ldots J\right\} $, $r=1,\ldots k$, allowing
repetition of indexes. Then it follows, analogously to (\ref{eq:cumulants_archetypal}),
that
\begin{eqnarray}
\mu^{j_{1},j_{2}} & = & \frac{m_{1}}{2}\left\{ \Sigma_{j_{1}j_{2}}+\Sigma_{j_{2}j_{1}}\right\} \label{eq:cumulants_archetypal-1}\\
\mu^{j_{1},j_{2},j_{3},j_{4}} & = & \frac{m_{2}}{2!}\frac{1}{2^{2}}\left\{ \Sigma_{j_{1}j_{2}}\Sigma_{j_{3}j_{4}}+\Sigma_{j_{1}j_{3}}\Sigma_{j_{2}j_{4}}+\Sigma_{j_{1}j_{4}}\Sigma_{j_{2}j_{3}}\right\} \\
 & \vdots\\
\mu^{j_{1},\ldots,j_{k}} & = & \frac{m_{\frac{k}{2}}}{\frac{k}{2}!}\frac{1}{2^{\frac{k}{2}}}\left\{ \sum_{j_{1},\ldots,j_{k}=1}^{J}\Gamma_{j_{1}j_{2}}\ldots\Gamma_{j_{k-1}j_{k}}\right\} 
\end{eqnarray}
where $m_{k}$ is as in (\ref{eq:cums_in_terms_moms-2}). These moment
equations will be useful for parameter estimation purposes, as seen
in section \ref{sec:Model-Implementation}.

We see then, for example by setting $c_{1}=1$ and $c_{r>1}=0$, that
we can have non-zero joint moments of orders greater than two, even
though no dependence of order greater than two is present in the distribution
of $\mathbf{X}$, according to our definition of high order dependence,
as justified by \citet{CumulantsRodriguezBardossy}.

\subsubsection*{The proposed c.g.f. as an extension to the covariance function}

Covariance functions, such as (\ref{eq:Power-exp_cov}) or (\ref{eq:Matern_cov})
have proved valuable tools for spatial statistics analysis. They define
the order-two joint cumulant of every pair of components, e.g., 
\begin{multline}
C\left(d_{ij}\mid\left(\theta_{1},\theta_{2},\sigma_{0}^{2},\sigma_{1}^{2}\right)\right)=\sigma_{0}^{2}.I\left(d=0\right)+\sigma_{1}^{2}\exp\left(-\left(d/\theta_{1}\right)^{\theta_{2}}\right)\\
=cov\left(X_{i},X_{j}\right)=\frac{\partial^{2}K_{\mathbf{X}}\left(\mathbf{t}\right)}{\partial t_{2}\partial t_{1}}\mid_{\mathbf{t}=\mathbf{0}}\label{eq:cov01-1}
\end{multline}
where $d_{ij}\in\left[0,+\infty\right)$ denotes the distance between
the sites, $\mathbf{s_{i}}$ and $\mathbf{s_{j}}$, to which $X_{i}$
and $X_{j}$ correspond. 

Let $D=\left\{ d_{ij}\right\} $ be the matrix of distances between
the sites corresponding to the different components of $\mathbf{X}$.
Then, one has the matrix equality $\left\{ \Sigma_{ij}\right\} =\left\{ C\left(d_{ij}\mid\left(\theta_{1},\theta_{2},\sigma_{0}^{2},\sigma_{1}^{2}\right)\right)\right\} $.
With slight abuse of notation, denote 
\[
C\left(D\mid\left(\theta_{1},\theta_{2},\sigma_{0}^{2},\sigma_{1}^{2}\right)\right):=\Sigma
\]
 The c.g.f. (\ref{eq:archetypal_cgf}) can then be written as a ``higher
order'' spatial covariance function as
\[
K_{\mathbf{X}}\left(\mathbf{t}\right)=c_{1}\frac{1}{2}\mathbf{t}C\left(D\right)\mathbf{t}+\frac{1}{2!}c_{2}\left[\frac{1}{2}\mathbf{t}C\left(D\right)\mathbf{t}\right]^{2}+\frac{1}{3!}c_{3}\left[\frac{1}{2}\mathbf{t}C\left(D\right)\mathbf{t}\right]^{3}+\ldots
\]
where the dependence on parameters $\left(\theta_{1},\theta_{2},\sigma_{0}^{2},\sigma_{1}^{2}\right)$
have been obviated to avoid cumbersome notation. This higher order
covariance function allows us to represent covariances in terms of
the distance separating the two sites in question, and higher order
(>2) joint cumulants in terms of distances among the sites involved
\emph{and} the coefficients $c_{r>2}$.

\subsubsection*{``Orthogonality'' in joint cumulants}

Joint cumulants of higher order do not affect lower ordered ones,
as follows from inspecting (\ref{eq:cumulants_archetypal}). After
fixing $\Sigma$, each \emph{r}-ordered joint cumulant depends on
a separate coefficient, $c_{\frac{k}{2}}$. Hence, one can have similar
joint cumulants up to a given order $K$, but then different coefficients
$c_{r>\frac{K}{2}}$ will lead to different joint cumulants of higher
order. This results in different association types that may go totally
unnoticed in the analysis of low dimensional marginal distributions,
such as 1 and 2-dimensional ones. Note that these marginal distributions
are all that is usually inspected to evaluate the goodness of fit
of a model, in current Spatial Statistics techniques. This topic is
explored in detail in section (\ref{sec:A-simulation-based-illustration}),
where two random fields are equal in terms of their one and second
order joint cumulants (i.e. mean and covariance structure), and in
terms of their one and two dimensional marginal distributions. Yet,
they exhibit very different clustering behaviors.

\section{\label{sec:Model-Implementation}Model Implementation in the context
of Spatial Statistics}

\subsection{Spatial Consistency}

An important issue when dealing with a probability distribution for
Spatial Data is that this distribution should be ``consistent''.
If we denote by $\mathbf{X}\in\mathbb{R}^{J}$ our modeling vector,
consistency means that any subvector $\left(X_{j_{1}},\ldots,X_{j_{K}}\right)\in\mathbb{R}^{K}$,
of $\mathbf{X}$, with $K\leq J$, will have the same type of distribution
distribution as $\mathbf{X}$. Equivalently, any extension of our
field to $J+1$ components must be such, that every sub-vector of
dimension $J$ has the original probability distribution. Gaussian
fields are of course of this type.

In order to be more specific, consider elliptically distributed vector
$\left(X_{1},\ldots,X_{J}\right)\in\mathbb{R}^{J}$ having a density
function. This density function can be written as 
\begin{equation}
\left\{ f\left(\left(X_{1},\ldots,X_{J}\right)\mid J\right)\mid J\in\mathbb{N}\right\} \label{eq:familia}
\end{equation}
where dependence on dimension $J$ has been made explicit. \citet{Kano1994139}
has given a definition that can be stated as follows: The family at
(\ref{eq:familia}) possesses the consistency property if and only
if 
\begin{equation}
\intop_{-\infty}^{+\infty}f\left(\left(x_{1},\ldots,x_{J+1}\right)\mid J+1\right)dx_{J+1}=f\left(\left(x_{1},\ldots,x_{J}\right)\mid J\right)
\end{equation}
for any $J\in\mathbb{N}$ and almost all $\left(x_{1},\ldots,x_{J}\right)\in\mathbb{R}^{J}$.
We also say that such a family is consistent. 

As \citet{Kano1994139} notes, not all members of the elliptical family
are consistent. He gives a necessary and sufficient condition for
a family such as (\ref{eq:familia}) to be consistent. The family
is consistent if and only if, for each $J\in\mathbb{N}$, random vector
$\mathbf{X}\in J$ can be stochastically written as 
\begin{equation}
\mathbf{X}=\sqrt{V}\times\mathbf{Z}\label{eq:Kano_dcomp}
\end{equation}
where $\mathbf{Z}\sim N_{J}\left(\mathbf{0},\Sigma\right)$ stands
for a normally distributed vector with the same covariance matrix
as $\mathbf{X}$, and $V>0$ is a univariate random variable independent
of $\mathbf{Z}$ and \emph{unrelated} to dimension $J$. 

By ``unrelated'' to $J$, we mean that the distribution of scaling
variable $V$ does not depend on $J$. This was part of the difficulty
of the elliptical family mentioned at the introduction of this paper:
the dependence of the distribution's generating variable on the dimension
$J$, making the family inconvenient for interpolation purposes, where
one must extend the field at least to $J+1$ locations (except for
the well known cases of the Gaussian and Student distributions). The
construction given by (\ref{eq:Kano_dcomp}), and the fact that $V$
is not related to $J$ is the key to circumvent this issue, for our
model.

As \citet{Kano1994139} reminds us, the construction at (\ref{eq:Kano_dcomp})
produces distributions with tails at least as heavy as the Normal
distribution, whereby Normal tail dependence (i.e. ``zero'' tail
dependence) can only be achieved for the case where $V$ is a positive
constant.

\subsection{Relation between $R^{2}$ and coefficients $c_{1},c_{2},c_{3},\ldots$ }

This relationship is important for estimation purposes. It also tells
us what kind random variable must be $V$ in order to ensure spatial
consistency of the model built as in (\ref{eq:Kano_dcomp}), i.e.
the model we advocate.

In  \ref{sec:Relation-between-moments-R-scaling}, it is shown
that if $\mathbf{X}\in\mathbb{R}^{J}$ has a c.g.f. as in (\ref{eq:archetypal_cgf}),
and consequently a m.g.f. as in (\ref{eq:archetype_mgf}), then the
following relation between the $k$-th order moments of its squared
generating variable, $R^{2}$, and coefficients $m_{1},m_{2},m_{3},\ldots$
exists: 
\begin{equation}
E\left(\left(R^{2}\right)^{k}\right)=\frac{m_{k}}{c_{1}^{k}}\frac{2^{k}\Gamma\left(k+\frac{J}{2}\right)}{\Gamma\left(\frac{J}{2}\right)}\label{eq:moments_of_R2_usefull-1}
\end{equation}

Where $\Gamma$ stands for the Gamma function. The expression is conveniently
expressed in terms of $m_{1},m_{2},m_{3},\ldots$, but it can be written
in terms of the $c_{r}$ coefficients by virtue of (\ref{eq:cums_in_terms_moms-2}),
\begin{eqnarray}
E\left(\left(R^{2}\right)^{1}\right) & = & \frac{c_{1}}{c_{1}}\frac{2^{1}\Gamma\left(1+\frac{J}{2}\right)}{\Gamma\left(\frac{J}{2}\right)}\\
E\left(\left(R^{2}\right)^{2}\right) & = & \frac{\left(c_{2}+c_{1}^{2}\right)}{c_{1}^{2}}\frac{2^{2}\Gamma\left(2+\frac{J}{2}\right)}{\Gamma\left(\frac{J}{2}\right)}\\
E\left(\left(R^{2}\right)^{3}\right) & = & \frac{\left(c_{3}+3c_{2}c_{1}+c_{1}^{3}\right)}{c_{1}^{3}}\frac{2^{3}\Gamma\left(3+\frac{J}{2}\right)}{\Gamma\left(\frac{J}{2}\right)}\\
\vdots & \vdots & \vdots
\end{eqnarray}
 and still further simplified by substituting 1 for $c_{1}$.

If we consider (\ref{eq:representation1}) and example \ref{ejamplo_Gaussian},
then construction (\ref{eq:Kano_dcomp}) indicates that the generating
variable of $\mathbf{X}$ can be represented as follows :
\begin{equation}
R\underset{d}{=}\sqrt{V\times\chi_{J}^{2}}
\end{equation}
and hence 
\begin{equation}
R^{2}\underset{d}{=}V\times\chi_{J}^{2}
\end{equation}
where $V$ and $\chi_{J}^{2}$ are independent (see item \emph{iii}
at theorem 1 of \citet{Kano1994139}). Due to this independence,
\begin{equation}
E\left(\left(R^{2}\right)^{k}\right)=E\left(V\right)\times E\left(\chi_{J}^{2}\right)
\end{equation}

Note that the moments of $\chi_{J}^{2}$ are given by 
\[
E\left(\left(\chi_{J}^{2}\right)^{k}\right)=\frac{2^{k}\Gamma\left(k+\frac{J}{2}\right)}{\Gamma\left(\frac{J}{2}\right)}
\]

Since $c_{1}=1$, equation (\ref{eq:moments_of_R2_usefull-1}) then
means that the moments of $V$ are given by $m_{1}=1,m_{2},m_{3},\ldots$,
whereas its cumulants are given by $c_{1}=1,c_{2},c_{3},\ldots$.
We have then identified a sufficient condition under which both the
expression at (\ref{eq:archetypal_cgf}) is a legitimate cumulant
generating function \emph{and} it produces a consistent model, useful
for spatial statistics: the coefficients $c_{1}=1,c_{2},c_{3},\ldots$
must be the cumulants of some random variable, $V>0$, whereas $m_{1}=1,m_{2},m_{3},\ldots$
must be its moments.
\paragraph{Remark} Note that a scaling variable $V>0$ having a very small variance,
$c_{2}$, will produce a random field very similar to a Gaussian random
field in its one and two dimensional marginal distributions (which
is all that current Geostatistical techniques fit and check for goodness
of fit). This is because the (common) kurtosis of each marginal distribution,
given by 
\[
\frac{\kappa_{4}\left(X_{j}\right)}{Var\left(X_{j}\right)^{2}}=\frac{\kappa^{j,j,j,j}}{Var\left(X_{j}\right)^{2}}=\frac{3c_{2}}{8Var\left(X_{j}\right)}
\]
will be very close to zero, as in the Normal model. But if coefficients
$c_{r>2}$ are relatively big, then (\ref{eq:cumulants_archetypal})
indicates that the joint cumulants of higher order, involving the
interaction of 4, 6 and more components of $\mathbf{X}$, will be
considerably altered. As the dimension of the field increases, important
characteristics of the field constructed via (\ref{eq:Kano_dcomp})
will be totally different from those of the Gaussian field (see example
below), though these differences \emph{will not be noticed} from the
one and two dimensional marginals. Additionally, conditional distributions
(i.e. at \textquotedbl{}ungauged sites\textquotedbl{}) will also be
different, particularly as the number of conditioning values increases.

\subsection{\label{sub:Parameter-Estimation}Parameter Estimation}

Apart from the estimation of covariance matrix $\Sigma$, estimation
of the model defined by (\ref{eq:archetypal_cgf}) amounts to estimating
coefficients $c_{2},c_{3},\ldots$, or equivalently, coefficients
$m_{2},m_{3},\ldots$. 

If we assign a flexible model to (squared) scaling variable $V$,
such as a mixture of gamma distributions, 
\begin{equation}
f_{V}\left(V\right)=\sum_{s=1}^{S}\pi_{s}\frac{\beta_{s}^{-\alpha_{s}}}{\Gamma\left(\alpha_{s}\right)}V^{\alpha_{s}-1}e^{-\frac{V}{\beta_{s}}}\label{eq:mixture_V}
\end{equation}
then parameter estimation for our model can be effected as follows: 
\begin{enumerate}
\item First, we estimate covariance matrix $\Sigma$, for which we may use
standard covariance function models, such as (\ref{eq:Power-exp_cov})
or (\ref{eq:Matern_cov}). We can do this in a first, independent
step, because of the \textquotedbl{}orthogonality\textquotedbl{} property
of the joint cumulants of $\mathbf{X}$ referred to in the remarks
of section \ref{sec:The-proposed-model}.
\item Second, we fit the density of $V$ conditioned on $E\left(V\right)=c_{1}=m_{1}=1$,
thus fitting the parameters present at density function (\ref{eq:mixture_V}).
One must impose some restrictions on these latter parameters, in order
to avoid lack of identifiability; we impose at the example below that
weights $\pi_{1},\ldots,\pi_{S-1}$ must be in decreasing order, whereas
$\pi_{S}:=1-\sum_{s=1}^{S-1}\pi_{s}$. 
\end{enumerate}
The parameters estimation at step 2 will be effected by computing
estimators $\hat{m}_{2},\hat{m}_{3},\ldots$ and then finding $\hat{\pi}_{1},\ldots,\hat{\pi}_{S-1},\hat{\beta}_{1},\ldots,\hat{\alpha}_{S}$,
such that 
\begin{equation}
\hat{m}_{k}\approx\sum_{s=1}^{S}\hat{\pi}_{s}\frac{\hat{\beta}_{s}^{k}\Gamma\left(\hat{\alpha}_{s}+k\right)}{\Gamma\left(\hat{\alpha}_{s}\right)}
\end{equation}
for $k=2,3,\ldots$, where the ``hat'' symbol can be read as ``estimator
of'' the parameter it covers. This is an instance of the method of
moments.

Note also that estimation at step 2 above does not alter in any manner
the already estimated covariance matrix, containing the joint cumulants
of order two. Step 2 is concerned with estimating coefficients, $c_{2},c_{3},c_{4},\ldots$,
affecting joint cumulants of higher orders, only. This is the reason
why our model can capitalize on the available low dimensional covariance
matrix estimation methods, via the covariance function. 

The estimation technique will be now explained in more detail. 

Assume one has a sample $\mathbf{x}_{1},\ldots,\mathbf{x}_{I}$ of
$\mathbf{X}\in\mathbb{R}^{J}$. This sample might represent, for example,
$I$ observations of the (spatially associated) residual process obtained
by applying a daily time series model to each of $J$ precipitation
gauging stations spread over sites with coordinates $\mathbf{s_{1}},\ldots,\mathbf{s_{J}}$,
$\mathbf{s_{j}}\in\mathbb{R}^{2}$. The fact that precipitation demands
a truncated model will be ignored for now, since this issue will be
briefly considered in section \ref{sec:Discussion-and-work}. Begin
by standardizing data, so that each component has mean zero. 

Perform the following estimating steps:
\begin{enumerate}
\item Apply any transformation to data that might be necessary (cf. \citet{sanso_venezuelan_1999}),
in order to make data approximately Gaussian in its one-dimensional
marginals.
\item Fit a multivariate Normal model to $\mathbf{X}$, on the basis of
$\mathbf{x_{1}},\ldots,\mathbf{x_{I}}$. A standard covariance model,
such as (\ref{eq:Power-exp_cov}), can be used to estimate covariance
structure of $\mathbf{X}$. The covariance between every two components
of $\mathbf{X}$ referred to locations $\mathbf{s_{j_{1}}}$ and $\mathbf{s_{j_{2}}}$,
are then estimated as a function of the distance between the locations
by $\hat{\Sigma}_{j_{1}j_{2}}=C\left(\left\Vert \mathbf{s_{j_{1}}}-\mathbf{s_{j_{2}}}\right\Vert \mid\hat{\theta}_{1},\hat{\theta}_{2},\hat{\sigma}_{0}^{2},\hat{\sigma}_{1}^{2}\right)$. 
\item Compute $r_{i}^{2}=\mathbf{x_{i}}\hat{\Sigma}^{-1}\mathbf{x_{i}}^{T}$,
for $i=1,\ldots,I$. These are approximate samples of $R^{2}$, the
squared generating variable of $\mathbf{X}$, as can be seen by an
argument similar to that presented in  \ref{sec:Relation-between-moments-R-scaling}.
\item Compute $\hat{\vartheta}_{k}=\frac{1}{I}\sum_{i=1}^{I}\left(r_{i}^{2}\right)^{k}$,
the estimates of the moments of squared generating variable $R^{2}$,
up to a prudent order, say $K=5$.
\item By virtue of (\ref{eq:moments_of_R2_usefull-1}) and remembering that
$c_{1}=m_{1}=1$, one has estimates for $m_{k}$, for $k=2,\ldots,K$,
given by
\begin{equation}
\hat{m}_{k}=\frac{\Gamma\left(\frac{J}{2}\right)}{2^{k}\Gamma\left(k+\frac{J}{2}\right)}\hat{\vartheta}_{k}
\end{equation}

\item Apply the method of moments to estimate the parameters of the density
of scaling variable $V$, which density is a mixture of $S$ gamma
densities. That is, solve the following minimization problem: 
\begin{equation}
\underset{\vec{\alpha},\vec{\beta},\vec{\pi}_{-S}.}{\min}\sum_{k=2}^{K}\left(\hat{m}_{k}-\sum_{s=1}^{S}\pi_{s}\frac{\beta_{s}^{k}\Gamma\left(\alpha_{s}+k\right)}{\Gamma\left(\alpha_{s}\right)}\right)^{2}\label{eq:method_of_moments}
\end{equation}
subject to 
\begin{eqnarray*}
\sum_{s=1}^{S}\pi_{s}\frac{\beta_{s}\Gamma\left(\alpha_{s}+1\right)}{\Gamma\left(\alpha_{s}\right)} & = & m_{1}=1\\
\pi_{s} & \geq & \pi_{s+1}\geq0\\
\sum_{s=1}^{S-1}\pi_{s} & \leq & 1
\end{eqnarray*}
where $\vec{\pi}_{-S}=\left(\pi_{1},\ldots,\pi_{S-1}\right)$, $\pi_{S}=1-\sum_{s=1}^{S}\pi_{s}$,
and the inequalities at the second constraint are valid for $1\leq s\leq S-2$. 
\end{enumerate}
For step 6 above, the Lagrange multipliers approach can be employed. 

As output to the procedure outlined by steps 1 through 6, one has
an estimation of the covariance model, and of the distribution of
the squared scaling variable, $V$. With these, simulation and interpolation
can be performed, as explained subsequently.

\paragraph{Remark} The representation of the density of scaling variable $V$ as a mixture
of gamma distributions, indicates that the model here presented can
approximate a wide spectrum of tail dependence association, which
includes that of the Normal and the Student-t distribution.

\section{\label{sec:Simulation,-interpolation-and}Simulation and interpolation }

\subsection{Random Simulation}

The decomposition (\ref{eq:Kano_dcomp}) can be conveniently used
both for simulation and for interpolation. 

In order to simulate a realization of vector $\mathbf{X}\in\mathbb{R}^{J}$:
\begin{enumerate}
\item Sample a realization $\mathbf{z}_{i}$ from a multivariate Normal
distribution with means vector $\mathbf{0}\in\mathbb{R}^{J}$ and
covariance matrix $\hat{\Sigma}$. 
\item Sample a realization $v_{i}$ of $V$. To this end sample an index,
$s\in\mathbb{N}$, from a multinomial distribution with class probabilities
$\left(\hat{\pi}_{1},\ldots,\hat{\pi}_{S}\right)$ and then sample
$v_{i}\sim Gamma\left(\hat{\alpha}_{s},\hat{\beta}_{s}\right)$.
\item The realization of $\mathbf{X}$ is given by $\mathbf{x}_{i}:=\sqrt{v_{i}}\times\mathbf{z}_{i}$.
Add a means vector, $\mu\in\mathbb{R}^{J}$, to $\mathbf{x}_{i}$,
if necessary.
\end{enumerate}
Note that a field of dimension $J^{*}\neq J$ can be simulated in
the same manner, since the distribution of $V$ does not depend on
$J$. Hence, one can simulate a big random field by obtaining (approximately)
a realization of a Gaussian random field using some fast generation
mechanism, such as turning bands (see, for example, \citet{ripley_spatial_1981}),
and then multiplying it by a realization of $\sqrt{V}$. This is done
for section \ref{sec:A-simulation-based-illustration}, and the consequences
on some manifestations of interaction, as compared to the original
Gaussian field, are there illustrated.

\subsection{\label{sub:Interpolation-to-ungauged}Interpolation to ungauged sites}

\subsubsection{Interpolation via the saddlepoint approximation }

The distribution of the environmental variable of interest at a new
location can be described with little additional inconvenience. This
is because we are building on the idea of the covariance function.
Hence, we can extend the covariance matrix $\Sigma$ to include the
covariance between the variable of interest at any gauged site and
any new location. Denote by $j_{1}$ any generic component of $\mathbf{X}$.
The correlation matrix components corresponding to site $\mathbf{s}_{j^{*}}$
are given by
\begin{equation}
\Sigma_{j^{*}j_{1}}=C\left(\left\Vert \mathbf{s}_{j_{1}}-\mathbf{s}_{j^{*}}\right\Vert \mid\hat{\theta}_{1},\hat{\theta}_{2},\hat{\sigma}_{0}^{2},\hat{\sigma}_{1}^{2}\right)
\end{equation}

For the subsequent discussion, we shall denote the extended covariance
matrix by $\Sigma^{*}\in\mathbb{R}^{J+1\times J+1}$.

Suppose that the distribution of the variable is desired for a new
location with coordinates $\mathbf{s}_{j^{*}}\in\mathbb{R}^{2}$,
given that one has observed a realization $\mathbf{x}$ of $\mathbf{X}$
at sites $\mathbf{s}_{1},\ldots,\mathbf{s}_{J}$. We present now a
method for obtaining the approximate distribution of $X_{j^{*}}=X\left(\mathbf{s}_{j^{*}}\right)$
given $\mathbf{x}\in\mathbb{R}^{J}$. 

Since the distribution of $\mathbf{X}$ is consistent, the cumulant
generating function of $\mathbf{Y}:=\left(X_{j^{*}},\mathbf{X}\right)$
is of the same form as that of $\mathbf{X}$ (cf. \citet{Kano1994139}).
This new c.g.f. can then be written as,
\begin{equation}
K_{\mathbf{Y}}\left(\mathbf{w}\right)=1\frac{1}{2}\mathbf{w^{T}\Sigma^{*}w}+\frac{1}{2!}\hat{c}_{2}\left[\frac{1}{2}\mathbf{w^{T}\Sigma^{*}w}\right]^{2}+\frac{1}{3!}\hat{c}_{3}\left[\frac{1}{2}\mathbf{w^{T}\Sigma^{*}w}\right]^{3}+\ldots\label{eq:archetypal_cgf_extended}
\end{equation}
where $\mathbf{w}\in\mathbb{R}^{J+1}$. As shown by \citet{skovgaard1987saddlepoint}
(see also: \citet{kolassa2006series,barndorff-nielsen_asymptotic_1990}),
we have that: 
\begin{equation}
\Pr\left(X_{j^{*}}\leq a\mid\mathbf{X}=\mathbf{x}\right)\approx\Phi\left(r\right)+\phi\left(r\right)\left(\frac{1}{r}-q\right)\label{eq:Skovgaard}
\end{equation}
where 
\begin{eqnarray}
r & = & sign\left(\hat{\mathbf{w}}_{1}\right)\sqrt{2\left\{ \hat{\mathbf{w}}^{T}\left(a,\mathbf{x}\right)-\hat{\mathbf{w}}_{-1}^{T}\mathbf{x}-K_{\mathbf{Y}}\left(\hat{\mathbf{w}}\right)+K_{\mathbf{X}}\left(\hat{\mathbf{w}}_{-1}\right)\right\} }\\
q & = & \frac{1}{\hat{\mathbf{w}}_{1}}\det\left(K_{\mathbf{X}}^{''}\left(\hat{\mathbf{w}}_{-1}\right)\right)\det\left(K_{\mathbf{Y}}^{''}\left(\text{\ensuremath{\hat{\mathbf{w}}}}\right)\right)^{-\frac{1}{2}}
\end{eqnarray}
and $\hat{\mathbf{w}}\in\mathbb{R}^{J+1}$, $\hat{\mathbf{w}}_{-1}\in\mathbb{R}^{J}$
are the solutions to equations 
\begin{eqnarray*}
\nabla K_{\mathbf{Y}}\left(\hat{\mathbf{w}}\right) & = & \left(a,\mathbf{x}\right)\\
\nabla K_{\mathbf{X}}\left(\hat{\mathbf{w}}_{-1}\right) & = & \mathbf{x}
\end{eqnarray*}

Additionally, $\hat{\mathbf{w}}_{1}$ is the first component of $\hat{\mathbf{w}}$,
and $K_{\mathbf{X}}^{''}\left(\text{\ensuremath{\hat{\mathbf{w}}}}\right)$
stands for the matrix of second derivatives on the c.g.f. evaluated
at $\hat{\mathbf{w}}$. 

We can apply this approximation to $\Pr\left(X_{j^{*}}\leq a\mid\mathbf{X}=\mathbf{x}\right)$
directly, using the extended c.g.f. given by (\ref{eq:archetypal_cgf_extended}).
This is done in section \ref{sub:Conditional-distributions-and-interpolation}
below.

In case one wishes the distribution of the environmental variable
at several new locations $\mathbf{s}_{j_{1}^{*}},\ldots,\mathbf{s}_{j_{K}^{*}}$
simultaneously,
\[
\Pr\left(X_{j_{1}^{*}}\leq a_{1},\ldots,X_{j_{K}^{*}}\leq a_{K}\mid\mathbf{X}=\mathbf{x}\right)
\]
one can apply the the extension to this approach presented by \citet{kolassa2010multivariate}.
A conceptually easier approach would be to run the Gibbs sampler repeatedly,
using (\ref{eq:Skovgaard}) to sample from each (approximate) full
conditional distribution (see \citet{kolassa1994approximate}). After
sufficiently many iterations, the samples obtained can be considered
approximate realizations of the conditional distribution desired.
However, a more efficient method for this task is given in the next
sub-section.

\subsubsection{Interpolation using the underlying Gaussian field}

The fact that our field $\mathbf{X}$ can be constructed as in equation
(\ref{eq:Kano_dcomp}) can also be used, jointly with the MCMC method,
to simulate conditional fields of arbitrary dimensions. Assume you
observe $\mathbf{x}\in\mathbb{R}^{J}$, which is a partial observation
of the whole field of interest, $\left(\mathbf{X},\mathbf{X}^{*}\right)\in\mathbb{R}^{J+M}$.
Here vector $\mathbf{X}^{*}\in\mathbb{R}^{M}$ comprises the values
of the random quantity at the $M$ ungauged sites. The construction
\[
\left(\mathbf{X},\mathbf{X}^{*}\right)=\left(\mathbf{\mu},\mu^{*}\right)+\sqrt{V}\times\left(\mathbf{Z},\mathbf{Z}^{*}\right)
\]
would constitute the complete field. The value of $\mu^{*}$ can be
found using the estimated ``drift'' function $\hat{\xi}$, as in
equation (\ref{eq:external_drift}). The covariance matrix for the
extended field $\left(\mathbf{Z},\mathbf{Z}^{*}\right)$ can be found
using the fitted covariance function. 

So, \emph{if we had} the value of $V$, our conditional simulation
method can proceed as follows. For $b=1,\ldots,B$, do:
\begin{enumerate}
\item Sample $\mathbf{z}^{*\left(b\right)}\in\mathbb{R}^{M}$ from the conditional
Gaussian vector $\mathbf{Z}^{*}\mid\text{\ensuremath{\mathbf{Z}}}=\mathbf{z}$,
with $\mathbf{z}=\left(\mathbf{x}-\mathbf{\mu}\right)/\sqrt{V}$.
\item Set $\mathbf{x}^{*\left(b\right)}\in\mathbb{R}^{M}$, the sought for
conditional vector, to $\mathbf{x}^{*\left(b\right)}:=\mu^{*}+\sqrt{V}\times\mathbf{z}^{*\left(b\right)}$.
\end{enumerate}
Since $V$ is not available, it can be considered a random variable
from which we have to sample. So, at each iteration $b$ above, we
shall have a realization $V^{\left(b\right)}$ instead of a single
value $V$. 

To sample from the distribution of $V$ given the already fitted $\mu$
and $\Sigma$, we use the Metropolis algorithm. For a given observed
$\mathbf{x}-\mu$, Bayes' theorem tells us that
\begin{equation}
p\left(V\mid\mathbf{x}-\mu\right)\propto p\left(\mathbf{x}-\mu\mid V\right)p\left(V\right)\label{eq:bayes_th}
\end{equation}
where we have used $p\left(*\right)$ as the respective densities,
in order to avoid cumbersome notation. Here, $p\left(V\right)=f_{V}\left(V\right)$
is the (fitted) distribution given by equation (\ref{eq:mixture_V}).
Since $\mathbf{x}-\mu=\sqrt{V}\times\mathbf{z}$, for $\mathbf{z}\sim N_{J}\left(\mathbf{0},\Sigma\right)$,
conditional density $p\left(\mathbf{x}-\mu\mid V\right)$ is just
$N_{J}\left(\mathbf{0},V\times\Sigma\right)$.

In  \ref{sec:MCMC-estimation-of} we show how we can obtain
samples from the conditional distribution of $V$ given a partial
observation of the field, $\mathbf{x}$, by using the Metropolis-Hastings
algorithm. This technique will be applied in section \ref{sec:A-glimpse-at}
in the conditional simulation of rainfall fields for June 1st 2013
over the Saalach river catchment.

\section{\label{sec:A-simulation-based-illustration}A simulation-based illustration}

In this section, we present a simulation study of the type of interdependence
that can be generated using a model having c.f.g. as (\ref{eq:archetypal_cgf}).
The study is built so as to mimic the model building process in Spatial
Statistics: from data obtained at a limited number of locations (\textquotedbl{}gauging
stations\textquotedbl{}), we want to infer a model for the interesting
variable over the whole region to which these locations belong.

It will be noted that the additional interdependence characteristics
the field possesses can be unnoticeable from the one and two dimensional
marginal distributions. In this example, they are indistinguishable
from those of a Gaussian field. However, specific characteristics
of the underlying field, which are relevant for applications, such
as rainfall modeling and mining geostatistics, will be considerably
different.

\subsection{\label{sub:Scaling-variable-used}Scaling variable used and simulated
fields employed}

We generated $n=3650$ realizations of a $J^{*}=300\times300$ Gaussian
field, using the circular embedding method as implemented in package
\emph{RandomFields} of the statistical software R. The covariance
function model used is the exponential one, given by setting $\theta_{2}=1$
at equation (\ref{eq:Power-exp_cov}). The specification of the field
is: $\mu=\mathbf{0}$, $\theta_{1}=20$, $\sigma_{0}^{2}=0$ and $\sigma_{1}^{2}=1$,
where $\mu$, $\theta_{1}$, $\sigma_{0}^{2}$ and $\sigma_{1}^{2}$
denote the field mean, the range parameter of the covariance function,
the nugget effect and the field variance, respectively. 

\begin{figure}
\begin{centering}
\includegraphics[scale=0.75]{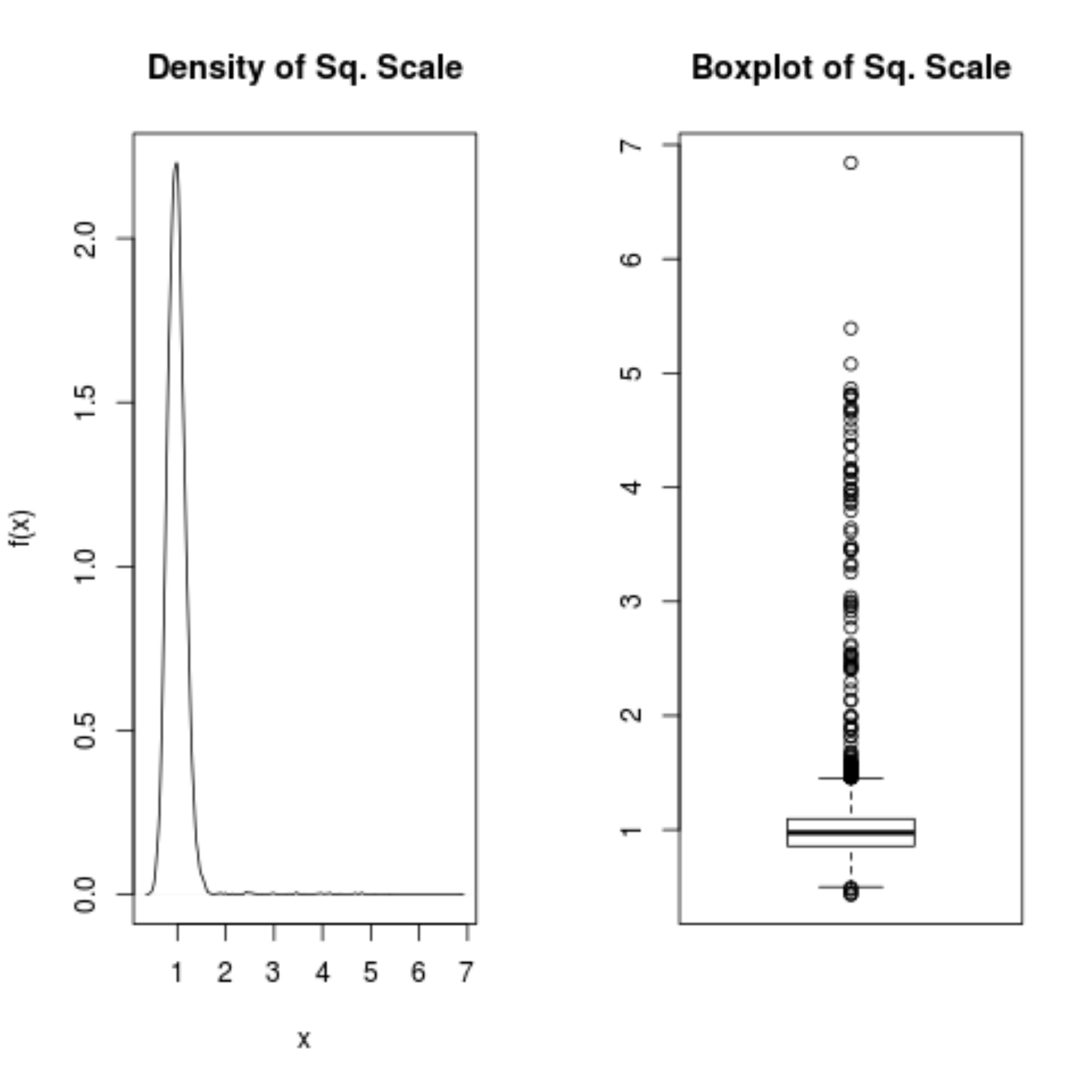}
\par\end{centering}

\caption{\label{fig:Density-and-simulation-based-V}Density and simulation-based
boxplot (n=10000) of the squared scaling variable, $V>0$, used for
the example in this section. This scaling variable helps to construct
fields that are very difficult to differentiate from Gaussian fields.}

\end{figure}

In order to apply (\ref{eq:Kano_dcomp}) we simulated 3650 realizations
of a mixture of 5 Gamma distributions, as in (\ref{eq:mixture_V}),
with the following parameters, rounded up to the fourth decimal place:
\begin{description}
\item [{Mixture~Weights}] $\vec{\pi}=\left(0.7137,0.1697,0.1094,<0.0000,<0.0000\right)$
\item [{Shape~Parameters}] $\left(\alpha_{1},\ldots,\alpha_{5}\right)=\left(32.5168,25.0004,27.4404,0.3582,11.3288\right)$
\item [{Scale~Parameters}] $\left(\beta_{1},\ldots,\beta_{5}\right)=\left(0.0302,0.0393,0.0357,0.6012,0.2975\right)$
\end{description}
This amounts to $V$ having moments $\left(m_{1},\ldots,m_{5}\right)=\left(0.9986,1.0766,1.3856,2.6163,8.0863\right)$,
and cumulants $\left(c_{1},\ldots,c_{5}\right)=\left(0.9986,0.0795,0.1519,0.5210,1.9712\right)$.
A plot of the density of $V$, together with a boxplot based on 10000
realizations, is presented at figure \ref{fig:Density-and-simulation-based-V}. 

The small second order cumulant of $V$, i.e. $c_{2}=0.0795$, will
produce only a very small kurtosis on the 1-dimensional marginal distribution
of the field, and a small 4-ordered joint cumulant on the 2-dimensional
marginals. This makes the field very difficult to differentiate from
a Guaussian field with equal covariance function; the parameters of
scaling variable, $V$, were selected precisely to produce that similarity
effect. Specifically, using equation (\ref{eq:cumulants_archetypal}),
we have
\begin{equation}
\kappa_{4}\left(X_{j}\right)=\frac{c_{2}}{8}\left(3\times\left(\sigma_{0}^{2}+\sigma_{1}^{2}\right)\right)=\frac{3c_{2}}{8}=0.0298
\end{equation}
for any 1-dimensional marginal distribution. And
\begin{equation}
\kappa^{j_{1},j_{1},j_{2},j_{2}}=\frac{c_{2}}{8}\left\{ \left(\sigma_{0}^{2}+\sigma_{1}^{2}\right)^{2}+2\times cov\left(X_{j_{1}},X_{j_{2}}\right)\right\} <0.0298
\end{equation}
for any 2-dimensional marginal.

In spite of this apparent similarity, some realizations from the original
Gaussian field will be very different from those of the Non-Gaussian
field built as in equation (\ref{eq:Kano_dcomp}). 

The non-Gaussian field $\mathbf{X}\in\mathbb{R}^{90000}$ will be
the multiplication of scaling variable $V$ depicted in figure \ref{fig:Density-and-simulation-based-V}
\emph{times} a Gaussian field $\mathbf{Z}\in\mathbb{R}^{90000}$.
The mean a covariance structure of both fields is the same, but some
realizations of field $\mathbf{X}$ will be realizations of a Gaussian
field \emph{times} values of the magnitude of $\sqrt{V}=\sqrt{7}\approx2.65$
(i.e. the maximum value displayed at figure \ref{fig:Density-and-simulation-based-V}). 

From figure \ref{fig:Density-and-simulation-based-V}, we see that
there is a non-negligible probability of getting $V>4$, which implies
that many realizations of $\mathbf{Z}$ will be multiplied by values
$\sqrt{V}>2$ to produce realizations of $\mathbf{X}$. Again, this
goes unnoticed in the one and two dimensional marginal distributions. 

In the modeling of atmospheric processes, such as rainfall, $\sqrt{V}>2$
might represent the presence of some large scale atmospheric process
triggering rainfall within a day of an intensity not expected in a
century (See the final illustration, in connection to the European
floods of May-June of 2013). We currently explore this modeling possibility. 

The behavior of scaling variable $V$ influences the tail behavior
of the resulting vector $\mathbf{X}$. A typical representation of
the multivariate Student distribution with correlation matrix $\Sigma$
and $\nu$ degrees of freedom is (see \citet{kotz2004multivariate}):
\[
\mathbf{X}=\sqrt{Q}\mathbf{Z}
\]
where $\mathbf{Z}$ is a normally distributed vector with vector of
means $\mathbf{0}$ and correlation matrix $\Sigma$, and 
\[
Q\sim\frac{\nu}{\chi_{\nu}^{2}}
\]

\begin{figure}
\begin{centering}
\includegraphics[width=0.5\textwidth]{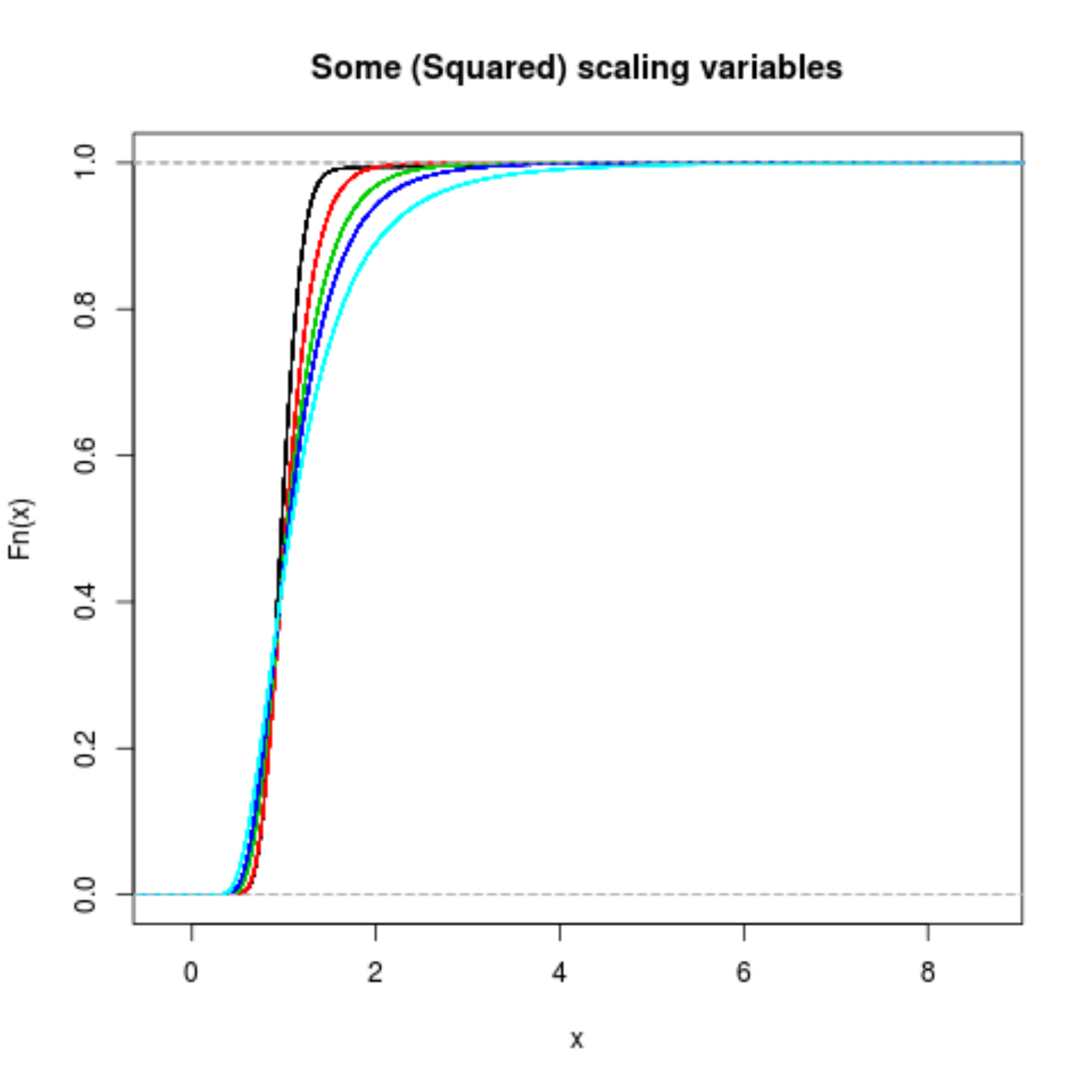}\includegraphics[width=0.5\textwidth]{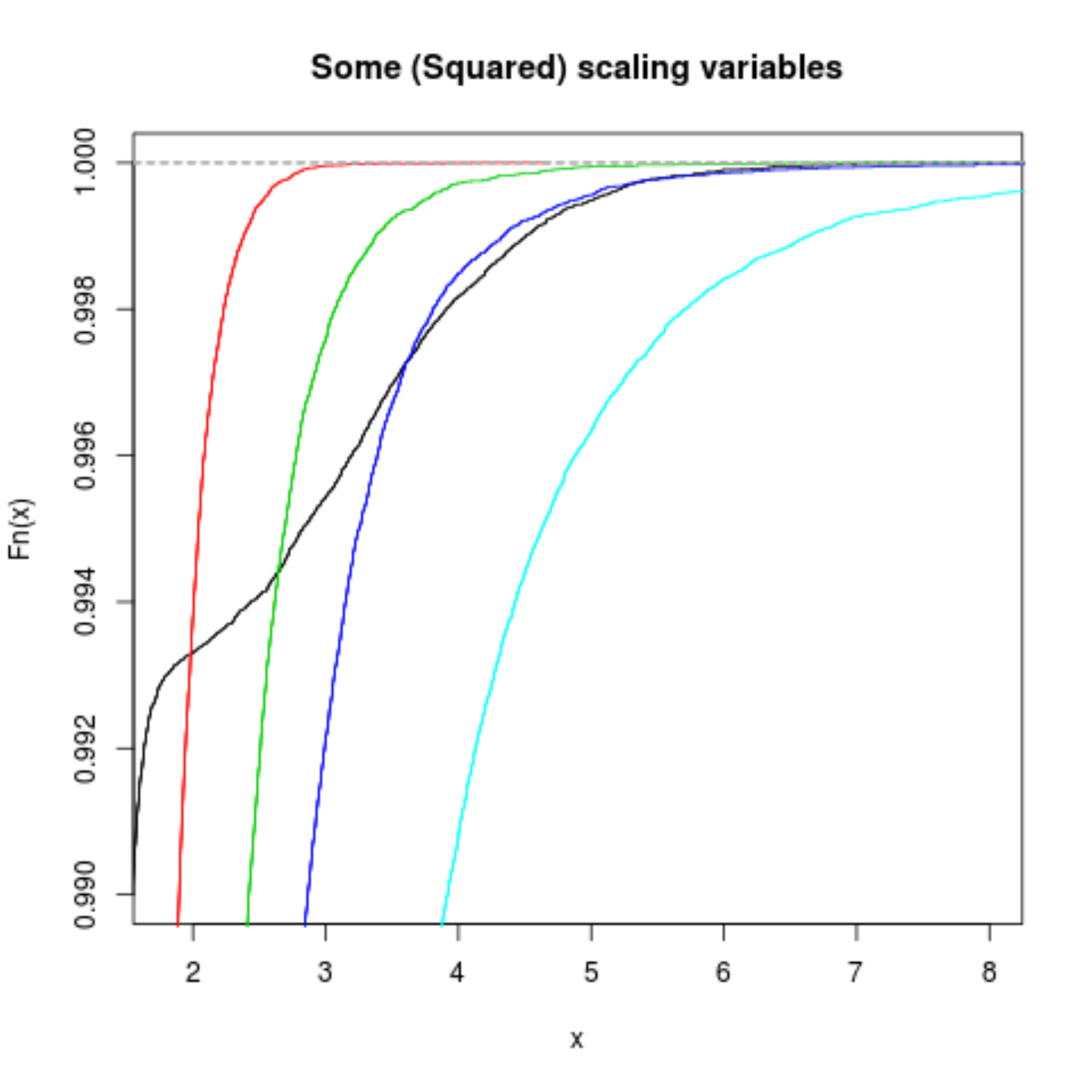}
\par\end{centering}

\begin{centering}
\caption{\label{fig:Comparison-of-the-scalings}Comparison of the distribution
of squared scaling variable $V$ (black) with distributions of the
scaling variable of the multivariate Student distribution for degrees
of freedom: 10 (light blue), 15 (dark blue), 20 (green), 35 (red).
The uppermost part of the distribution of $V$ produces a tail behaviour
similar to that of a multivariate Student distribution with 15 degrees
of freedom. }

\par\end{centering}

\end{figure}

Hence we can compare the distribution of squared scaling variable
$V$, presented at figure \ref{fig:Density-and-simulation-based-V}
with the distribution of a multivariate Student distribution, for
various degrees of freedom. The distributions of the squared scaling
variables are presented at figure \ref{fig:Comparison-of-the-scalings},
for a multivariate Student distribution with $\nu\in\left\{ 10,15,20,35\right\} $
degrees of freedom.

It is noteworthy that scaling variable $V$ seems to have the lightest
tail, if you focus on the left hand panel of figure \ref{fig:Comparison-of-the-scalings}.
However, the uppermost part of the distribution of $V$ is similar
to that of $\frac{\nu}{\chi_{\nu}^{2}}$ with $\nu=15$. That is,
the tail dependence of our model is actually similar to that of the
multivariate Student distribution with $\nu=15$ degrees of freedom.
This fact goes completely unnoticed in the 1 and 2-dimensinal marginal
distributions, as we shall show.

\subsection{Partial observation of the fields: A network of 30 stations}

\begin{figure}
\begin{centering}
\includegraphics[width=0.75\textwidth]{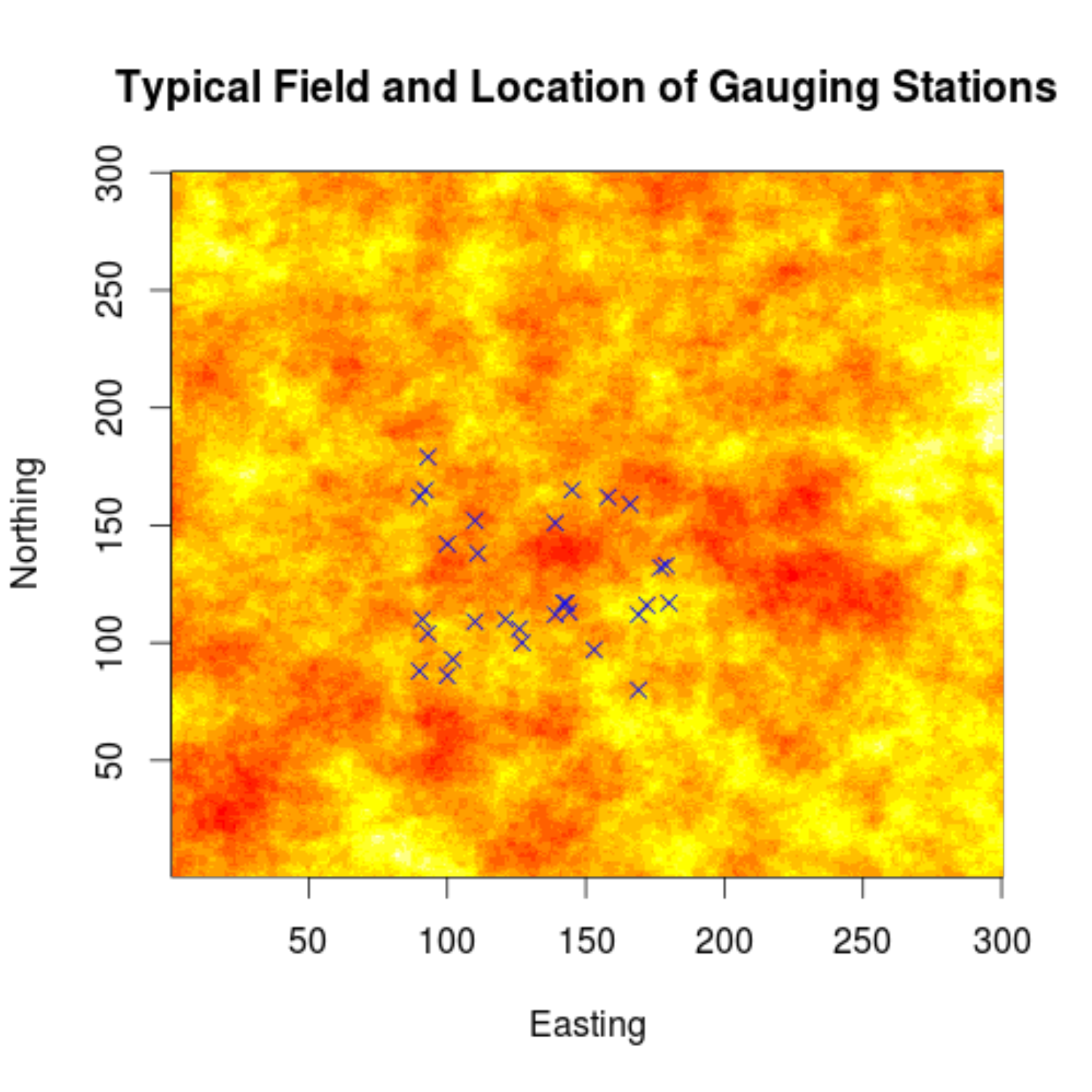}\caption{\label{fig:Typical-(non-Gaussian)-field}Typical (non-Gaussian) field
of the n=3650 generated, and the 30 locations at which data was recorded
to form $\mathbf{X}$ and $\mathbf{Z}$.}

\par\end{centering}

\end{figure}

Let us denote by $\mathbf{Z}^{*}\in\mathbb{R}^{J^{*}}$ and $\mathbf{X}^{*}\in\mathbb{R}^{J^{*}}$
the random fields generated as a Gaussian field, and by multiplication
of the latter by $\sqrt{V}$, respectively. In this example, $J^{*}=300\times300=90000$.
We selected 30 components of the field, corresponding in a Spatial
context to 30 locations on the plane, and stored the data of these
components. The setting is illustrated in figure \ref{fig:Typical-(non-Gaussian)-field}.

The n=3650, 30-dimensional observations thus available from field
$\mathbf{Z}^{*}$ are in the following considered as realizations
from sub-vector $\mathbf{Z}\in\mathbb{R}^{30}$ of $\mathbf{Z}^{*}$,
whereas those from field $\mathbf{X}^{*}$ are considered as realizations
of sub-vector $\mathbf{X}\in\mathbb{R}^{30}$.

Data from these vectors, $\mathbf{Z}$ and $\mathbf{X}$, represent
the data available at a limited number of gauging stations. As usual
in Spatial Statistics, we intend to identify characteristics of the
whole fields, $\mathbf{Z}^{*}$ and $\mathbf{X}^{*}$, on the basis
of the partial observations provided by $\mathbf{Z}$ and $\mathbf{X}$. 

A third vector dealt with in this section is $\mathbf{W}\in\mathbb{R}^{30}$,
of which each component is given by
\begin{equation}
W_{j}=F_{Z_{j}}^{-1}\left(F_{X_{j}}\left(X_{j}\right)\right)
\end{equation}
that is, $\mathbf{W}$ is the vector resulting from applying the quantile-quantile
transformation to each component of $\mathbf{X}$, mapping these into
the quantiles of the components of $\mathbf{Z}$. Hence, each marginal
distribution of $\mathbf{W}$ is exactly standard normal, like those
of $\mathbf{Z}$, but the joint distribution of its ranks (the copula),
is like that of $\mathbf{X}$. 

\subsection{Invisibility of differences for standard Spatial Statistics diagnostics,
and how to avoid this problem}

A detailed analysis of the one and two dimensional marginal distributions
(and copulas) of $\mathbf{X}$, $\mathbf{Z}$ and $\mathbf{W}$ on
the basis of the data simulated is performed at  \ref{sec:Similarity-of-one-and-two-dimensional}. 

The analysis implies that all three random vectors can be modeled
with a Gaussian distribution. In terms of current Spatial Statistics
techniques, this means that the three fields, from which data collected
are partial observations, can be safely modeled as a Gaussian field,
with the same mean and covariance function parameters. This is of
course wrong, but is all we can say if we just focus on one and two
dimensional distributions.

In  \ref{sec:Analysis-of-Aggregating-statistics}, we present
some aggregating statistics which can be employed to discriminate
between the complete fields $\mathbf{X}^{*}\in\mathbb{R}^{90000}$
and $\mathbf{Z}^{*}\in\mathbb{R}^{90000}$, on the basis of the whole
30-dimensional data-sets available (not just its 2-dimensional marginals).
We suggest that these statistics should be considered for model validation,
in addition to statistics for one and two dimensional marginal distributions
(including the covariance function). This will help to avoid missing
important characteristics of data, which can have important implications
for the inferred complete field, as seen in the following. 

We have relegated these topics to the mentioned appendixes, in order
to improve the readability of this paper.

\subsection{\label{sub:Discrepancies-in-the-big-fields}Applications-relevant
discrepancies in the underlying fields}

The object of this section and of section (\ref{sub:Conditional-distributions-and-interpolation})
is to show what kind of inference about the complete fields can go
wrong and unnoticed, if one does not pay attention to the discrepancies
pointed out by the aggregating statistics shown in  \ref{sec:Analysis-of-Aggregating-statistics}.
Please keep in mind that, according to the analysis of  \ref{sec:Similarity-of-one-and-two-dimensional},
the three fields,$\mathbf{Z}^{*}$, $\mathbf{X}^{*}$ and $\mathbf{W}^{*}$,
can be modeled by one and the same Gaussian model.

We focus on characteristics of the whole underlying fields, relevant
for hydrological applications, in this section. In section \ref{sub:Conditional-distributions-and-interpolation}
we deal with conditional distributions, more relevant for mining geostatistics.

We present two aggregating statistics of the complete fields, $\mathbf{Z}^{*}$,
$\mathbf{X}^{*}$ and $\mathbf{W}^{*}$, portrayed in figure \ref{fig:Typical-(non-Gaussian)-field}.
We indicate the kind of global statistics that would go totally unnoticed,
if we were to check only the one and two dimensional marginal distributions
of the data available, and fit a Gaussian field for the whole geographical
region.

\subsubsection*{Sums of positive values of the whole field}

The first interaction manifestation we investigate is the distribution
of
\begin{equation}
S_{\mathbf{X}^{*}}^{+}=\sum_{j=1}^{J=90000}\max\left(X_{j},0\right)\label{eq:manifestations_all_field_sum}
\end{equation}
that is, the sum of positive values of the whole field, $\mathbf{X}^{*}$.
Similarly, we define $S_{\mathbf{Z}^{*}}^{+}$ and $S_{\mathbf{W}^{*}}^{+}$
for fields $\mathbf{Z}^{*}$ and $\mathbf{W}^{*}$, respectively. 

The distributions of the sums, $S_{\mathbf{X}^{*}}^{+}$, $S_{\mathbf{Z}^{*}}^{+}$
and $S_{\mathbf{W}^{*}}^{+}$, are investigated in terms of their
sample quantiles. These are important statistics for rainfall modeling
over a basin, for example, since a value proportional to this sum
must find its way through the outlet of the basin, possibly causing
a flood.

\begin{figure}
\begin{centering}
\includegraphics[width=0.75\textwidth]{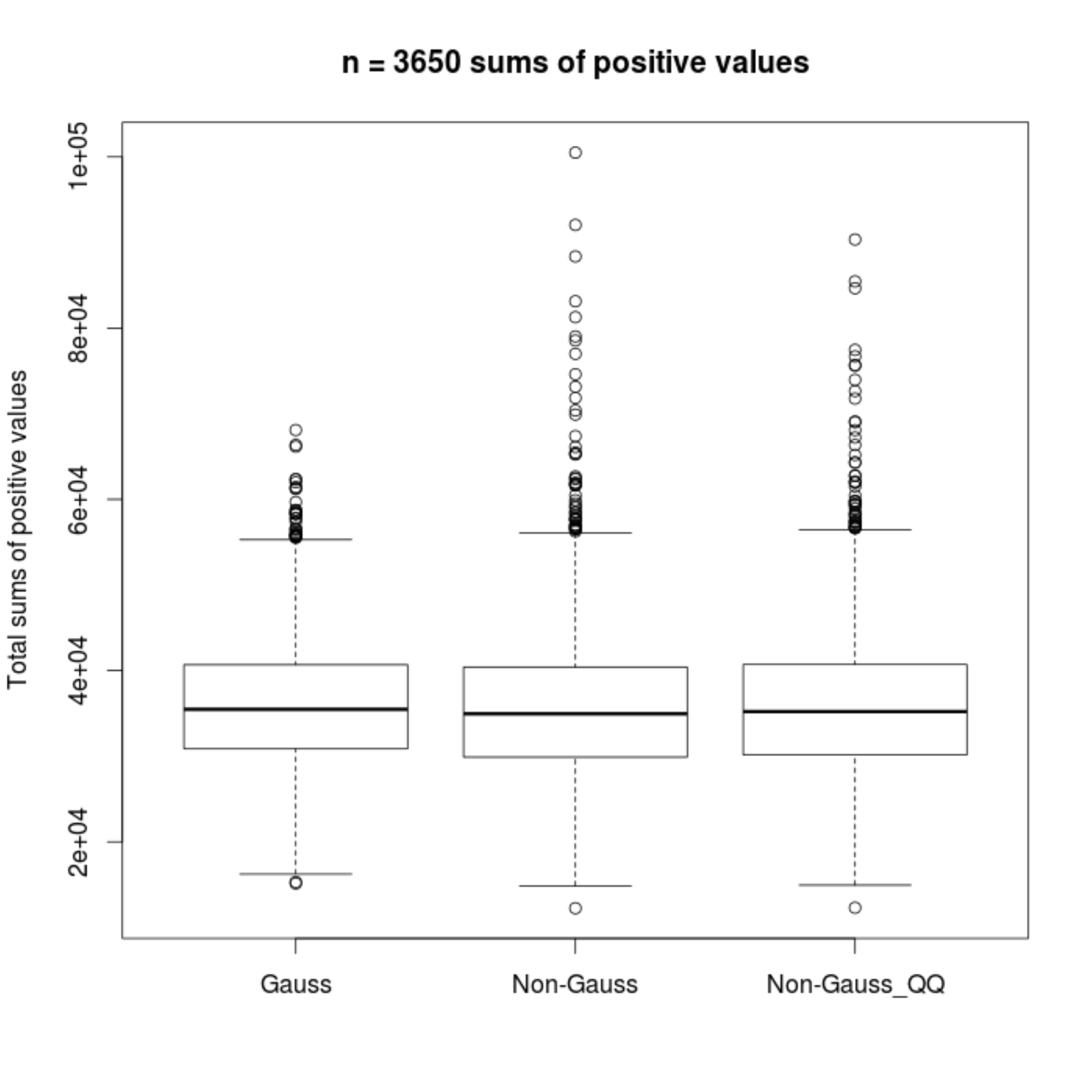}
\par\end{centering}

\caption{\label{fig:From-left-to-right_boxplots_fields_sums}Boxplots of sums
of positive values for \emph{n} realizations of $\mathbf{Z}^{*}$,
$\mathbf{X}^{*}$ and $\mathbf{W}^{*}$, with $n=3650$.}
\end{figure}

Boxplots illustrating the distribution of the sum of positive values
are given in figure \ref{fig:From-left-to-right_boxplots_fields_sums},
whereas a table with some important sample quantiles of $S_{\mathbf{Z}^{*}}^{+}$,
$S_{\mathbf{X}^{*}}^{+}$ and $S_{\mathbf{W}^{*}}^{+}$ are given
in table \ref{tab:From-left-to-quantiles_sums}. Notice that the sample
quantiles of $S_{\mathbf{Z}^{*}}^{+}$ begin to deviate from those
of $S_{\mathbf{X}^{*}}^{+}$ and $S_{\mathbf{W}^{*}}^{+}$ from the
99\% quantile on. The relative percentage increase of the quantiles
of $S_{\mathbf{X}^{*}}^{+}$ and $S_{\mathbf{W}^{*}}^{+}$ with respect
to those of $S_{\mathbf{Z}^{*}}^{+}$ are given within parentheses
in table \ref{tab:From-left-to-quantiles_sums}. 

\begin{table}
\begin{centering}
\begin{tabular}{|c||c||c||c|}
\hline 
Quantile & Gauss & Non-Gauss & Non-Gauss\_QQ\tabularnewline
\hline 
\hline 
80\% & 41901.96 & 41992.56 (+0.22\%) & 42288.41 (+0.92\%)\tabularnewline
\hline 
\hline 
90\% & 45342.97 & 46334.13 (+2.19\%) & 46576.23 (+2.72\%)\tabularnewline
\hline 
\hline 
95\% & 48609.21 & 49678.96 (+2.20\%) & 49910.30 (+2.68\%)\tabularnewline
\hline 
\hline 
99\% & 54906.90 & 57634.93 (+4.97\%) & 57583.49 (+4.87\%)\tabularnewline
\hline 
\hline 
99.5\% & 57660.03 & 62627.25 (+8.61\%) & 62794.43 (+8.90\%)\tabularnewline
\hline 
\hline 
99.9\% & 62331.09 & 81939.63 (+31.46\%) & 76972.03 (+23.49\%)\tabularnewline
\hline 
\hline 
100\% & 68099.17 & 100503.12 (+47.58\%) & 90353.53 (+32.68\%)\tabularnewline
\hline 
\end{tabular}
\par\end{centering}

\caption{\label{tab:From-left-to-quantiles_sums}From left to right: Sample
quantiles ($n=3650$) for $S_{\mathbf{Z}^{*}}^{+}$, $S_{\mathbf{X}^{*}}^{+}$
and $S_{\mathbf{W}^{*}}^{+}$. Percentages within parentheses indicate
percentage increase with respect to data from the Gaussian field.}
\end{table}

The sample size $n=3650$ would amount to a 10-year period, if data
were to represent some daily measured variable. If the field $\mathbf{X}^{*}$
were to represent daily rainfall over a catchment, the maximum total
rainfall would be 47.58\% higher than one would expect by fitting
a Gaussian model with adequate one and two dimensional marginal distributions
and covariance function. By letting the simulation run up to $n=10000$
(roughly thirty years data), the increase in the maximum sum ascends
to 61.11\% for $\mathbf{X}^{*}$ and to $50.36\%$ for $\mathbf{W}^{*}$,
as compared to the maximum sum produced by field $\mathbf{Z}^{*}$.
These possibilities are completely missed by an analysis based on
one and two dimensional marginal distributions, and the field's covariance
function.

\subsubsection*{Number of components of the whole field trespassing a given threshold}

A second interaction manifestation we shall investigate for the complete
fields, is the distribution of the number of components trespassing
a given threshold. Analogously to (\ref{eq:definicion_componentes_umbral}),
we define for $\mathbf{X}^{*}\in\mathbb{R}^{J^{*}}$, 
\[
L^{\mathbf{X}^{*}}=\sum_{j=1}^{J^{*}}1\left\{ X_{j}^{*}>a\right\} 
\]
where $J^{*}=300\times300$, and 
\begin{equation}
1\left\{ X_{j}^{*}>a\right\} =\begin{cases}
1, & X_{j}^{*}>a\\
0, & X_{j}^{*}\leq a
\end{cases}\label{eq:definicion_componentes_umbral-1}
\end{equation}

\begin{figure}
\begin{centering}
\includegraphics[width=1\textwidth]{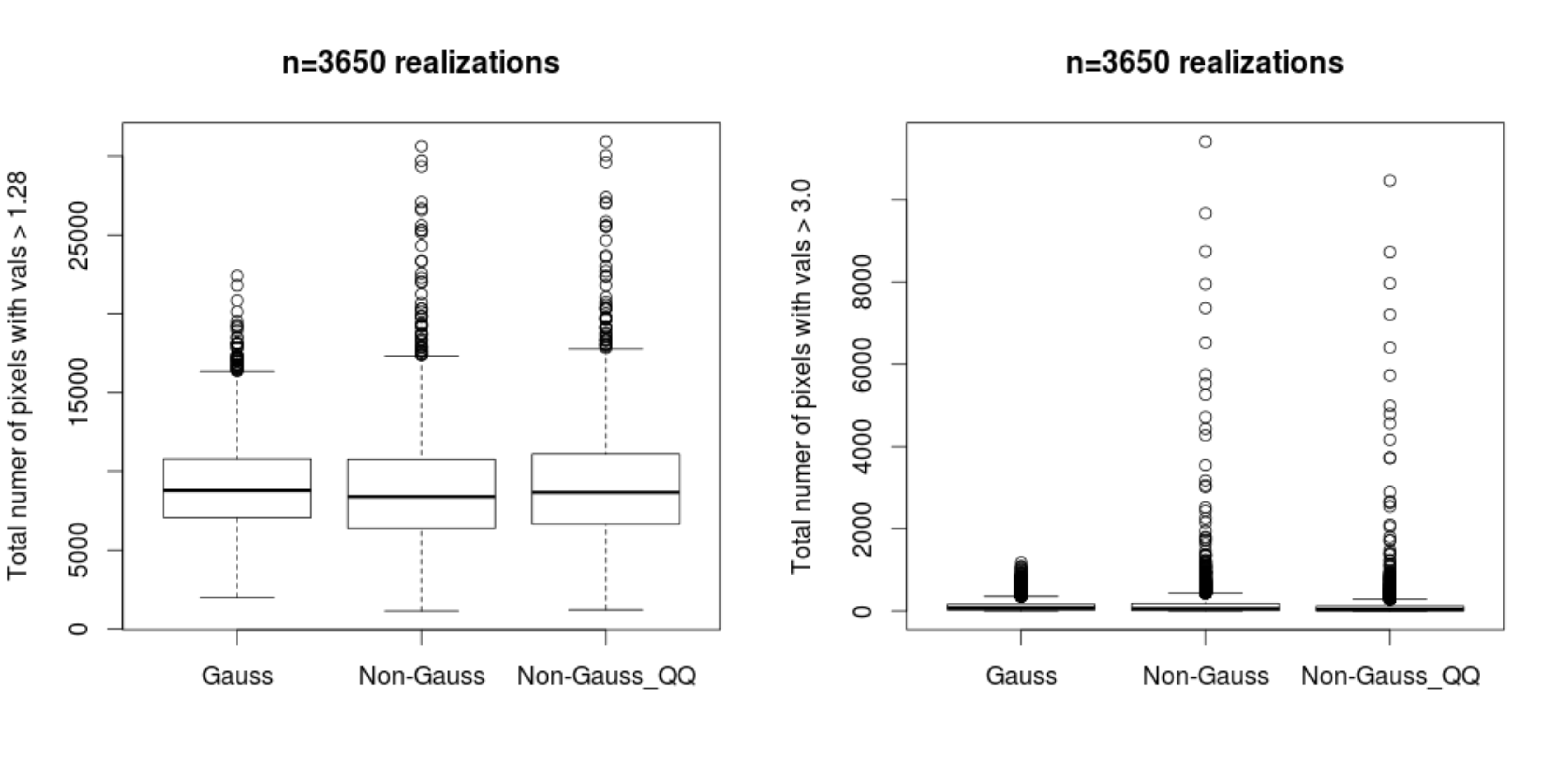}
\par\end{centering}

\caption{\label{fig:Boxplots-of-the-128-25}Boxplots of the number of components
above thresholds 1.28 (left) and 3.0 (right). The two non-gaussian
fields are very different from the gaussian one with respect to this
interaction manifestation. The difference is exacerbated as the threshold
is pulled upwards. More extensive areas with very high values are
to be expected for the non-gaussian fields.}
\end{figure}

In the context of spatial statistics, $L^{\mathbf{X}^{*}}$ can be
interpreted as the total area over which the environmental variable
of interest realizes \textquotedbl{}extreme\textquotedbl{} values.
Similar constructions define $L^{\mathbf{Z}^{*}}$ and $L^{\mathbf{W}^{*}}$.
We have a total of $n=3650$ samples from each of these three random
variables, which are plotted at figure \ref{fig:Boxplots-of-the-128-25}
for thresholds 1.28 (left) and 2.5 (right). 

\begin{figure}
\begin{centering}
\includegraphics[width=0.5\textwidth]{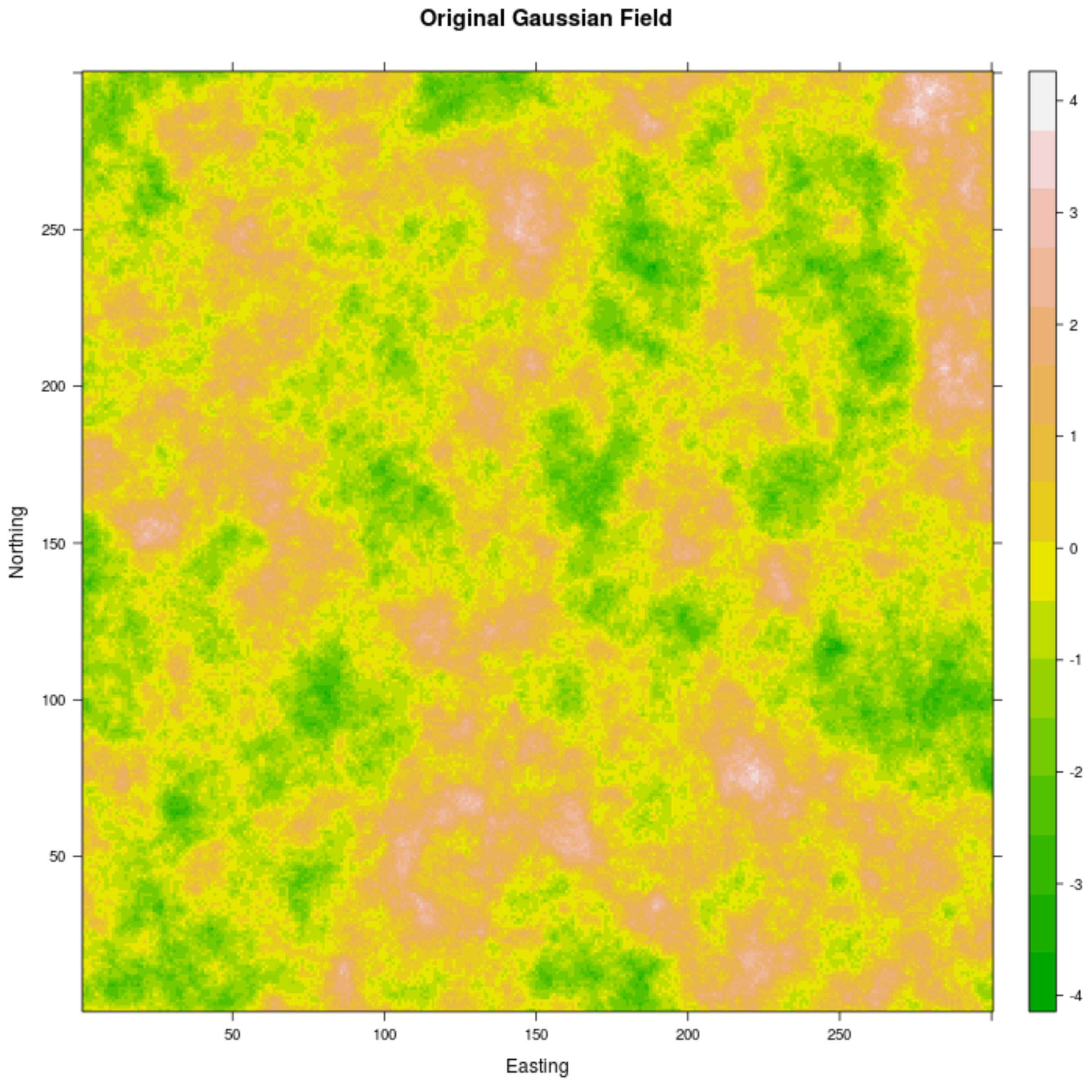}\includegraphics[width=0.5\textwidth]{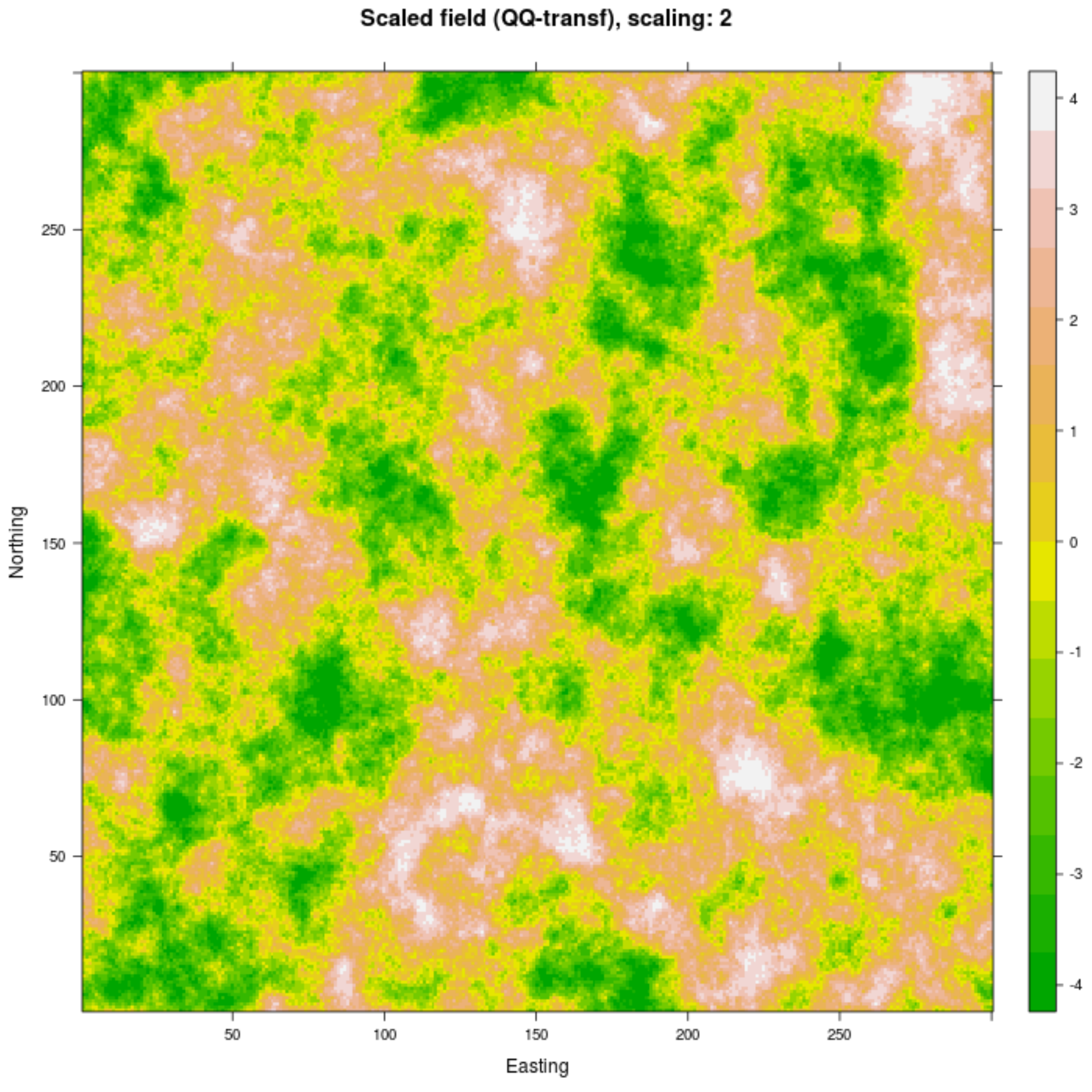}
\par\end{centering}

\caption{\label{fig:Different-clustering-characteris}Different clustering
characteristics between one realization of the gaussian (left) and
non-gaussian (right) fields $\mathbf{Z}^{*}$ and $\mathbf{W}^{*}$.
Scaling originally corresponding to non-gaussian field is $\sqrt{V}=2$.
Field $\mathbf{W}^{*}$ can exhibit big clusters of values around
4, even though marginally and in terms of the covariance function
it has the same specification as $\mathbf{Z}^{*}$. }

\end{figure}

Notice the great difference between the samples of $L^{\mathbf{Z}^{*}}$
and $L^{\mathbf{W}^{*}}$ (labeled \textquotedbl{}Gauss\textquotedbl{}
and \textquotedbl{}Non-Gauss\_QQ\textquotedbl{}, respectively) when
we use 2.5 as threshold. This occurs even though marginally fields
$\mathbf{Z}^{*}$ and $\mathbf{W}^{*}$ have exactly the same distribution,
and the covariance function of both fields is the same. In more practical
terms, the difference in this variable amounts to $\mathbf{Z}^{*}$
and $\mathbf{W}^{*}$ having very different types of clusters of very
high values, as illustrated in figure \ref{fig:Different-clustering-characteris}.
Field $\mathbf{W}^{*}$ can exhibit much bigger clusters of values
above 4 (99.99683\% quantile of its marginal distribution), even though
marginally and in terms of its covariance function it has the same
specification as $\mathbf{Z}^{*}$.

\subsection{\label{sub:Conditional-distributions-and-interpolation}Conditional
distributions and interpolation}

The conditional distribution of the random quantity at a new location,
given a partial observation of the field will now be analyzed, by
using the approximation given at equation (\ref{eq:Skovgaard}). This
is an important type of distribution in mining geostatistics, where
inference on the random quantity investigated is necessary at unbored
locations. We shall illustrate the type of discrepancy between the
conditional distribution arising from a Gaussian model, as compared
with that of the model given by equation  (\ref{eq:archetypal_cgf}).
To this end, we focus on a realization of a sub-vector, $\mathbf{X}_{I}$,
of $\mathbf{X}$. The set of indexes (i.e. locations) considered is
$\left\{ 3,28,19,16,25,9,21\right\} $.

Vector 
\begin{equation}
\mathbf{z}_{I}=\left(-1.489,-0.626,-0.050,0.068,0.491,0.832,-0.666\right)\label{eq:realization_z_for_conditional}
\end{equation}
constitutes the first realization of $\mathbf{Z}_{I}=\left(Z_{3},Z_{28},Z_{19},Z_{16},Z_{25},Z_{9},Z_{21}\right)$
of the random field shown at figure \ref{fig:Different-clustering-characteris},
left panel. Due to the mechanism depicted by equation (\ref{eq:Kano_dcomp}),
one can have immediately a realization, $\mathbf{x}_{I}$, of $\mathbf{X}_{I}=\left(X_{3},\ldots,X_{21}\right)$
by multiplying $\mathbf{Z}_{I}$ by probable values of scaling variable
$\sqrt{V}$. 

For our subsequent analysis we employ the following values as realizations
of $\sqrt{V}$: 0.64, 1 and 2; hence obtaining three different realizations
of $\mathbf{X}_{I}$. This will help us to understand why the two
conditional fields of section \ref{sec:A-glimpse-at} are so different:
they correspond to a value of around $\sqrt{V}\approx\sqrt{6}$, as
inferred from the data available.

We analyze the distribution of $Z_{3}$ and of $X_{3}$, conditioned
on an increasing number of components of the vector. Such conditioning
values are given by multiplying (\ref{eq:realization_z_for_conditional})
times 0.64, 1 and 2. We plot percentiles: 80\%, 90\%, 95\%, 99\%,
99.5\%, 99.9\%, 99.99\% and 99.999\%.

\begin{figure}
\begin{centering}
\includegraphics[width=1\textwidth]{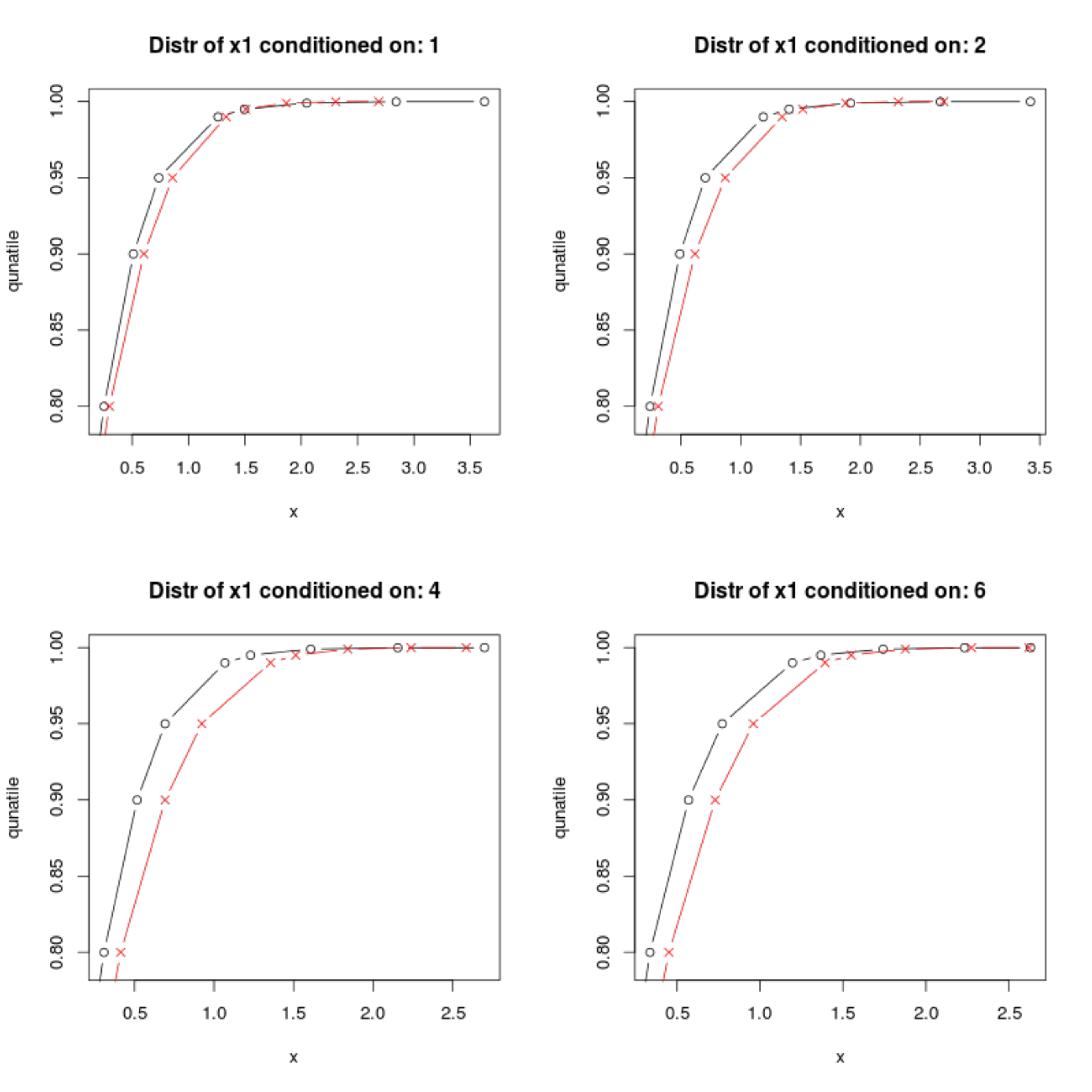}
\par\end{centering}

\caption{\label{fig:conditional_small_x}From left to right and downwards:
Comparison of upper parts of conditional distributions for $Z_{3}$
(red) and $X_{3}$ (black), for $n=1,2,4,6$ conditioning values.
Realization of $\mathbf{X}_{I}$ is given by $0.64\times\mathbf{z}_{I}$
(small scaling variable).}
\end{figure}

In figure \ref{fig:conditional_small_x} we show the conditional distributions
using $\sqrt{v}=0.64$ for $\mathbf{X}_{I}$. A moderate increase
in the discrepancy between the conditional distributions is seen as
the number of conditioning values increases, while the tail of the
non-gaussian distribution becomes lighter and lighter as compared
with that of the conditional Gaussian one.

\begin{figure}
\begin{centering}
\includegraphics[width=1\textwidth]{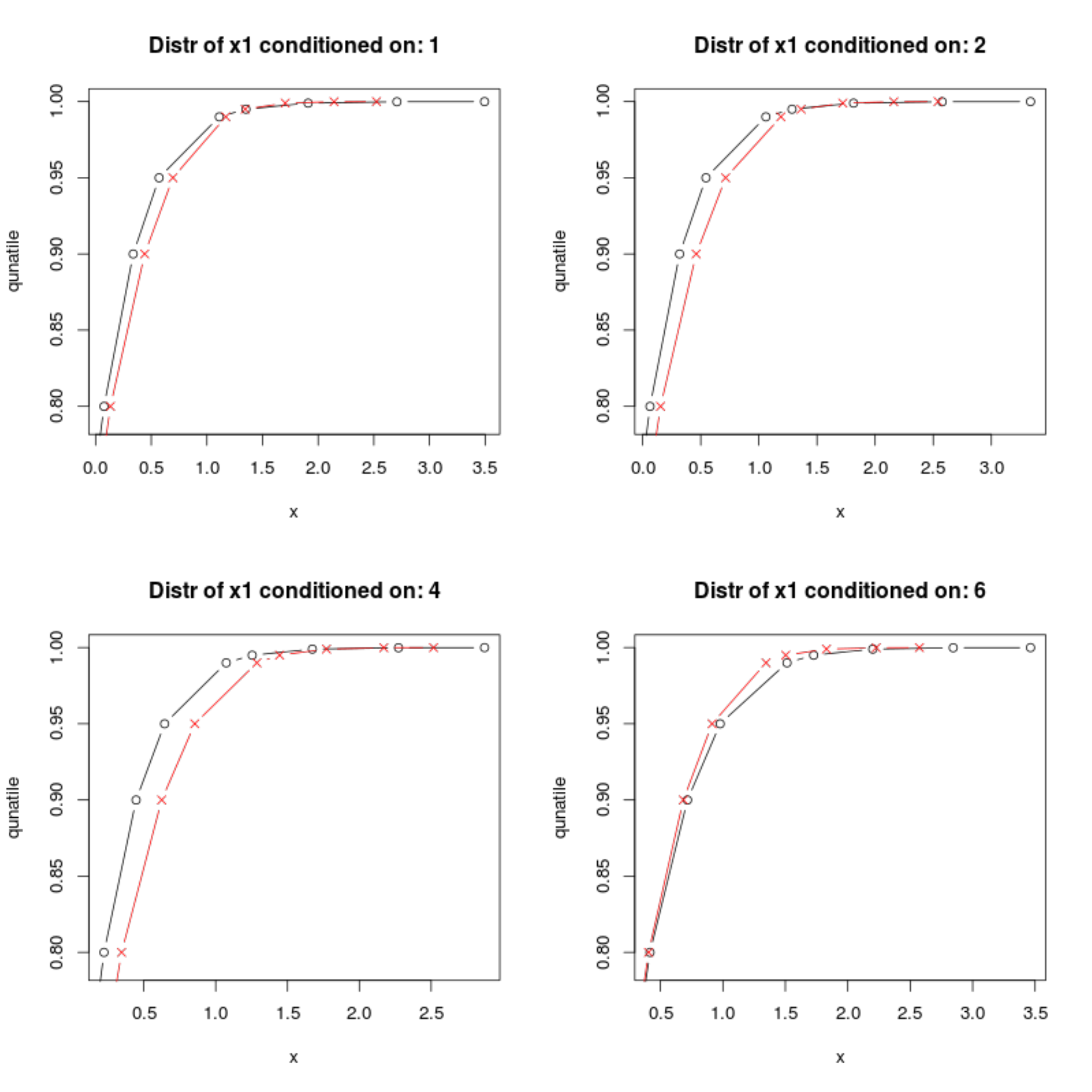}
\par\end{centering}

\caption{\label{fig:conditional_middle_x}From left to right and downwards:
Comparison of upper parts of conditional distributions for $Z_{3}$
(red) and $X_{3}$ (black), for $n=1,2,4,6$ conditioning values.
Realization of $\mathbf{X}_{I}$ is given by $1\times\mathbf{z}_{I}$
(middle-valued scaling variable).}
\end{figure}

In figure \ref{fig:conditional_middle_x} we show the conditional
distributions using $\sqrt{v}=1$ for $\mathbf{X}_{I}$. Note that
the non-gaussian conditional distribution keeps its similarity to
the gaussian conditional, though it has higher quantiles for all conditioning
schemes. 

In figure \ref{fig:conditional_small_x} we show the conditional distributions
using $\sqrt{v}=2$ for $\mathbf{X}_{I}$. The conditional distribution
given one conditioning value is very similar for both models, but
this situation quickly changes, as more conditioning values are considered.
The quantiles of the upper part of the conditional distribution for
$X_{3}$ become sensibly bigger already for 2 conditioning values.
Note that the values one may expect for the conditioned (\textquotedbl{}ungauged\textquotedbl{})
variable are considerably greater for $X_{3}$ than for $Z_{3}$.
This is a relevant issue for applications.

\begin{figure}
\begin{centering}
\includegraphics[width=1\textwidth]{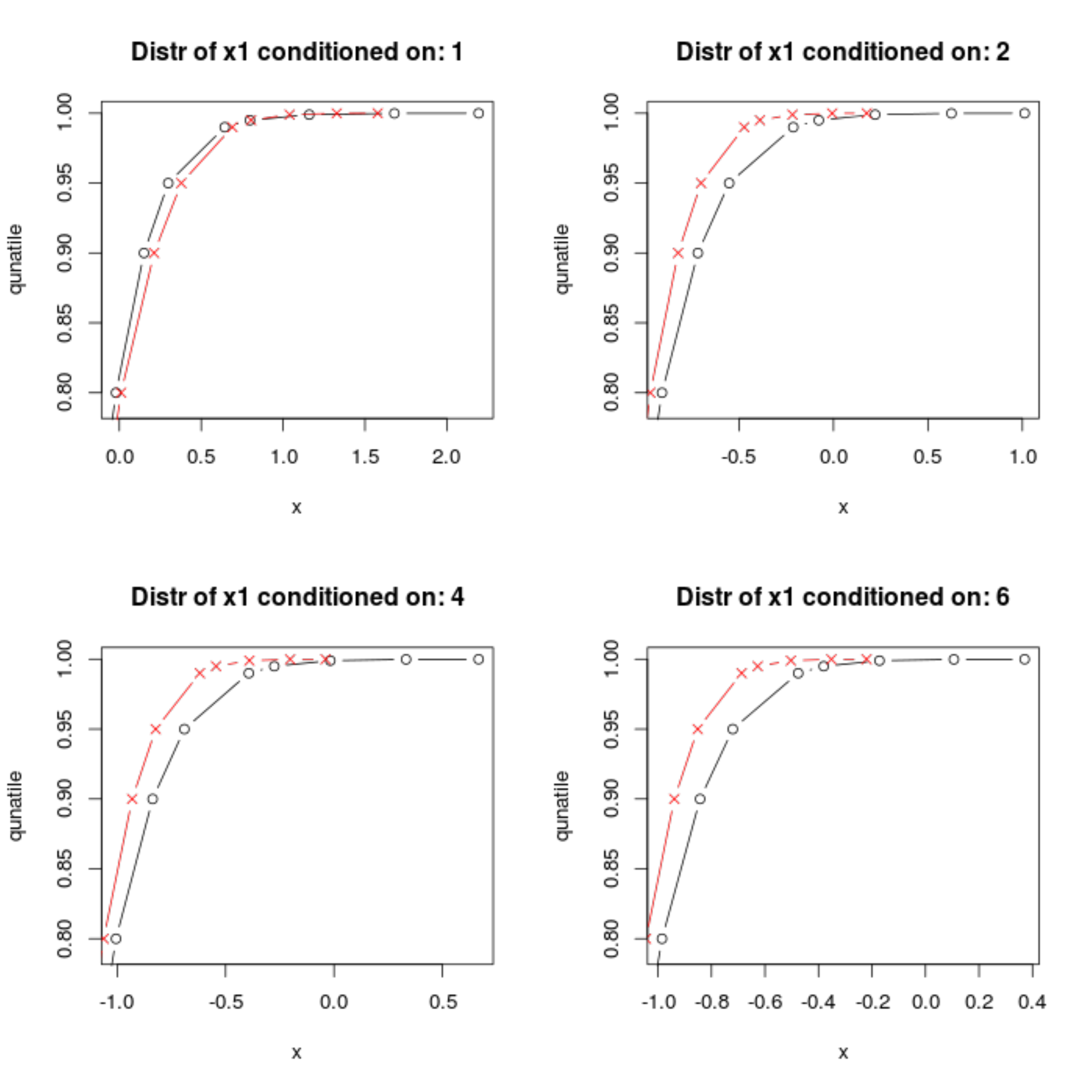}
\par\end{centering}

\caption{\label{fig:conditional_big_x}From left to right and downwards: Comparison
of upper parts of conditional distributions for $Z_{3}$ (red) and
$X_{3}$ (black), for $n=1,2,4,6$ conditioning values. Realization
of $\mathbf{X}_{I}$ is given by $2\times\mathbf{z}_{I}$ (high-valued
scaling variable).}
\end{figure}

\subsection{Estimated parameters}

We employ now the simple technique given in section \ref{sub:Parameter-Estimation}
to estimate the additional parameters, $c_{2},\ldots,c_{5}$, corresponding
to the cumulants of a non-degenerate (i.e. non-constant) scaling variable
$V$. The estimated cumulants and moments of squared scaling variable
$V$ are presented in table \ref{tab:Coefficients-estimated-by}. 

\begin{figure}
\begin{centering}
\includegraphics[width=0.75\textwidth]{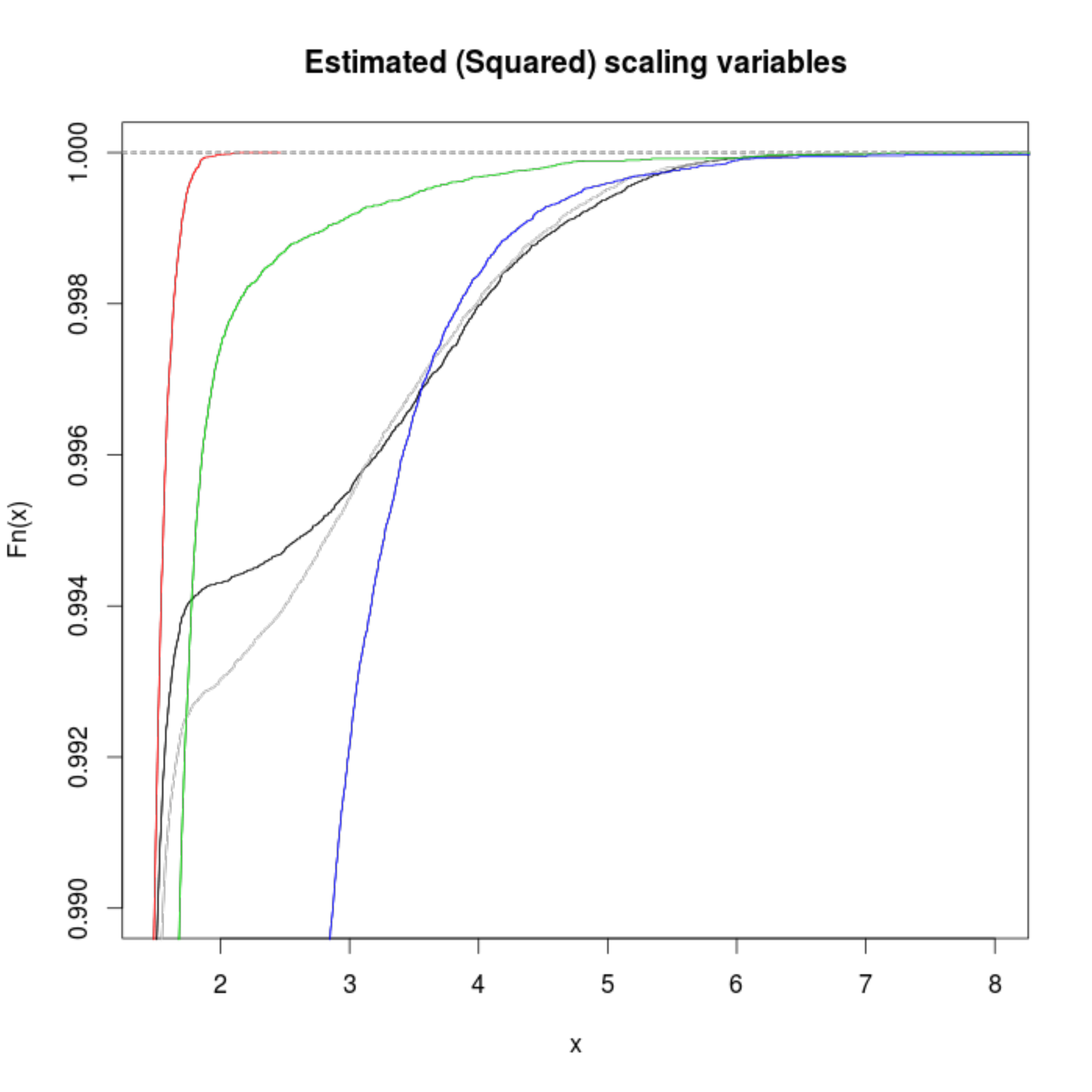}
\par\end{centering}

\caption{\label{fig:Estimated-squared-scaling}Estimated squared scaling variables,
uppermost part: for $\mathbf{X}$ (black), $\mathbf{Z}$ (red), $\mathbf{W}$
(green), and a Student with 15 degrees of freedom (blue). The squared
scaling variable originally employed for the simulation case study
is shown in gray. The method of moments estimation was successful
in capturing the uppermost behavior of the scaling variable.}

\end{figure}

Using the method of moments, and the n=3650 data values, we fitted
a mixture of 5 gamma distributions to each of the series of moments
shown in table \ref{tab:Coefficients-estimated-by}. The resulting
distributions, together with the distribution of the original squared
scaling variable $V>0$ are shown and compared in figure \ref{fig:Estimated-squared-scaling}.
The distribution of the scaling variable of a student multivariate
distribution with 15 degrees of freedom, shown in blue, has been added
for comparison.

\begin{table}
\begin{centering}
\begin{tabular}{|c||c||c||c|}
\hline 
Estimated & Gauss & Non-Gauss & Non-Gauss\_QQ\tabularnewline
\hline 
\hline 
m.1 & 1 & 1 & 1\tabularnewline
\hline 
\hline 
m.2 & 0.9998 & 1.0792 & 1.0547\tabularnewline
\hline 
\hline 
m.3 & 0.9988 & 1.4098 & 1.2378\tabularnewline
\hline 
\hline 
m.4 & 0.9958 & 2.8072 & 1.8129\tabularnewline
\hline 
\hline 
m.5 & 0.9897 & 9.0952 & 3.6831\tabularnewline
\hline 
\hline 
c.1 & 1 & 1 & 1\tabularnewline
\hline 
\hline 
c.2 & -2e-04 & 0.0792 & 0.0547\tabularnewline
\hline 
\hline 
c.3 & -7e-04 & 0.1722 & 0.0738\tabularnewline
\hline 
\hline 
c.4 & -3e-04 & 0.6244 & 0.1806\tabularnewline
\hline 
\hline 
c.5 & 2e-04 & 2.2289 & 0.4100\tabularnewline
\hline 
\end{tabular}
\par\end{centering}

\caption{\label{tab:Coefficients-estimated-by}Coefficients estimated by the
method of moments, rounded up to four decimal places.}

\end{table}

We notice that the scaling variable is approximately recovered by
this technique. However this technique cannot be used, for example,
in the context of rainfall modeling, where data is constrained to
be positive. Even though a latent variable approach (cf. \citet{sanso_venezuelan_1999})
can be employed for fitting the best Gaussian model to data (step
1 of estimation), the step effecting the estimation of additional
parameters $c_{2},\ldots,c_{5}$ which determine important interaction
manifestations of the field, cannot be executed via the method of
moments: the latent imputed data correspond to a Normal distribution
and hence does not produce valid realizations of squared generating
variable, $R^{2}$.

A paper describing an alternative estimation method, which circumvents
this difficulty, is already in preparation. For now, we present in
a real context, that of the May-June of 2013 extreme rainfall over
central Europe, what kind of inference may be unrealistic, if one
validates one's model only on the basis of one a two dimensional marginal
distributions.

\section{\label{sec:A-glimpse-at}A glimpse at the June 2013 extreme central
Europe rainfall events}

We show in this section the implications of fitting a model that considers
interactions beyond correlations, for modeling rainfall. We shall
see that the probability of extreme rainfall over a whole catchment
increases dramatically, even though, again, this is not noticed on
the 1 and 2-dimensional marginal validation of the model. This has
implications for forecasting and risk assessment.

\begin{figure}
\centering{}\includegraphics[width=0.75\textwidth]{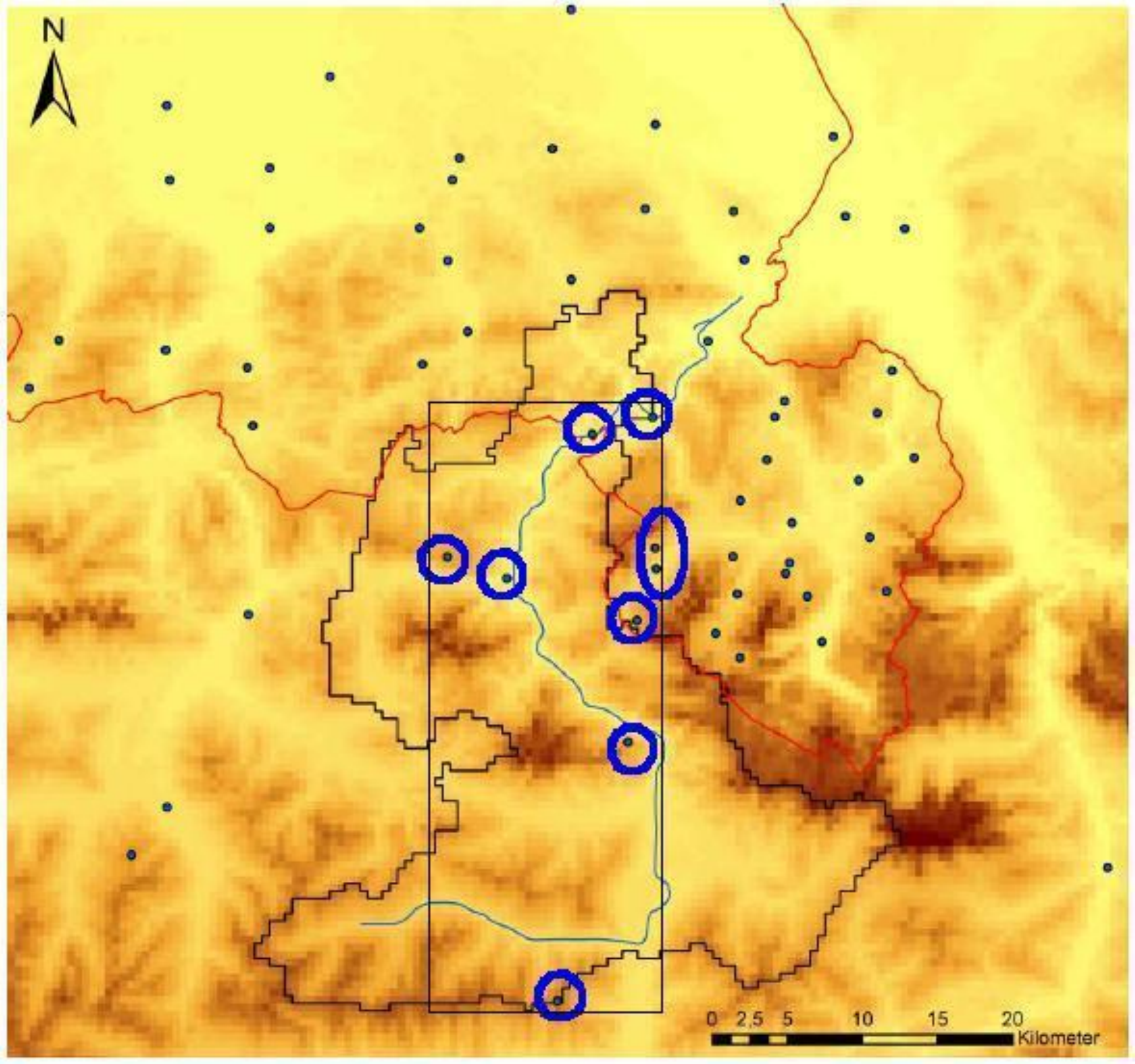}\caption{\label{fig:Catchment-of-the-Saalach}Catchment of the Saalach river.
The rainfall gauging stations used for the analysis are shown encircled.}
\end{figure}

At figure \ref{fig:Catchment-of-the-Saalach} we show a map of the
Saalach river catchment, in southeast Germany. The catchment is relatively
small, with an area of ca. 1043 $km^{2}$. Darker colors indicate
higher elevations. The points plotted represent the locations of gauging
stations recording total daily precipitation. The superimposed rectangle
indicates the area to which our subsequent conditionally simulated
fields refer. 

We selected nine stations for our analysis, which are encircled in
the map, since most of the stations are outside the catchment. However,
these nine stations will suffice to make clear our argument about
the need to consider multivariate interactions in Spatial Statistics
modeling, for example, through the model we propose in this paper. 

We fit to the daily data record from the 1st of January 2004 till
the 31th of December 2009 a space-time model very similar to the one
proposed by \citet{sanso_venezuelan_1999}. The following analysis
constitutes by no means an attack on that model; we could have selected
any other model which uses latent Gaussian fields, for example, that
of \citet{WRCR:WRCR13288}. The model selected is convenient because
it can easily accommodate missing data, of which we have some in our
record. Estimation is performed in a Bayesian framework.

At the core of the model of \citet{sanso_venezuelan_1999} lies a
latent Gaussian field, providing the spatial structure of the modeled
rainfall field. We present in this section the implications of replacing
this latent Gaussian field by a non-Gaussian one, built as in equation
(\ref{eq:Kano_dcomp}), but which is indistinguishable from a Gaussian
field in its one a two dimensional marginals, as studied in the 
\ref{sec:Similarity-of-one-and-two-dimensional}.

On a given day, $t=1,\ldots,2192$, the data of the nine gauging stations
are represented by vector $\mathbf{Y}_{t}=\left(Y_{t,1},\ldots,Y_{t,9}\right)$.
We decompose the generic data vector into 
\[
\mathbf{Y}_{t}=\left(Y_{t,obs},Y_{t,zero},Y_{t,miss}\right)
\]
where the components $Y_{t,obs}$, $Y_{t,zero}$ and $Y_{t,miss}$
represent the non-zero observed precipitation part, the no-precipitation
observed part, and the missing part of vector $\mathbf{Y}_{t}$, respectively. 

To avoid the problem posed by the vector parts $Y_{t,zero}$ and $Y_{t,miss}$,
we use an ``extended data'' approach, whereby these parts are replaced
by random vectors $W_{t}$ and $U_{t}$, respectively. All components
of $W_{t}$ are constrained to be negative, whereas those of $U_{t}$
are not constrained.

Hence our typical data vector is given by

\[
\mathbf{Y}_{t}=\left(Y_{t,obs},W_{t},U_{t}\right)
\]

These vectors $W_{t}$ and $U_{t}$, for $t=1,\ldots,2192$, are then
considered unknown parameters, and the posterior distributions found
as part of the MCMC output. 

Consider a time-evolving Gaussian field, $\mathbf{Z}_{t}$, connected
to $\mathbf{Y}_{t}$ by the transformation $T^{-1}:\mathbf{Y}_{t}\rightarrow\mathbf{Z}_{t}$,
with
\begin{equation}
\left(\begin{array}{c}
Y_{t,obs}\\
W_{t}\\
U_{t}
\end{array}\right)\underset{T^{-1}}{\rightarrow}\left(\begin{array}{c}
Y_{t,obs}^{1/\beta_{m\left(t\right)}}\\
W_{t}\\
U_{t}
\end{array}\right):=\mathbf{Z}_{t}\label{eq:transformacion}
\end{equation}
where $\beta_{m\left(t\right)}$ is a positive real number, and $m\left(t\right):t\rightarrow\left\{ 1,\ldots,12\right\} $
is a function mapping $t$ to its corresponding month of the year.
Furthermore,
\begin{equation}
\mathbf{Z}_{t}\sim N_{9}\left(\mu_{t},\sigma_{m\left(t\right)}^{2}\Sigma\right)
\end{equation}
where, for $j=1,\ldots,9$, we have
\begin{equation}
\mu_{t,j}=\alpha_{0}+\alpha_{1}h_{j}+\gamma_{m\left(t\right)}
\end{equation}
with $h_{j}$ standing for elevation above sea level (i.e. an ``external
drift'') at the location of station $j$, and $\gamma_{m\left(t\right)}$
represents a monthly temporal effect. Function $m\left(t\right)$
is as before. This temporal effect is modeled by three harmonics,
\begin{equation}
\gamma_{m\left(t\right)}=\sum_{r=1}^{3}\left\{ A_{r}\cos\left(\frac{2\pi r}{12}m\left(t\right)\right)+B_{r}\sin\left(\frac{2\pi r}{12}m\left(t\right)\right)\right\} 
\end{equation}
to allow for variability within the year's cycle. 

Correlation matrix, $\Sigma\in\mathbb{R}^{9\times9}$, is assumed
to follow an exponential correlation function, 
\begin{equation}
\Sigma_{j_{1},j_{2}}=\exp\left(-\lambda\left\Vert \mathbf{s}_{j_{1}}-\mathbf{s}_{j_{2}}\right\Vert \right)
\end{equation}
where $\lambda>0$ is an unknown scale parameter, $\mathbf{s}_{j_{1}}$
and $\mathbf{s}_{j_{2}}$ stand for the locations on $\mathbb{R}^{2}$
of stations $j_{1}$ and $j_{2}$, and the symbol $\left\Vert *\right\Vert $
represents the Euclidean distance. The (spatially) common variance
$\sigma_{m\left(t\right)}^{2}$ is allowed to change with the month
on which $t$ falls. 

We assume flat prior distributions on all parameters, and a priori
independence among the parameters, so that
\begin{equation}
p\left(\alpha_{0},\alpha_{1},A_{1},\ldots,B_{3},\lambda,\sigma_{1}^{2},\ldots,\sigma_{12}^{2},W_{1},\ldots,U_{2192}\right)\propto1_{\left\{ \lambda>0,W_{t}<0\right\} }
\end{equation}
where $1_{\left\{ A\right\} }$ is the indicator variable for the
event $\left\{ A\right\} $. We refer to all parameters collectively
as $\Phi$. 

That the issues of zero valued observations and missing data have
been conveniently solved, can be seen from the relative simplicity
of the resulting likelihood function of the extended data, on which
our inference is based. Defining $J_{t}$ as the set of indexes of
$Y_{t,obs}$ for each $t=1,\ldots,2192:=T$, one has:

\begin{multline}
L_{data}\left(\Phi\right)\propto\frac{\prod_{t=1}^{T}\left(\prod_{J_{t}}\frac{1}{\beta_{m\left(t\right)}}y_{j\in J_{t}}^{\frac{1}{\beta_{m\left(t\right)}}-1}\right)}{\left(\prod_{t=1}^{T}\left(\sigma_{m\left(t\right)}^{2}\right)^{\frac{J}{2}}\right)\left|\Sigma\right|^{\frac{T}{2}}}\times\\
\exp\left(-\frac{1}{2}\sum_{t=1}^{T}\left\{ \frac{1}{\sigma_{m\left(t\right)}^{2}}\left(T^{-1}\left(\mathbf{y}_{t}\right)-\mu_{t}\right)^{'}\Sigma^{-1}\left(T^{-1}\left(\mathbf{y}_{t}\right)-\mu_{t}\right)\right\} \right)\label{eq:likelihood_latent}
\end{multline}

For example, no integration is required for (\ref{eq:likelihood_latent}).
We refer the reader to \citet{sanso_venezuelan_1999} for details
on this type of model. 

For our purposes, it suffices to present here the estimated parameters,
computed as the mean values of the respective Markov Chains, after
letting sufficiently many burn-in iterations of the Metropolis Hastings
algorithm run (11000, in our case). The estimates are given on table
\ref{tab:Parameters-fitted-for-space-time}, rounded up to the fourth
decimal place.

\begin{table}
\begin{centering}
\begin{tabular}{|c|c|c|c|c|c|}
\hline 
$\alpha_{i}$,$i=0,1$ & $\beta_{i}$, $i=1,\ldots,12$ & $\sigma_{i}^{2}$, $i=1,\ldots,12$ & $A_{i}$, $i=1,2,3$ & $B_{i}$, $i=1,2,3$ & $\lambda$\tabularnewline
\hline 
\hline 
-0.0644 & 1.8503 & 15.5086 & -0.8199 & -0.0214 & 0.0221\tabularnewline
\hline 
0.0001 & 1.6930 & 13.1596 & 0.3916 & 0.4156 & \tabularnewline
\hline 
 & 1.8016 & 11.1473 & -0.1907 & -0.3885 & \tabularnewline
\hline 
 & 1.7018 & 14.0445 &  &  & \tabularnewline
\hline 
 & 1.6415 & 18.7905 &  &  & \tabularnewline
\hline 
 & 1.5893 & 15.1922 &  &  & \tabularnewline
\hline 
 & 1.4565 & 31.2774 &  &  & \tabularnewline
\hline 
 & 1.6544 & 22.8418 &  &  & \tabularnewline
\hline 
 & 1.7055 & 20.4347 &  &  & \tabularnewline
\hline 
 & 1.5693 & 11.5969 &  &  & \tabularnewline
\hline 
 & 1.6610 & 20.4527 &  &  & \tabularnewline
\hline 
 & 1.6910 & 10.0384 &  &  & \tabularnewline
\hline 
\end{tabular}\caption{\label{tab:Parameters-fitted-for-space-time}Parameters fitted for
the Space-Time model. The indexes ``i'' increase downwards. }

\par\end{centering}

\end{table}

As mentioned earlier, we are currently working on a coherent estimation
method for estimating all parameters simultaneously. For the moment,
in order to show the implications of considering higher order interdependences,
we multiply the latent Gaussian field fitted using the MCMC method
times the scaling variable of section (\ref{sub:Scaling-variable-used}).
Thereby we obtain a latent field of the form (\ref{eq:Kano_dcomp}).
Note that, according to the analysis of  \ref{sec:Similarity-of-one-and-two-dimensional},
these two latent fields are not distinguishable by analyzing their
one and two dimensional marginal distributions. 

The rectangle superimposed on figure \ref{fig:Catchment-of-the-Saalach}
is formed of a $33\times83$ grid, in which each square side represents
a 500 meter length. Using the fitted parameters of table \ref{tab:Parameters-fitted-for-space-time},
we obtain the mean value $\mu_{t,j}$ at each location $j=1,\ldots,2739$,
and we get also the correlation matrix for the whole $33\times83$
grid. 

We then simulated for each month of the year 3000 realizations, $\mathbf{z}_{t}\in\mathbb{R}^{2739}$,
of the latent Gaussian field using the parameters extended to the
$33\times83$ grid. We set the negative values of these Gaussian vectors
to zero, and ten applied the transformation $\mathbf{z}_{t}\rightarrow\mathbf{z}_{t}^{\beta_{m\left(t\right)}}:=\mathbf{y}_{t}$.
Vectors $\mathbf{y}_{t}$ are our simulated precipitations fields
with latent Gaussian structure. 

To obtain vectors $\tilde{\mathbf{y}_{t}}$ which consider interdependence
beyond correlation, we simulated $3000\times12=36000$ realizations,
$v_{t}$, of the scaling variable $V$ from section \ref{sub:Scaling-variable-used},
3000 for each month of the year. We set $\tilde{\mathbf{z}}_{t}=\mu_{t}+\left(\mathbf{z}_{t}-\mu_{t}\right)\times\sqrt{v_{t}}$
; the negative components of these vectors were set to zero, and then
we applied the transformation $\tilde{\mathbf{z}}_{t}\rightarrow\tilde{\mathbf{z}}_{t}^{\beta_{m\left(t\right)}}:=\tilde{\mathbf{y}}_{t}$.
Vectors $\tilde{\mathbf{y}}_{t}$ are our simulated precipitation
fields with latent non-Gaussian structure. 

\begin{figure}
\begin{centering}
\includegraphics[width=0.9\textwidth]{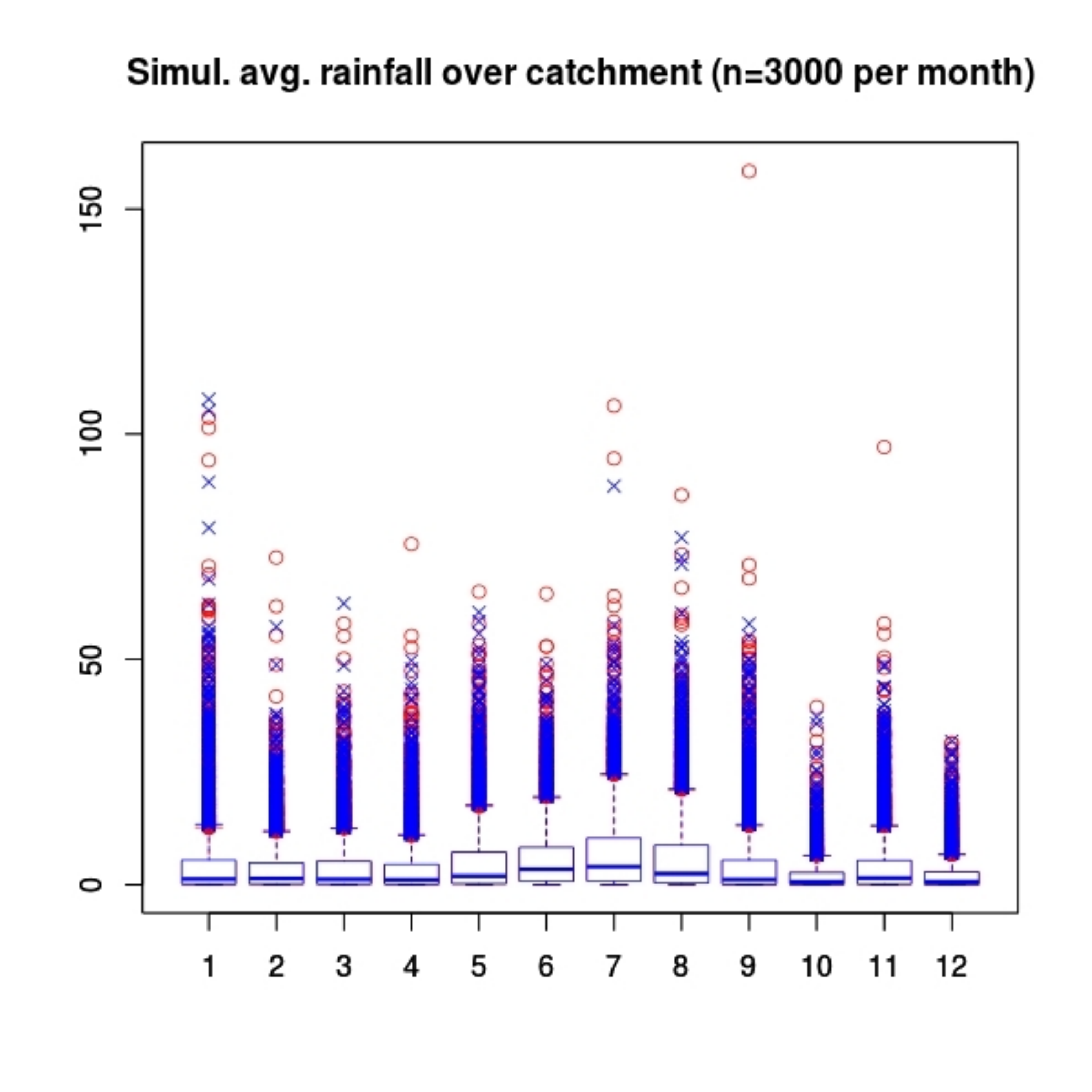}\caption{\label{fig:Boxplots-of-the-real-fields}Boxplots of the 3000 per month
unconditional simulations from the field with latent Gaussian structure
(blue), and the field with non-Gaussian latent structure (red). Values
are in millimeters. The interaction of order greater than two among
components can trigger very high simultaneous values in the components
of the random field.}

\par\end{centering}

\end{figure}

By averaging the values of the components of vectors $\mathbf{y}_{t}$
and $\tilde{\mathbf{y}}_{t}$, we get for each type of field 3000
average precipitation values per month, over the rectangular area
shown at figure \ref{fig:Catchment-of-the-Saalach}. These values
are plotted in figure \ref{fig:Boxplots-of-the-real-fields}. Note
that the distribution of the average values of both fields is very
similar, except that some values of the field with non-Gaussian latent
structure are much bigger than those expectable from a model with
latent Gaussian structure. This is the effect of interactions among
more than two variables. 

Is these simulations were to be included as forecasts in a model for
flood risk assessment, for example, the forecast based on the space-time
model with Gaussian latent structure would suggest much longer flood
return periods.

The entropy-based graphical technique of  \ref{sec:Analysis-of-Aggregating-statistics}
can be used for validation of the model with Gaussian latent structure.
We focus on the three stations having less missing values, out of
the 9 stations (labeled 1,2 and 3 in figure \ref{fig:Two-conditionally-simulated-fields}). 

\begin{figure}
\begin{centering}
\includegraphics[width=0.75\textwidth]{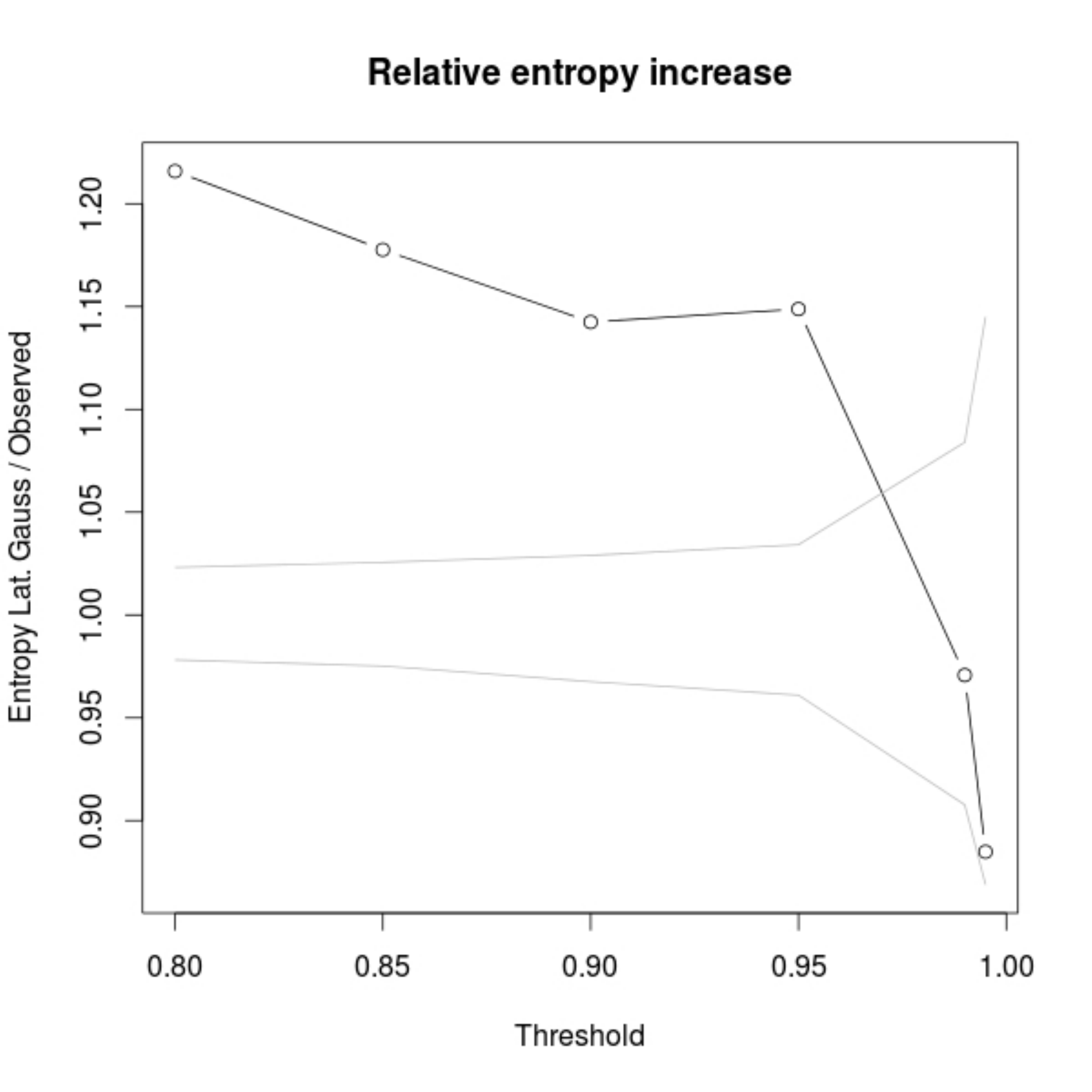}
\par\end{centering}

\caption{\label{fig:Ratio-of-congregation-body}Ratio of congregation measures
and 90\% confidence interval, as explained in  \ref{sec:Analysis-of-Aggregating-statistics},
for the three stations labeled 1, 2 and 3. Three-wise association
of data is considerable higher up to threshold 0.95. }
\end{figure}

In figure \ref{fig:Ratio-of-congregation-body}, the graphical validation
tool is presented for thresholds $a\in\left\{ .80,.85,.90,.95,.99,.995\right\} $.
The 3-wise observed association is considerably bigger, up to a threshold
of 0.95. A simulation-based 90\% confidence interval is also shown.
This is an indicator that the model is systematically underestimating
3-wise association. Note that this technique is robust to non-decreasing
transformations on the marginal distributions.

We are currently working on techniques to systematically estimate
scaling variable, $V$, of the latent non-Gaussian field, such that
the resulting 3-wise association is more similar to the observed one.

\subsection{Conditional simulation for the 1st of June 2013}

Although our model was fitted with data from 2004-2009, we now show
that the probability of very intense precipitations, such as those
of early June 2013 over the Saalach river catchment, can be more realistically
evaluated if we consider higher order interactions in our space-time
model.

Taking a close look at the available data of the nine stations selected,
one finds very high values at virtually all stations for June the
1st 2013. The next day, June the 2nd, there was a tremendous increase
in the water flow of river Saalach, according to discharge measurements
at the village of Unterjettenberg, very near to the town of Bad Reichenhall,
in southeast Germany. 

Hence in this section we produce rainfall fields, conditional on the
observed data of June 1st 2013, for the rectangular area presented
in figure \ref{fig:Catchment-of-the-Saalach}. This is a good proxi
for the average precipitation over the whole Saalach catchment. To
produce the conditional simulations, we used the second technique
presented at section \ref{sub:Interpolation-to-ungauged}. 

\begin{figure}

\centering{}\includegraphics[width=0.5\textwidth]{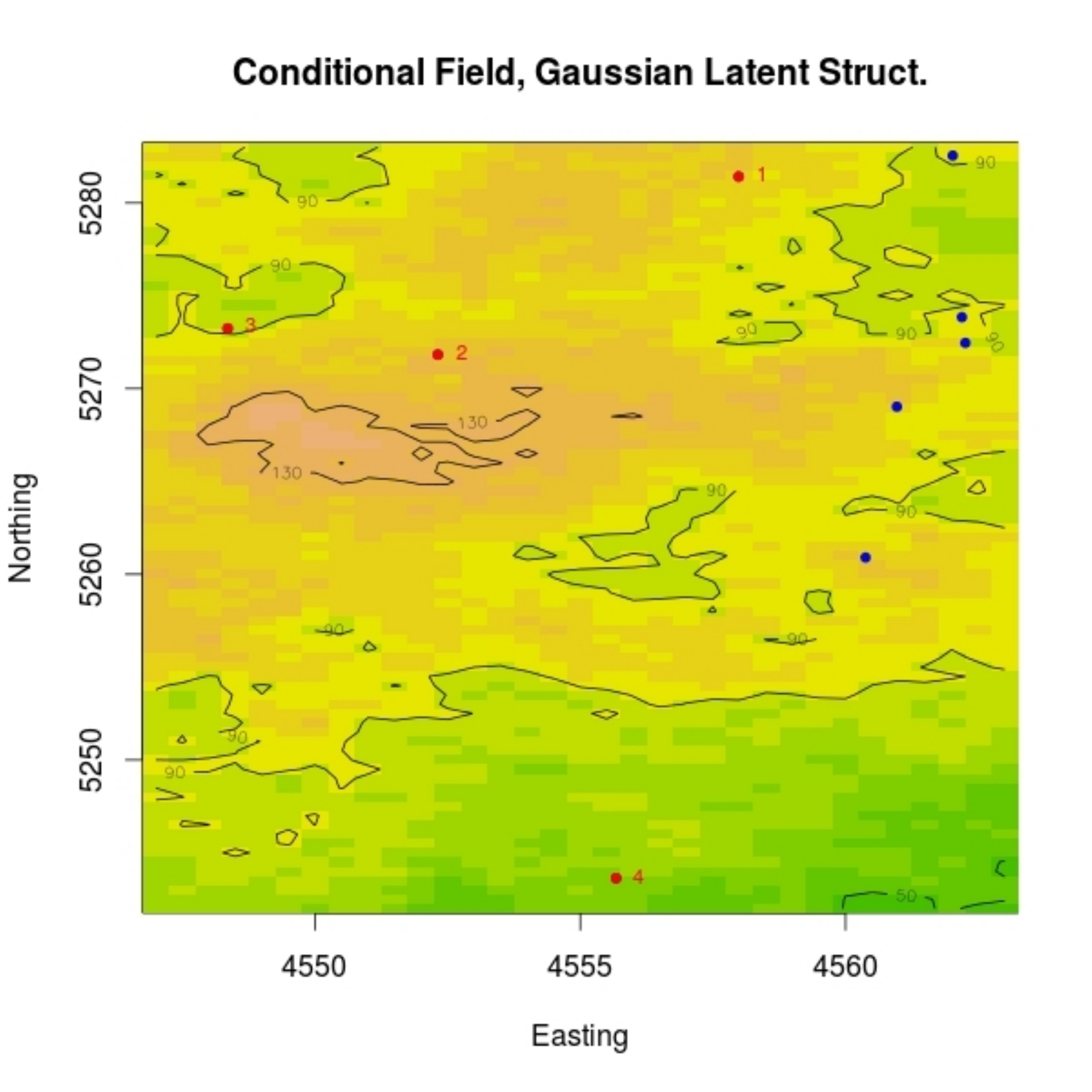}\includegraphics[width=0.5\textwidth]{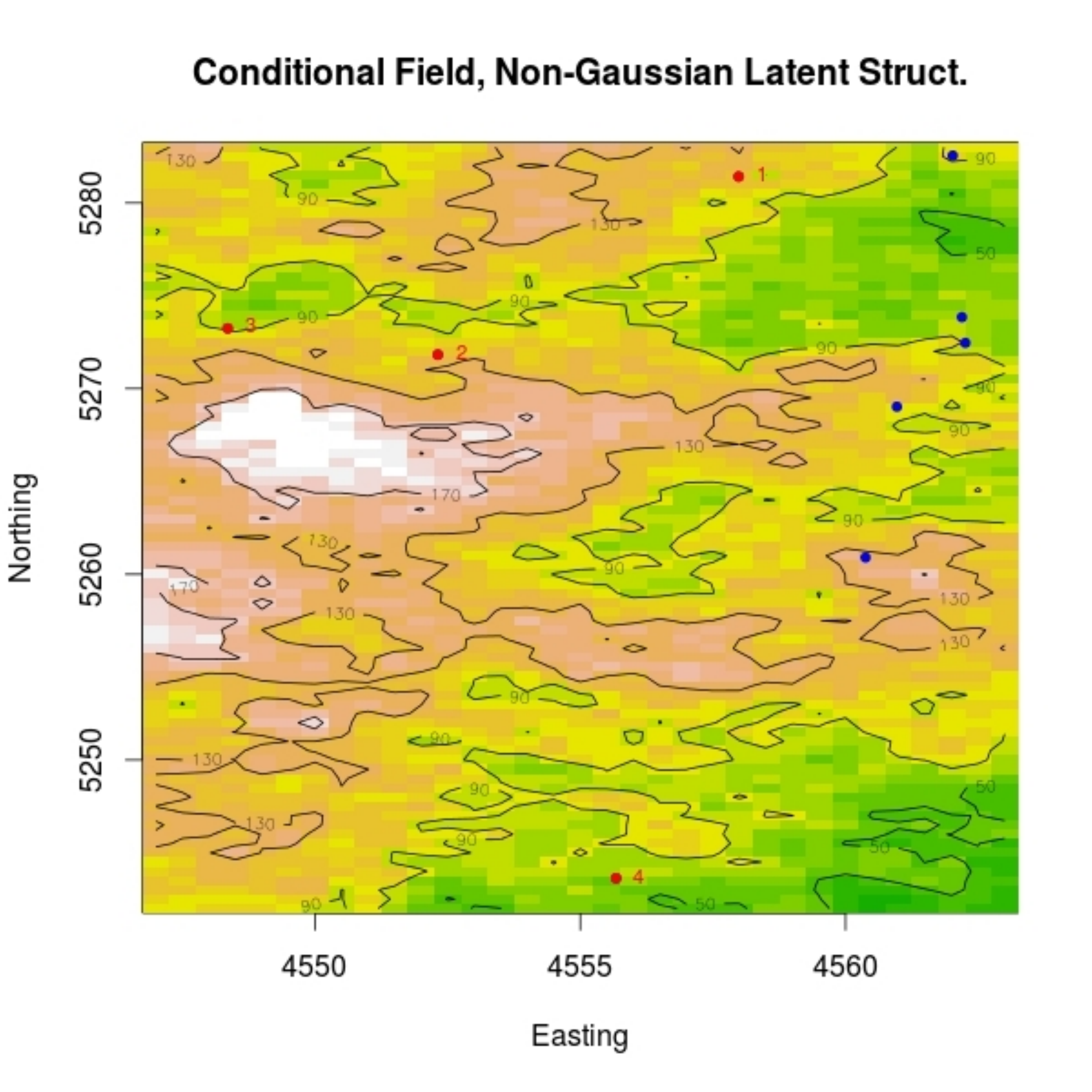}\caption{\label{fig:Two-conditionally-simulated-fields}Two conditionally simulated
fields for June 1st 2013: Field with Gaussian latent structure (left),
and field with non-Gaussian latent structure (right). Stations providing
the observed data are indicated in red.}
\end{figure}

Only four gauging stations have data for June 1st 2013. These stations
are shown in red in figure \ref{fig:Two-conditionally-simulated-fields};
we shall address this figure shortly. The four stations with observed
data are also labeled 1,2,3 and 4, in the figure. Their data values
(in mm) are: 104.1, 120.0, 85.1 and 65.1, respectively. 

\begin{figure}
\begin{centering}
\includegraphics[width=0.5\textwidth]{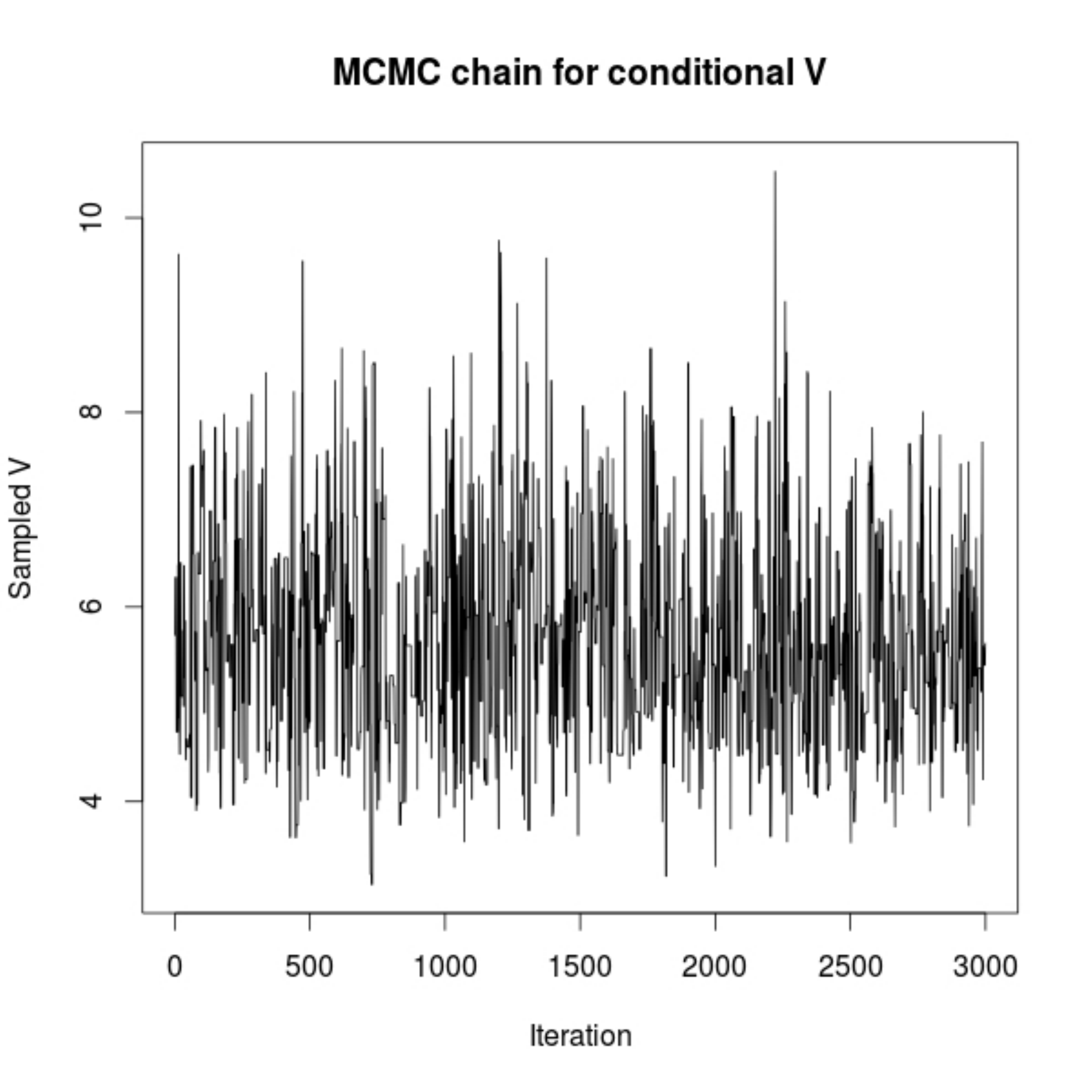}\includegraphics[width=0.5\textwidth]{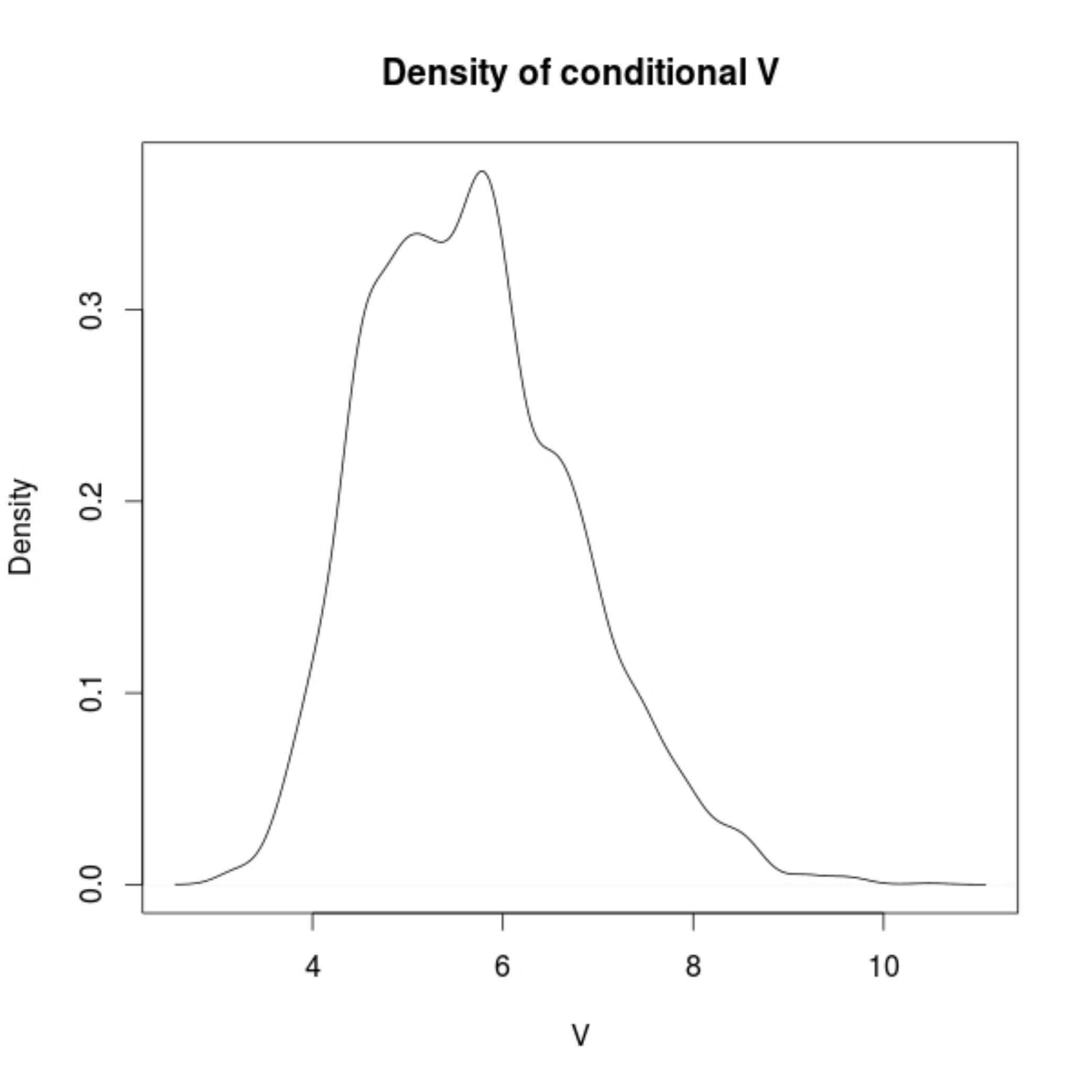}
\par\end{centering}

\begin{centering}
\caption{\label{fig:Conditionally-sampled-scaling}Sampled scaling variable,
$V$, for June 1st 2013, conditional on observed values: MCMC chain
after 500 burn-in iterations (left), and estimated probability density
(right). }

\par\end{centering}

\end{figure}

The first step in generating the conditional fields, according to
the technique in section \ref{sub:Interpolation-to-ungauged}, is
to obtain a sample of the scaling variable, $V$, conditional on the
observed data. The sampled scaling variable is shown at figure \ref{fig:Conditionally-sampled-scaling}.
Note the high values for $V$ (up to $V=10$) that are consistent
with the observed high rainfall values.

\begin{figure}
\begin{centering}
\includegraphics[width=0.75\textwidth]{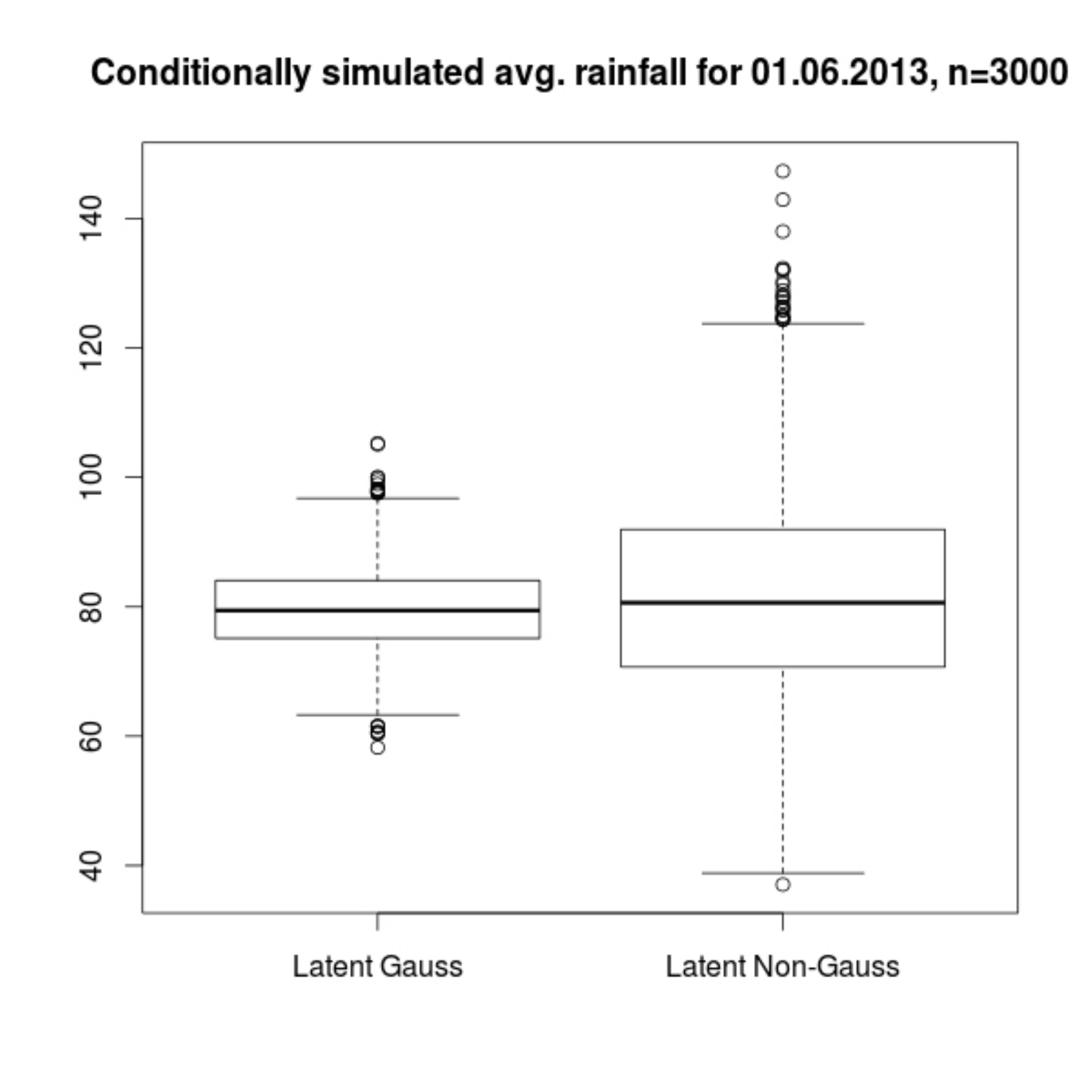}
\par\end{centering}

\caption{\label{fig:Boxplots-of-the-conditional}Boxplots of the average of
the conditionally simulated random fields for June 1st 2013, in millimeters.
The field with high oder interacting latent structure shows much more
variability. In particular, average precipitation over the catchment
above 120 mm are quite probable under the new model.}

\end{figure}

Using these sampled $V$'s, we generated the conditional fields, 3000
in total. Two realizations of these fields are presented in figure
\ref{fig:Two-conditionally-simulated-fields}. The contrast between
them is by no means atypical. 

Note the two intense clusters, with values of over 170 mm each, which
one encounters in the realization of the field with multivariate interactions
(right panel). On the other hand, other regions of the map exhibit
lower values than the field with the Gaussian latent structure; for
example, the southeast region has slightly smaller values.

Using the 3000 conditional simulations for each field, we have an
idea of the kind extreme event we can expect over the catchment, according
to each of the models. We focus on the mean catchment precipitation,
as before. In figure \ref{fig:Boxplots-of-the-conditional} we show
boxplots of the average of the simulated fields. Note that values
above 120 mm seem quite probable for the model with non-Gaussian latent
structure, whereas they seem to be almost improbable for the model
with Gaussian latent structure. 

Whether the true average rainfall over the Saalach river catchment
was 120 mm or more on June 1st 2013, is not yet clear; that statement
would require a detailed analysis of the river discharge, and of the
meteorological conditions during the days immediately before, and
up to that day. But with this example we hope at least to show the
need to develop models that consider explicitly interactions among
groups of variables. Such interactions have the potential, as we have
seen, to increase greatly the probability of very high values at several
locations simultaneously. 

Considering these interactions might lead to more realistic estimated
flood return periods for the towns in the catchment which lie near
the river.

\section{\label{sec:Discussion-and-work}Discussion and work in progress}

The model introduced by (\ref{eq:archetypal_cgf}) allows for the
explicit consideration of joint cumulants of order greater than two
(i.e. not just covariances) in a manner that is convenient for spatial
modeling: building on available geostatistical techniques, requiring
a minimum of extra parameters, and respecting the principle of spatial
consistency. The range of tail dependence intensity of this model
goes from zero (i.e. Gaussian) to that of a Student-t, as in the synthetic
example presented.

The need in Spatial Statistics to consider interactions among more
than two variables at a time was illustrated using a thorough synthetic
case study. We also analyzed the possible implications of high order
interdependence for the forecasting of the total volume of precipitation
over the Saalach river catchment, in Germany.

The presence of interactions, not noticeable from the one and two
dimensional marginals, can be assessed using statistics that aggregate
information of higher dimensional marginal distributions of the data.
We presented some of such statistics. To make sure that data simulated
from the model reproduces those statistics (i.e. those interaction
manifestations) is an important complementary goodness of fit procedure
for a spatial model.

This paper has provided the theoretical basis for a model with which
interactions among more than two variables can be explicitly considered.
But there is much work to do in order to exploit the full power of
the model. A parameter estimation procedure more convenient than the
one presented here, usable also for truncated data (e.g. for precipitation
modeling) is under development. We also intend to connect the scaling
variable used to build our model, and which determines the additional
interaction characteristics, with large scale atmospheric processes,
in a hierarchical manner. In this way we expect to produce more realistic
forecasts of intense precipitation over large areas, for the sake
of risk assessment.

\subsection*{Acknowledgments}

This research forms part of the Ph.D thesis of the first author, which
was funded by a scholarship of the German Academic Exchange Service
(DAAD), and realized in the context of the ENWAT program of the University of Stuttgart.

We also  thank Prof. Lelys Bravo de Guenni 
for her critical review of this work and her very useful comments. The responsibility for any mistake in the text is of course entirely of the authors.

\appendix

\section{\label{sec:Derivation-of-joint}Derivation of joint cumulants of
the model}

Our object of study is the cumulant generating function of a random
variable $\mathbf{X}\in\mathbb{R}^{J}$. We shall be interested in
joint cumulants such as
\begin{equation}
cum\left(X_{j_{1}},\ldots,X_{j_{r}}\right)\label{eq:joint_cumlant_appendix}
\end{equation}
where some, or all, of the indexes can be repeated. Hence it is convenient
to refer to a random vector $\mathbf{X^{*}}\in\mathbb{R}^{J^{*}}$
having the components of $\mathbf{X}$, even repeated, and then find
the joint cumulants that appear with degree at most one, of this ``new''
random vector. Thus we can, without loss of generality, focus on finding
the joint cumulants with degree not greater than one, given by
\begin{equation}
\frac{\partial^{r}}{\partial t_{j_{r}}\ldots\partial t_{j_{1}}}K_{\mathbf{X^{*}}}\left(\mathbf{t}\right)\mid_{\mathbf{t}=\mathbf{0}}:=cum\left(X_{j_{1}},\ldots,X_{j_{r}}\right)\label{eq:Joint_cumulants_appendix_2}
\end{equation}
where no $t_{j}$, for $j\in\left\{ j_{1},\ldots,j_{r}\right\} $,
is repeated.

For example, when computing the variance of a component, $X_{j}$,
of $\mathbf{X}$, one would rather compute the covariance of vector
$\mathbf{X}^{*}=\left(X_{j},X_{j}\right)$, namely $\kappa_{11}\left(\mathbf{X^{*}}\right)$.

The archetypal dependence structure advocated for in this work is
given by
\begin{equation}
K_{\mathbf{X^{*}}}\left(\mathbf{t}\right)=c_{1}\frac{1}{2}\mathbf{t^{T}\Gamma t}+\frac{1}{2!}c_{2}\left[\frac{1}{2}\mathbf{t^{T}\Gamma t}\right]^{2}+\frac{1}{3!}c_{3}\left[\frac{1}{2}\mathbf{t^{T}\Gamma t}\right]^{3}+\ldots\label{eq:archetypal_cgf-1}
\end{equation}
for some coefficients $c_{1},c_{2},c_{3},\ldots$ and covariance matrix
$\Gamma_{J^{*}\times J^{*}}$, and $\mathbf{t}\in\mathbb{R}^{J^{*}}$. 

By expansion, the above expression can be written as
\begin{multline}
K_{\mathbf{X^{*}}}\left(\mathbf{t}\right)=\frac{c_{1}}{1!}\frac{1}{2}\sum_{j_{1},j_{2}=1}^{J}t_{j_{1}}t_{j_{2}}\Gamma_{j_{1}j_{2}}+\frac{c_{2}}{2!}\frac{1}{2^{2}}\sum_{j_{1},\ldots,j_{4}=1}^{J}t_{j_{1}}\ldots t_{j_{4}}\Gamma_{j_{1}j_{2}}\Gamma_{j_{3}j_{4}}+\\
\frac{c_{3}}{3!}\frac{1}{2^{3}}\sum_{j_{1},\ldots,j_{6}=1}^{J}t_{j_{1}}\ldots t_{j_{6}}\Gamma_{j_{1}j_{2}}\Gamma_{j_{3}j_{4}}\Gamma_{j_{5}j_{6}}+\ldots\label{eq:archetypal_cgf_expansion}
\end{multline}

For each coefficient $c_{\frac{r}{2}}$, for $r$ even, there appears
a sum of the form
\begin{equation}
\frac{c_{\frac{r}{2}}}{\frac{r}{2}!}\frac{1}{2^{\frac{r}{2}}}\sum_{j_{1}=1}^{J}\ldots\sum_{j_{2r}=1}^{J}t_{j_{1}}\ldots t_{j_{r}}\Gamma_{j_{1}j_{2}}\ldots\Gamma_{j_{r-1}j_{r}}\label{eq:joint_cumulants_block}
\end{equation}

This is the only block-summand of (\ref{eq:archetypal_cgf_expansion})
that does not vanish upon differentiation with respect to each variable
and equation to zero, as in (\ref{eq:Joint_cumulants_appendix_2}).
Other blocks will vanish either upon differentiation with respect
to a variable that does not appear in them, or upon equation to zero,
since such blocks become a sum of zeroes. So, it suffices to focus
on this block, to differentiate it and equate it with zero.

Let each member of the (\ref{eq:joint_cumulants_block}) be labeled
\[
s_{j_{1}\ldots,j_{r}}=t_{j_{1}}\ldots t_{j_{r}}\Gamma_{j_{1}j_{2}}\ldots\Gamma_{j_{r-1}j_{r}}
\]
then, we have stated that, 
\begin{multline}
\frac{\partial^{r}}{\partial t_{j_{r}}\ldots\partial t_{j_{1}}}K_{\mathbf{X^{*}}}\left(\mathbf{t}\right)\mid_{\mathbf{t}=\mathbf{0}}=\frac{c_{\frac{r}{2}}}{\frac{r}{2}!}\frac{1}{2^{\frac{r}{2}}}\sum_{j_{1}=1}^{J}\ldots\sum_{j_{2r}=1}^{J}\frac{\partial^{r}}{\partial t_{j_{r}}\ldots\partial t_{j_{1}}}s_{j_{1},\ldots,j_{r}}\label{eq:equation_cgf_appendix}
\end{multline}

Partial differentiation of $s_{j_{1},\ldots,j_{r}}$ is readily found
to be
\begin{equation}
\frac{\partial^{r}}{\partial t_{j_{r}}\ldots\partial t_{j_{1}}}s_{j_{1},\ldots,j_{r}}=\Gamma_{j_{1}j_{2}}\ldots\Gamma_{j_{r-1}j_{r}}\label{eq:cumlants_block_diff_Appendix}
\end{equation}

Sub-indexes appearing in the factors, $\Gamma_{j_{1}j_{2}}$, $\Gamma_{j_{3}j_{4}},\ldots$
constitute a partition of size $\frac{r}{2}$ of the set $A=\left\{ j_{1},j_{2},\ldots,j_{r}\right\} $.
That is, the union of the $\frac{r}{2}$ non-overlapping sets 
\[
\left\{ j_{1},j_{2}\right\} ,\left\{ j_{3},j_{4}\right\} ,\ldots,\left\{ j_{r-1},j_{r}\right\} 
\]
formed with elements of set $A=\left\{ j_{1},j_{2},\ldots,j_{r}\right\} $,
is equal to that set: 
\[
\left\{ j_{1},j_{2}\right\} \cup\left\{ j_{3},j_{4}\right\} \cup\ldots\cup\left\{ j_{r-1},j_{r}\right\} =A
\]

Since the sum at (\ref{eq:equation_cgf_appendix}) runs over all indexes
in $A$, the sum returning the joint cumulant in question comprises
all partitions of size two of $A$. How many different partitions
of size two can be obtained for $A$, by forming sets out of different
combinations of indexes? In general, a set with $n$ elements, $n$
even, can be seen to have 
\[
1\times3\times\ldots\times\left(n-1\right)
\]
such partitions.

We have shown that joint cumulants of the archetypal dependence structure
are given by
\begin{equation}
cum\left(X_{j_{1}},\ldots,X_{j_{r}}\right)=\frac{c_{\frac{r}{2}}}{\frac{r}{2}!}\frac{1}{2^{\frac{r}{2}}}\sum_{j_{1},\ldots,j_{r}=1}^{J}\Gamma_{j_{1}j_{2}}\ldots\Gamma_{j_{r-1}j_{r}}
\end{equation}

\section{\label{sec:Relation-between-moments-R-scaling}Relation between moments
of squared scaling variable and generating variable}

Assume that we have random vector $\mathbf{Z}\in\mathbb{R}^{J}$ with
c.g.f (\ref{eq:archetypal_cgf}), with $\mathbf{\mu}=\mathbf{0}$
and covariance matrix equal to the identity matrix, $\Sigma=I_{J\times J}$.
For this special case, in agreement with representation (\ref{eq:representation1}),
we have 
\[
\left\Vert \mathbf{Z}\right\Vert _{2}=\sqrt{\left\Vert \mathbf{Z}\right\Vert _{2}\left\Vert \mathbf{Z}\right\Vert _{2}}=\sqrt{\left\Vert R\times\mathbf{U}^{J-1}\right\Vert _{2}\left\Vert R\times\mathbf{U}^{J-1}\right\Vert _{2}}=R\times1
\]
since $\left\Vert \mathbf{U}^{J-1}\right\Vert _{2}=1$. Then,
\begin{equation}
R^{2}=\sum_{j=1}^{J}Z_{j}^{2}\label{eq:rel_1_Rsq_Z}
\end{equation}
which in turn means that,

\begin{multline}
E\left(\left(R^{2}\right)^{k}\right)=E\left(\left(\sum_{j_{1}=1}^{J}Z_{j_{1}}^{2}\right)\times\ldots\times\left(\sum_{j_{k}=1}^{J}Z_{j_{k}}^{2}\right)\right)=\sum_{j_{1}=1}^{J}\ldots\sum_{j_{k}=1}^{J}E\left(Z_{j_{1}}^{2}\ldots Z_{j_{k}}^{2}\right)\label{eq:moments-Rsq-1}
\end{multline}

Since $\mathbf{Z}$ has c.g.f. given by 
\[
K_{\mathbf{Z}}\left(\mathbf{t}\right)=\frac{c_{1}}{1!}\left(\frac{1}{2}\mathbf{t^{'}t}\right)+\frac{c_{2}}{2!}\left(\frac{1}{2}\mathbf{t^{'}t}\right)^{2}+\frac{c_{3}}{3!}\left(\frac{1}{2}\mathbf{t^{'}t}\right)^{3}+\ldots
\]
it follows, as seen in section \ref{sec:The-proposed-model}, that

\[
M_{\mathbf{Z}}\left(\mathbf{t}\right)=1+\frac{m_{1}}{1!}\left(\frac{1}{2}\mathbf{t^{'}t}\right)+\frac{m_{2}}{2!}\left(\frac{1}{2}\mathbf{t^{'}t}\right)^{2}+\frac{m_{3}}{3!}\left(\frac{1}{2}\mathbf{t^{'}t}\right)^{3}+\ldots
\]
with coefficients given by 
\begin{eqnarray}
m_{1} & = & c_{1}\nonumber \\
m_{2} & = & c_{2}+c_{1}^{2}\nonumber \\
m_{3} & = & c_{3}+3c_{2}c_{1}+c_{1}^{3}\nonumber \\
m_{4} & = & c_{4}+4c_{3}c_{1}+3c_{2}^{2}+6c_{2}c_{1}^{2}+c_{1}^{4}\label{eq:cums_in_terms_moms-2-1-1}
\end{eqnarray}
and so on. A particular case of this function is the Gaussian moment
generating function, for which all $c_{r>1}$ are set to zero. In
particular, for $\xi\sim N_{J}\left(\mathbf{0},I_{J\times J}\right)$,
\begin{equation}
M_{\mathbf{\xi}}\left(\mathbf{t}\right)=1+\frac{c_{1}}{1!}\left(\frac{1}{2}\mathbf{t^{'}t}\right)+\frac{c_{1}^{2}}{2!}\left(\frac{1}{2}\mathbf{t^{'}t}\right)^{2}+\frac{c_{1}^{3}}{3!}\left(\frac{1}{2}\mathbf{t^{'}t}\right)^{3}+\ldots\label{eq:moms-esph-gauss-1}
\end{equation}
with $c_{1}=1$. Hence joint moments of $\mathbf{Z}$ and $\mathbf{\xi}$
are similar, except for what pertains to coefficients $c_{2},c_{3},\ldots$.
In fact, calling 
\[
h_{r}\left(\mathbf{t}\right)=\left(\frac{1}{2}\mathbf{t^{'}t}\right)^{r}
\]
one has 
\[
\begin{cases}
\frac{\partial^{r_{1}+\ldots+r_{k}}}{\partial t_{j_{1}}\ldots\partial t_{j_{k}}}M_{\mathbf{\xi}}\left(\mathbf{t}\right)=\frac{c_{1}}{1!}\frac{\partial^{r_{1}+\ldots+r_{k}}}{\partial t_{j_{1}}\ldots\partial t_{j_{k}}}h_{1}\left(\mathbf{t}\right)+\frac{c_{1}^{2}}{2!}\frac{\partial^{r_{1}+\ldots+r_{k}}}{\partial t_{j_{1}}\ldots\partial t_{j_{k}}}h_{2}\left(\mathbf{t}\right)+\ldots\\
\frac{\partial^{r_{1}+\ldots+r_{k}}}{\partial t_{j_{1}}\ldots\partial t_{j_{k}}}M_{\mathbf{Z}}\left(\mathbf{t}\right)=\frac{m_{1}}{1!}\frac{\partial^{r_{1}+\ldots+r_{k}}}{\partial t_{j_{1}}\ldots\partial t_{j_{k}}}h_{1}\left(\mathbf{t}\right)+\frac{m_{2}}{2!}\frac{\partial^{r_{1}+\ldots+r_{k}}}{\partial t_{j_{1}}\ldots\partial t_{j_{k}}}h_{2}\left(\mathbf{t}\right)+\ldots
\end{cases}
\]

Hence, for odd orders joint moments of both random vectors are zero,
and for even orders
\begin{eqnarray*}
E\left(\xi_{i}\xi_{j}\right) & = & \frac{c_{1}}{m_{1}}E\left(Z_{i}Z_{j}\right)\\
E\left(\xi_{i}\xi_{j}\xi_{k}\xi_{l}\right) & = & \frac{c_{1}^{2}}{m_{2}}E\left(Z_{i}Z_{j}Z_{k}Z_{l}\right)\\
 & \vdots\\
E\left(\xi_{j_{1}}^{r_{1}}\ldots\xi_{j_{k}}^{r_{k}}\right) & = & \frac{c_{1}^{\frac{1}{2}\sum_{j=1}^{k}r_{j}}}{m_{\frac{1}{2}\sum_{j=1}^{k}r_{j}}}E\left(Z_{j_{1}}^{r_{1}}\ldots Z_{j_{k}}^{r_{k}}\right)
\end{eqnarray*}
whenever $order=\sum_{i=1}^{k}r_{i}$ is an even integer. It is then
clear that the following relation holds, for joint moments of even
order:
\begin{eqnarray}
\frac{m_{1}}{c_{1}}E\left(\xi_{i}\xi_{j}\right) & = & E\left(Z_{i}Z_{j}\right)\nonumber \\
\frac{m_{2}}{c_{1}^{2}}E\left(\xi_{i}\xi_{j}\xi_{k}\xi_{l}\right) & = & E\left(Z_{i}Z_{j}Z_{k}Z_{l}\right)\nonumber \\
 & \vdots\nonumber \\
\frac{m_{\frac{1}{2}\sum_{j=1}^{k}r_{j}}}{c_{1}^{\frac{1}{2}\sum_{j=1}^{k}r_{j}}}E\left(\xi_{j_{1}}^{r_{1}}\ldots\xi_{j_{k}}^{r_{k}}\right) & = & E\left(Z_{j_{1}}^{r_{1}}\ldots Z_{j_{k}}^{r_{k}}\right)\label{eq:moms_ellipt_normal-1}
\end{eqnarray}

Product moments appearing on the left hand side of equation (\ref{eq:moms_ellipt_normal-1})
can be readily found, since they are the moments of a multivariate
Gaussian distribution with covariance matrix equal to identity matrix
$I_{J\times J}$. 

Coefficients $m_{1},m_{2},m_{3},\ldots$ are given in terms of $c_{1},c_{2},c_{3},\ldots$
(and vice versa). Hence we have, by virtue of (\ref{eq:moments-Rsq-1}),
identified requirements on all moments of (squared) generating variable
$R^{2}$, so that the resulting multivariate distribution $\mathbf{X}$
has cumulant generating function (\ref{eq:archetypal_cgf}).

\paragraph*{Summarizing these results:}

First, since the multivariate Gaussian distribution referred to at
equation \ref{eq:moms_ellipt_normal-1} has covariance matrix equal
to identity, one can write for any set of components $\left(j_{1},\ldots,j_{k}\right)$,
\begin{equation}
\frac{m_{k}}{c_{1}^{k}}E\left(\xi_{j_{1}}^{2}\ldots\xi_{j_{k}}^{2}\right)=E\left(Z_{j_{1}}^{2}\ldots Z_{j_{k}}^{2}\right)\label{eq:bloque_R2-1}
\end{equation}
where $\xi$ is a $J$-dimensional normally distributed vector with
mean vector $\mathbf{0}$ and covariance matrix $I_{J\times J}$,
the identity matrix on $\mathbb{R}^{J\times J}$. Equation (\ref{eq:moments-Rsq-1})
holds in particular for vector $\xi$, in which case $R^{2}\sim\chi_{J}^{2}$,
and
\[
\sum_{j_{1}=1}^{J}\ldots\sum_{j_{k}=1}^{J}E\left(\xi_{j_{1}}^{2}\ldots\xi_{j_{k}}^{2}\right)=E\left(\left(\chi_{J}^{2}\right)^{k}\right)=\frac{2^{k}\Gamma\left(k+\frac{J}{2}\right)}{\Gamma\left(\frac{J}{2}\right)}
\]

Second and more importantly, by virtue of (\ref{eq:bloque_R2-1}),
one can re-write (\ref{eq:moments-Rsq-1}) as

\begin{equation}
E\left(\left(R^{2}\right)^{k}\right)=\sum_{j_{1}=1}^{J}\ldots\sum_{j_{k}=1}^{J}\frac{m_{k}}{c_{1}^{k}}E\left(\xi_{j_{1}}^{2}\ldots\xi_{j_{k}}^{2}\right)=\frac{m_{k}}{c_{1}^{k}}\frac{2^{k}\Gamma\left(k+\frac{J}{2}\right)}{\Gamma\left(\frac{J}{2}\right)}\label{eq:moments_of_R2_usefull-1-1}
\end{equation}
which expresses the moments of $R^{2}$ in terms of parameters $m_{k}$
(hence indirectly of $c_{k}$) and the dimension of the random vector
$\mathbf{X}$.

\section{\label{sec:Similarity-of-one-and-two-dimensional}Similarity of one
and two dimensional marginal distributions }

In this section, we show that the one and two dimensional marginal
distributions of the data collected from random vectors $\mathbf{X}\in\mathbb{R}^{30}$,
$\mathbf{W}\in\mathbb{R}^{30}$ and $\mathbf{Z}\in\mathbb{R}^{30}$
at section \ref{sec:A-simulation-based-illustration} are practically
indistinguishable. They all seem to be Guassian random vectors.

\subsection{Analysis of one dimensional marginal distributions}

\begin{figure}
\begin{centering}
\includegraphics[scale=0.75]{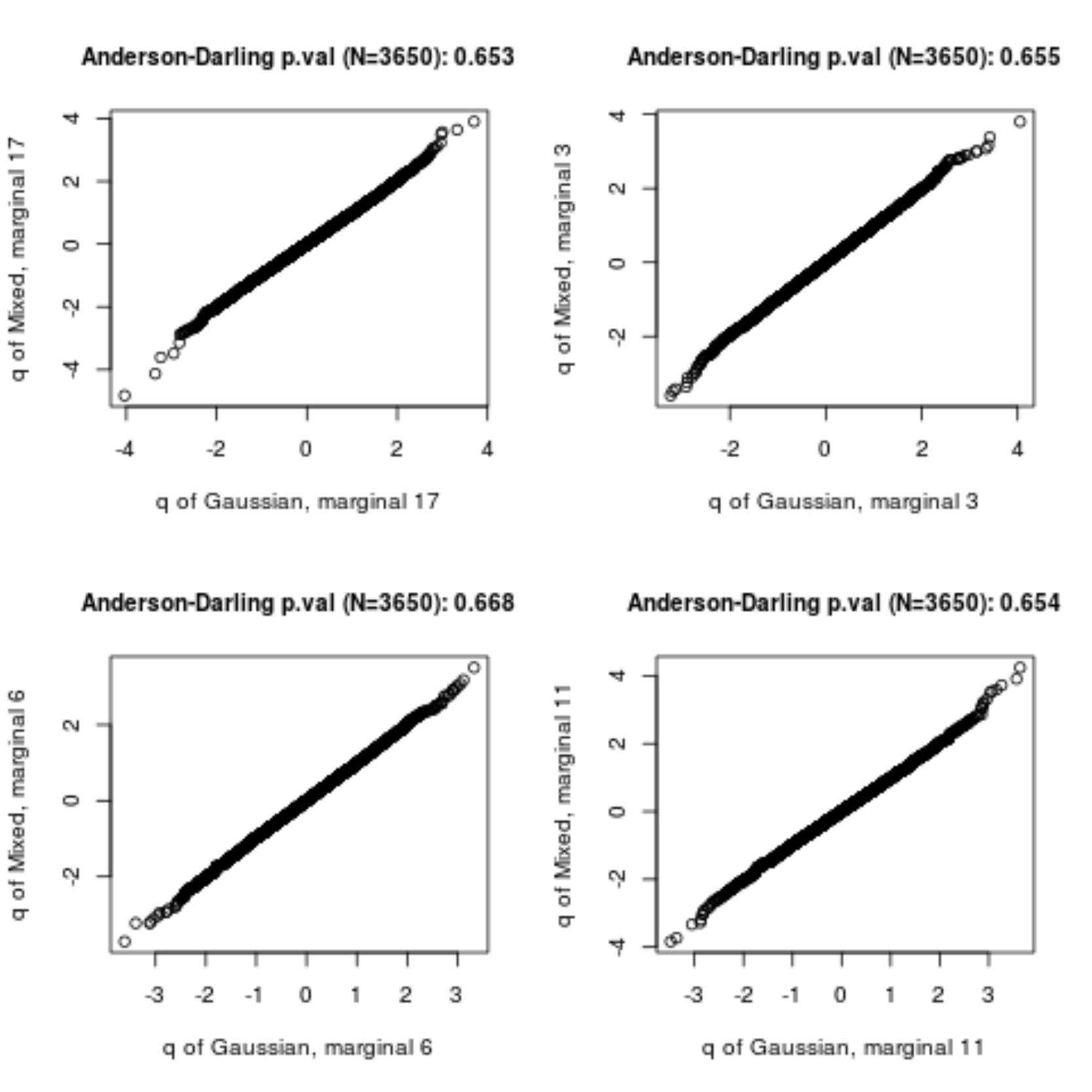}
\par\end{centering}

\caption{\label{fig:Quantile-quantile-plots-of-1D}Quantile-quantile plots
of four randomly selected components of $\mathbf{Z}$ and $\mathbf{X}$.
The p-values of the Anderson-Darling test for equality in distribution
(n=3650) are given. The marginal distributions illustrated can be
reasonably accepted to be equal.}
\end{figure}

Comparison of the 1-dimensional marginal distributions of $\mathbf{Z}$
and $\mathbf{X}$ is performed in this sub-section. At figure \ref{fig:Quantile-quantile-plots-of-1D}
we present four quantile-quantile plots. Each of these plots corresponds
to data from $\left(Z_{j},X_{j}\right)$, where $j$ has been randomly
selected from $\left\{ 1,\ldots,30\right\} $. The Anderson-Darling
test for equality in distributions was applied to data involved in
each plot, and the resulting p-value (n=3650) has been written on
each plot title. Both visually and from the testing viewpoint, the
marginal distributions considered at each plot seem to be the same.

Additionally, the Anderson-Darling test was applied to data from every
pair $\left(Z_{j},X_{j}\right)$, for $j=1,\ldots,30$, n=3650. The
minimum p-value obtained from all 30 tests was 0.642. Hence $\mathbf{X}$
and $\mathbf{Z}$ can be considered to have the same 1-dimensional
marginals, namely, standard normal marginal distributions.

\subsection{Analysis of two dimensional marginal distributions}

\begin{figure}
\begin{centering}
\includegraphics[width=0.75\textwidth]{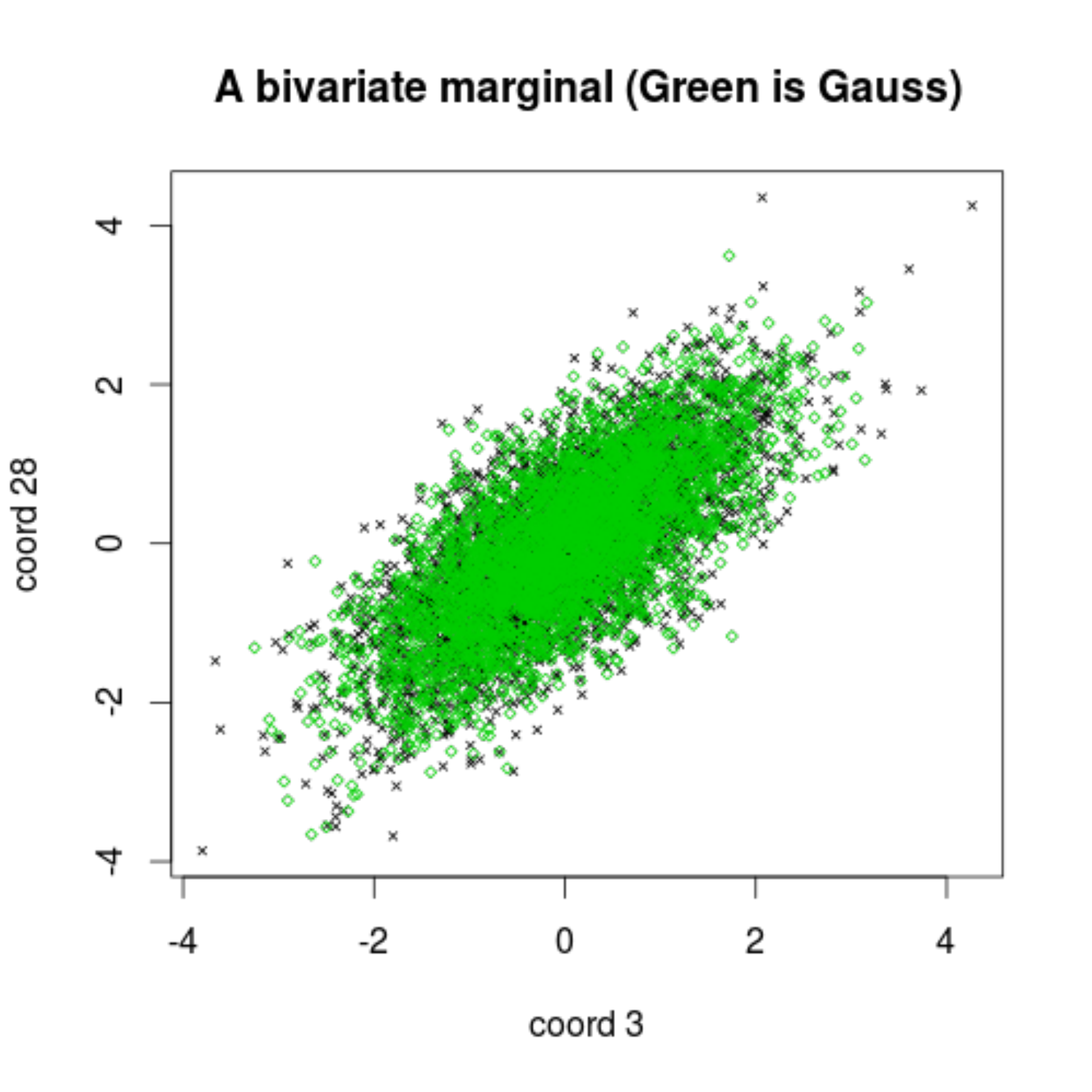}
\par\end{centering}

\caption{\label{fig:Dispersion-corrds}Dispersion plot of two typical components
of $\mathbf{Z}$ and $\mathbf{X}$. The p-value of the multivariate
Shapiro-Wilks test applied to 500 randomly selected samples of $\left(X_{3},X_{28}\right)$
test is 0.191.}
\end{figure}

Data corresponding to two components of both $\mathbf{X}$ and $\mathbf{Z}$,
namely 3 and 28, are shown at figure \ref{fig:Dispersion-corrds}
for illustration. The multivariate version of Shapiro-Wilks test for
normality introduced by \citet{villasenor2009generalization}, as
implemented in the R package \emph{mvShapiro.Test}, was applied to
a randomly selected sample (n=500) of $\left(X_{3},X_{28}\right)$.
This test resulted in a p-value of $0.191$, whereby $\left(X_{3},X_{28}\right)$
can be considered a Gaussian 2-dimensional vector%
\footnote{This procedure was repeated several times, and some of the p-values
obtained were rightly under 0.05. %
}. The same test procedure was performed on all ${30 \choose 2}=435$
pairs of marginals, for $\mathbf{Z}$, $\mathbf{X}$ and $\mathbf{W}$.
The results are summarized at table \ref{tab:Summary-of-Gaussianity}.
It can be seen that non-Gaussian vectors, $\mathbf{X}$ and $\mathbf{W}$,
exhibit Gaussian bivariate marginals most of the time. Results are
qualitatively similar to those of $\mathbf{Z}$, in particular for
$\mathbf{W}$.

\begin{table}
\centering{}%
\begin{tabular}{|c|c|c|c|}
\hline 
$\alpha$-Level & $\left(Z_{j_{1}},Z_{j_{2}}\right)$ & $\left(X_{j_{1}},X_{j_{2}}\right)$ & $\left(W_{j_{1}},W_{j_{2}}\right)$\tabularnewline
\hline 
\hline 
0.01 & 433 (99.54\%) & 402 (92.41\%) & 433 (99.54\%)\tabularnewline
\hline 
0.05 & 419 (96.32\%) & 360 (82.76\%) & 421 (96.78\%)\tabularnewline
\hline 
0.10 & 398 (91.49\%) & 323 (74.25\%) & 405 (93.10\%)\tabularnewline
\hline 
\end{tabular}\caption{\label{tab:Summary-of-Gaussianity}Summary of multivariate Shapiro-Wilks
test applied on all bivariate marginal distributions of $\mathbf{Z}$,
$\mathbf{X}$ and $\mathbf{W}$. A random sub-sample (n=500) from
the available data was used for each testing. Out of the total ${30 \choose 2}=435$
bivariate combinations, the total number (and percentage) of combinations
by which the Normality hypothesis cannot be rejected at the respective
$\alpha$-level are shown. }
\end{table}

\begin{figure}
\begin{centering}
\includegraphics[width=1\textwidth]{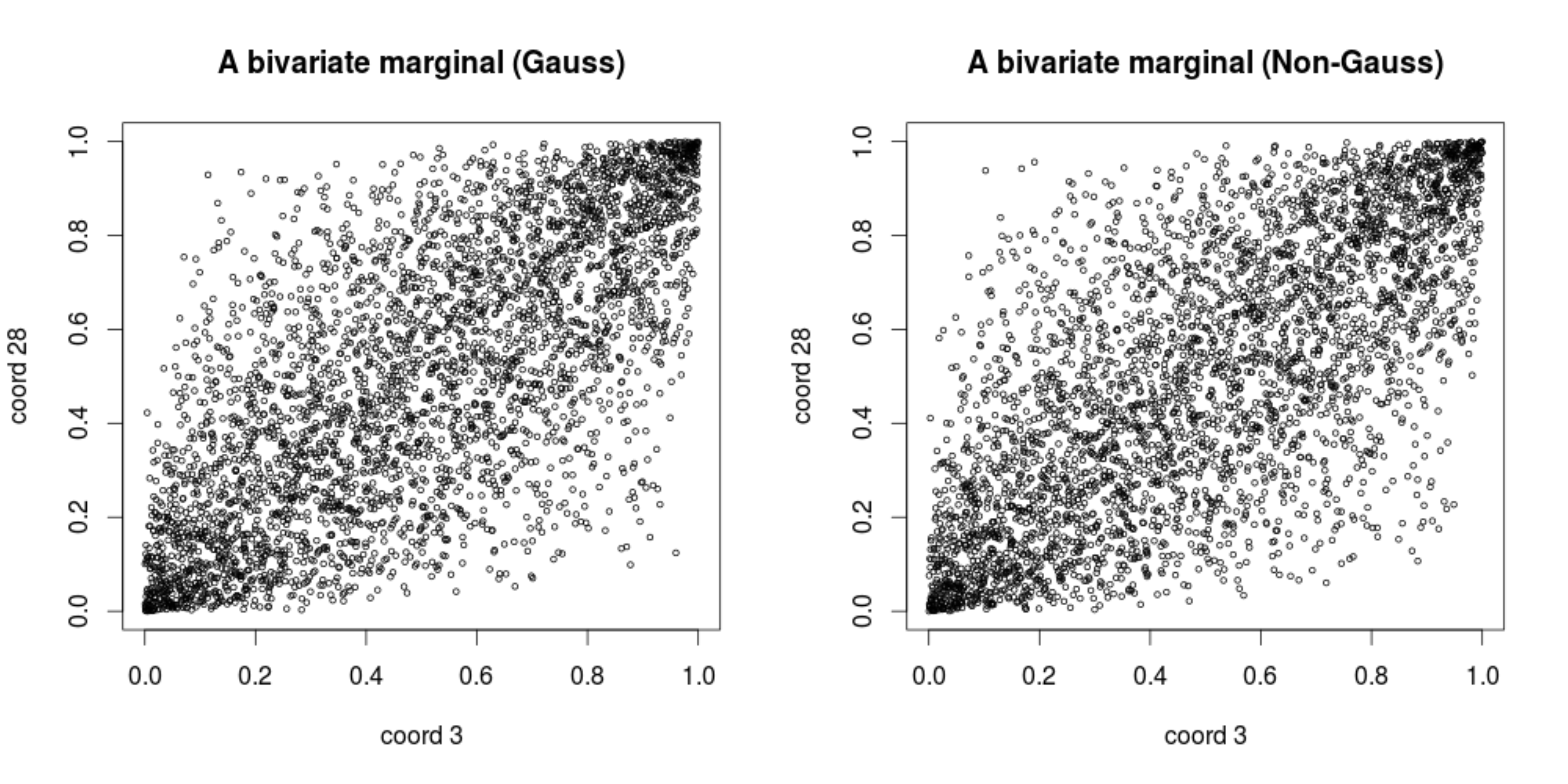}
\par\end{centering}

\caption{\label{fig:Empirical-copula-plots}Empirical copula plots for: (left)
data from $\left(Z_{3},Z_{28}\right)$, and (right) data from $\left(X_{3},X_{28}\right)$.
The p-value of goodness of fit test for Gaussianity on a randomly
selected subsample (n=500) is 0.955.}
\end{figure}

A more detailed analysis of the 2-dimensional components of $\mathbf{Z}$
and $\mathbf{X}$, comprises the study of their respective empirical
copulas. Data plotted at figure \ref{fig:Empirical-copula-plots}
is given, exemplifying for data of vector $\mathbf{X}$, by 
\[
u_{i,j}=\hat{F}_{j}\left(x_{i,j}\right)
\]
where 
\[
\hat{F}_{j}\left(a\right):=\frac{\#\left\{ x_{i,j}:x_{i,j}\leq a\right\} }{n+1}
\]
stands for the empirical cumulative distribution function of component
$X_{j}$, for $j=1,\ldots,30$, and $i=1,\ldots,n$. Visually, both
data sets seem to have the same empirical copula. The test proposed
by \citet{kojadinovic2011goodness} and implemented for package \emph{copula}
of R, was applied to a randomly selected sub-sample of size n=500
of data from $\left(X_{3},X_{28}\right)$, with the number of multipliers
replications set to N=1000. The resulting p-value is 0.955, whereby
gaussianity in the underlying copula seems an acceptable hypothesis.
Note that this test is already very efficient under sample sizes of
n=300 (see \citet{kojadinovic2011goodness}).

The same testing procedure was applied to all possible pair-wise combinations
of components of $\mathbf{Z}$ and $\mathbf{X}$, as had been done
with the multivariate Shapiro-Wilks test. Results are summarized at
table \ref{tab:Summary-of-goodness-copula}. Again, the bivariate
sub-vectors of $\mathbf{X}$ are most of the time considered to have
the Gaussian copula, in a qualitatively similar proportion as the
2-dimensional sub-vectors of $\mathbf{Z}$.

\begin{table}
\centering{}%
\begin{tabular}{|c|c|c|}
\hline 
$\alpha$-Level & $\left(Z_{j_{1}},Z_{j_{2}}\right)$ & $\left(X_{j_{1}},X_{j_{2}}\right)$\tabularnewline
\hline 
\hline 
0.01 & 433 (99.54\%) & 433 (99.54\%)\tabularnewline
\hline 
0.05 & 413 (94.94\%) & 416 (95.63\%)\tabularnewline
\hline 
0.10 & 392 (90.11\%) & 397 (91.26\%)\tabularnewline
\hline 
\end{tabular}\caption{\label{tab:Summary-of-goodness-copula}Summary of goodness of fit
test for the Gaussian copula applied on all bivariate marginal distributions
of $\mathbf{Z}$ and $\mathbf{X}$. A random sub-sample (n=500) from
the available data was used for each testing. Out of the total ${30 \choose 2}=435$
bivariate combinations, the total number (and percentage) of combinations
by which the Normality hypothesis cannot be rejected at the respective
$\alpha$-level are shown. }
\end{table}

We also fitted T-copulas to the data of all 435 pairs of components,
using the data available (n=3650). The idea is to find out how many
degrees of freedom would be an optimal assignment for each pair of
components, both of $\mathbf{Z}$ and of $\mathbf{X}$. The fitting
method employed is described at section 4.2 of \citet{INSR:INSR111},
and named \textquotedbl{}method of moments using Kendall's Tau\textquotedbl{}. 

The 5\%, 50\% and 95\% quantiles of the fitted degrees of freedom
are shown at table \ref{tab:Quantiles-of-the-dofs}. By fitting all
data one gets to a T-copula with 500 and 34.62 degrees of freedom
for $\mathbf{Z}$ and $\mathbf{X}$, respectively%
\footnote{500 degrees of freedom were the highest possible attainable with the
employed fitting algorithm.%
}. As seen in section \ref{sub:Scaling-variable-used}, however, the
tail dependence of $\mathbf{X}$ is comparable to that of a multivariate
T distribution with 15 degrees of freedom, a fact totally invisible
for the T-copula fitting procedure, even with a sample size of n=3650.
Such a tail behavior, which has gone mostly unnoticed in the one and
two dimensional marginals (what Geostatistics check!), may have a
great impact on the wider field, of which the data from $\mathbf{X}$
constitute but a partial observation. See section \ref{sub:Discrepancies-in-the-big-fields}.

\begin{table}
\begin{centering}
\begin{tabular}{|c|c|c|}
\hline 
D.o.f quantile (\%) & $\left(Z_{j_{1}},Z_{j_{2}}\right)$ & $\left(X_{j_{1}},X_{j_{2}}\right)$\tabularnewline
\hline 
\hline 
5\% & 35.75 & 17.97\tabularnewline
\hline 
50\% & 500.00 & 34.67\tabularnewline
\hline 
95\% & 500.00 & 500.00\tabularnewline
\hline 
\end{tabular}
\par\end{centering}

\caption{\label{tab:Quantiles-of-the-dofs}Quantiles of the degrees of freedom
fitted to each of the 435 pairs combinations $\left(Z_{j_{1}},Z_{j_{2}}\right)$
and $\left(X_{j_{1}},X_{j_{2}}\right)$. Using all data, the fitted
degrees of freedom are 500 and 45.75 for $\mathbf{Z}$ and $\mathbf{X}$,
respectively.}
\end{table}

\subsection{\label{sub:The-fitted-covariance}The fitted covariance function}

On the basis of the analysis of the one and two dimensional marginal
distributions, we deem adequate to fit a multivariate Normal distribution
to $\mathbf{Z}$, $\mathbf{X}$ and $\mathbf{W}$. 

\begin{figure}
\begin{centering}
\includegraphics[width=0.75\textwidth]{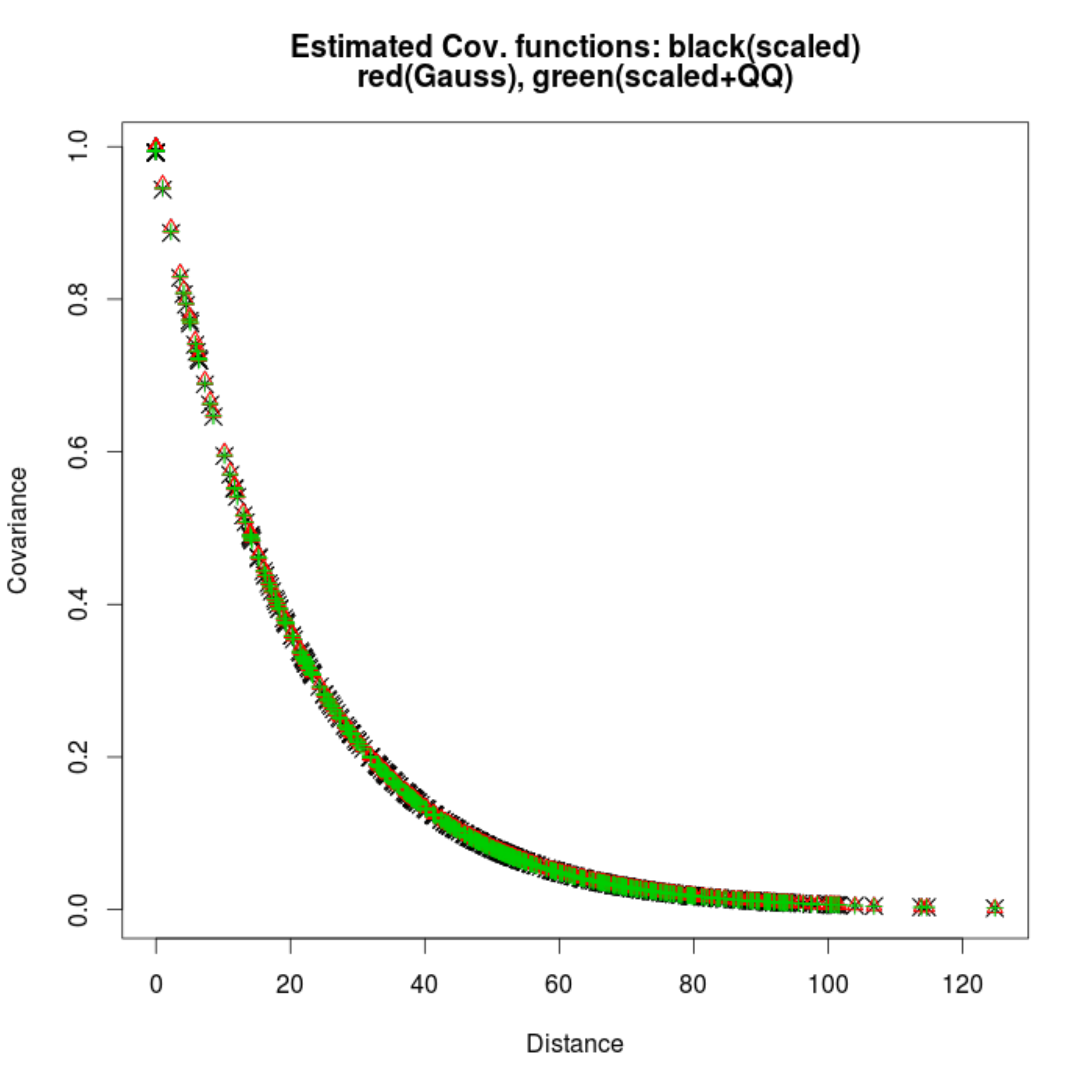}
\par\end{centering}

\caption{\label{fig:Plots-of-fitted-cova-fun}Plots of fitted exponential covariance
functions for $\mathbf{Z}$ (red), $\mathbf{X}$ (black) and $\mathbf{W}$
(green). Plots are practically identical.}
\end{figure}

Since data comes from the Spatial context illustrated at figure \ref{fig:Typical-(non-Gaussian)-field},
we fit covariance matrices, $cov\left(\mathbf{Z}\right)$, $cov\left(\mathbf{X}\right)$
and $cov\left(\mathbf{W}\right)$, using an exponential covariance
function. The estimation method was maximum likelihood using the Normal
distribution as model. Estimated parameters are shown in table \ref{tab:Gaussian-field-specification_table},
whereas plots of the resulting covariance functions appear at figure
\ref{fig:Plots-of-fitted-cova-fun}.

\begin{table}
\begin{centering}
\begin{tabular}{|c|c|c|c|c|}
\hline 
Vector & Mean & Range par. & Nugget & Var\tabularnewline
\hline 
\hline 
$\mathbf{Z}$ & 0.007 & 19.966 & 0.000 & 1.000\tabularnewline
\hline 
$\mathbf{X}$ & 0.006 & 19.992 & 0.000 & 0.992\tabularnewline
\hline 
$\mathbf{W}$ & 0.006 & 19.876 & 0.000 & 0.994\tabularnewline
\hline 
\end{tabular}
\par\end{centering}

\caption{\label{tab:Gaussian-field-specification_table}Gaussian field specification,
as estimated by maximum likelihood (n=3650), and using the exponential
covariance function model.}
\end{table}

Note that both the parameter estimates and the covariance function
plots are practically identical. As was true during the analysis of
the one and two dimensional marginal distributions, there is little
evidence that the distributions of $\mathbf{Z}$, $\mathbf{X}$ and
$\mathbf{W}$ are not the same. However, the complete fields $\mathbf{X}^{*}\in\mathbb{R}^{90000}$
and $\mathbf{W}^{*}\in\mathbb{R}^{90000}$ are very different from
$\mathbf{Z}^{*}\in\mathbb{R}^{90000}$, in terms of important manifestations
of interaction.

\section{\label{sec:Analysis-of-Aggregating-statistics}Analysis of Aggregating
statistics: statistics to notice the difference}

We have seen that both the 1-dimensional and the 2-dimensional marginal
distributions of $\mathbf{X}$ and $\mathbf{W}$ seem to indicate
that these vectors can be safely modeled by a multivariate Normal
model, like the one suitable for $\mathbf{Z}$. We know, however,
that the distributions of $\mathbf{Z}$ and $\mathbf{X}$ are not
the same.

In this sub-section we compute some statistics that can indicate that
the probability distributions of $\mathbf{X}$ and $\mathbf{W}$ may
actually be different from that of $\mathbf{Z}$. They aggregate data
beyond that of the 2-dimensional marginals. In \citet{CumulantsRodriguezBardossy}
these statistics are called \emph{interactions manifestations}.

The first aggregating statistic we consider is the number of components
trespassing a given threshold $a$. Data observed from $\mathbf{X}$
lead to realizations of random variable $L_{\mathbf{X}}$, defined
as 
\[
L^{\mathbf{X}}=\sum_{j=1}^{J}1\left\{ X_{j}>a\right\} 
\]
where 
\begin{equation}
1\left\{ X_{j}>a\right\} =\begin{cases}
1, & X_{j}>a\\
0, & X_{j}\leq a
\end{cases}\label{eq:definicion_componentes_umbral}
\end{equation}

\begin{figure}
\begin{centering}
\includegraphics[width=1\textwidth]{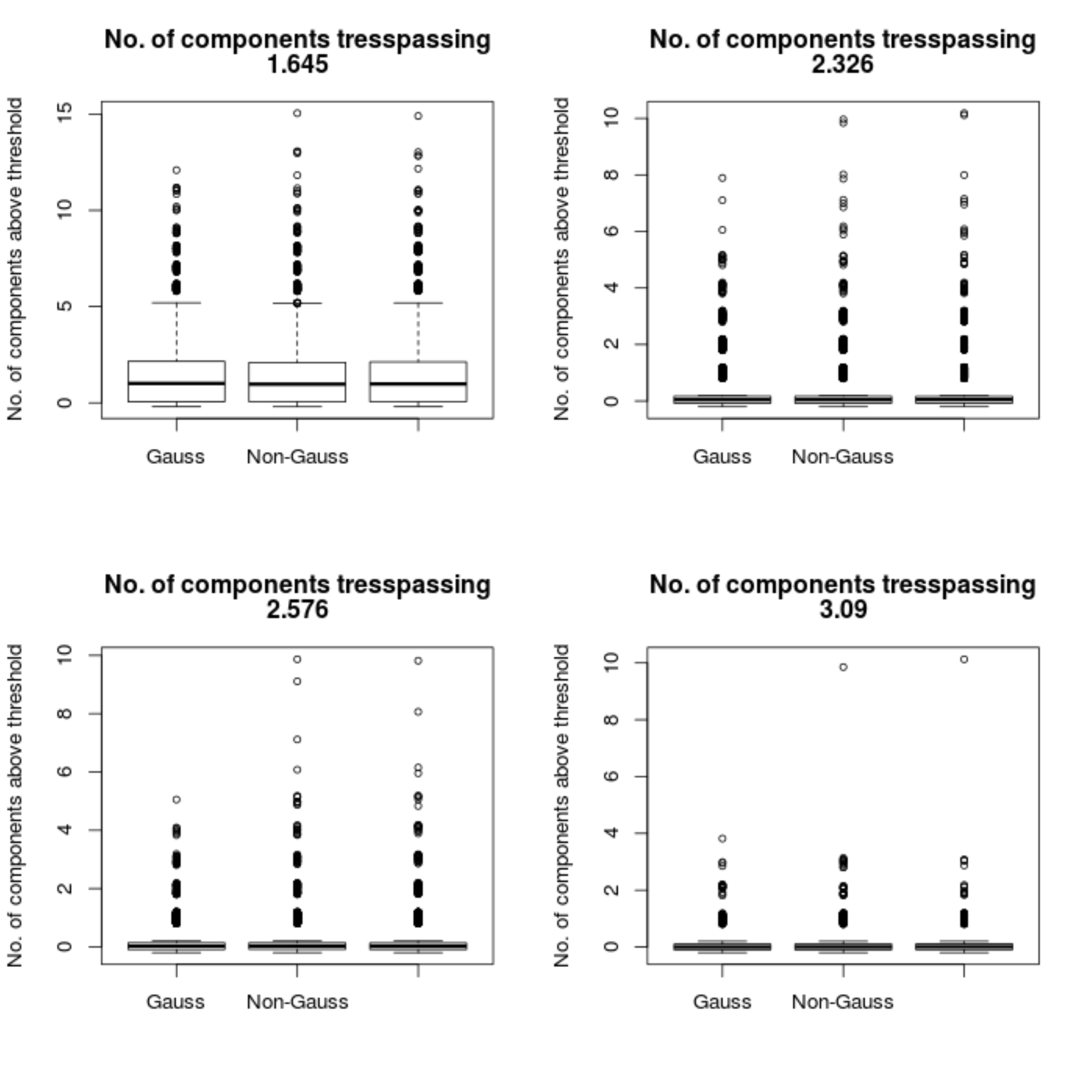}
\par\end{centering}

\caption{\label{fig:Boxplot-of-number-tresspassing}Boxplots of number of components
trespassing the four thresholds indicated (1.645, 2.326, 2.576 and
3.09), for samples from $\mathbf{Z}$ (left), $\mathbf{X}$ (middle)
and $\mathbf{Z}$ (right). Data has been jittered for visualization
purposes. The difference among the plots becomes most apparent as
the threshold is pulled up.}
\end{figure}

Similar constructions lead to $L^{\mathbf{Z}}$ and $L^{\mathbf{W}}$
from $\mathbf{Z}$ and $\mathbf{W}$, respectively. Denote by $l_{1}^{\mathbf{Z}},\ldots,l_{n}^{\mathbf{Z}}$;
$l_{1}^{\mathbf{X}},\ldots,l_{n}^{\mathbf{X}}$ and $l_{1}^{\mathbf{W}},\ldots,l_{n}^{\mathbf{W}}$
the samples of $L^{\mathbf{Z}}$, $L^{\mathbf{X}}$ and $L^{\mathbf{W}}$.
These are plotted in figure \ref{fig:Boxplot-of-number-tresspassing}.
Note that the difference among the plots begins to be quite apparent
for thresholds 2.326 through 3.09. As opposed to what was seen when
analyzing the one and two dimensional marginals separately, there
seems to be a difference among the distributions of $L^{\mathbf{Z}}$,
$L^{\mathbf{X}}$ and $L^{\mathbf{W}}$, and hence of $\mathbf{X}$,
$\mathbf{Z}$ and $\mathbf{W}$.

The second statistic we mention, is the \textquotedbl{}congregation
measure\textquotedbl{} used by \citet{bardossy_copula_2009} and \citet{bardossy_multiscale_2012},
for the sake of model validation. This is a measure not affected by
monotonic transformations on the components of the vector analyzed. 

The congregation measure referred to is constructed as follows. Set
a threshold percentile, $b\in\left(0,1\right)$. Select a set of indexes
$\left(j_{i_{1}},\ldots j_{i_{K}}\right)$, with $1\leq j_{i_{1}}<\ldots<j_{i_{K}}\leq J$.
For the analysis of the components of $\mathbf{X}$, define binary
random variables 
\begin{equation}
\varsigma_{b}\left(j_{i_{k}}\right)=\begin{cases}
1, & F_{j_{i_{k}}}\left(X_{j_{i_{k}}}\right)>b\\
0, & F_{j_{i_{k}}}\left(X_{j_{i_{k}}}\right)\leq b
\end{cases}
\end{equation}

This results in a discrete random vector $\varsigma_{b}=\left(\varsigma_{b}\left(j_{i_{1}}\right),\ldots,\varsigma_{b}\left(j_{i_{K}}\right)\right)$.
The congregation measure referred to is defined to be the entropy
of a sub-vector of $\varsigma$,
\begin{multline}
congr_{b}\left(X_{j_{i_{1}}},\ldots,X_{j_{i_{K}}}\right)=\\
-\sum_{j_{i_{1}},\ldots,j_{i_{K}}}\Pr\left(\varsigma_{b}\left(j_{i_{1}}\right),\ldots,\varsigma_{b}\left(j_{i_{K}}\right)\right)\log\left(\Pr\left(\varsigma_{b}\left(j_{i_{1}}\right),\ldots,\varsigma_{b}\left(j_{i_{K}}\right)\right)\right)\label{eq:entropy_measure}
\end{multline}

That is, the measure is defined as the entropy of the joint distribution
of the binary variables just defined. A higher value of this measure
indicates less association. A similar definition applies to $congr_{b}\left(Z_{j_{i_{1}}},\ldots,Z_{j_{i_{K}}}\right)$.
Note that this measure is not affected by the marginal distributions
of the components employed, hence 
\[
congr_{b}\left(W_{j_{i_{1}}},\ldots,W_{j_{i_{K}}}\right)=congr_{b}\left(X_{j_{i_{1}}},\ldots,X_{j_{i_{K}}}\right)
\]

We applied this measure to three components of vectors $\mathbf{Z}$
and $\mathbf{X}$, namely components 3, 28 and 19, of which the former
two were visualized at figure \ref{fig:Dispersion-corrds}. We used
percentiles $b\in\left\{ 0.6,0.7,0.8,0.85,0.9,0.95,0.99,0.995,0.999\right\} $,
and computed 
\[
r_{b}=\frac{congr_{b}\left(Z_{j_{i_{1}}},\ldots,Z_{j_{i_{K}}}\right)}{congr_{b}\left(X_{j_{i_{1}}},\ldots,X_{j_{i_{K}}}\right)}
\]

\begin{figure}
\begin{centering}
\includegraphics[width=0.75\textwidth]{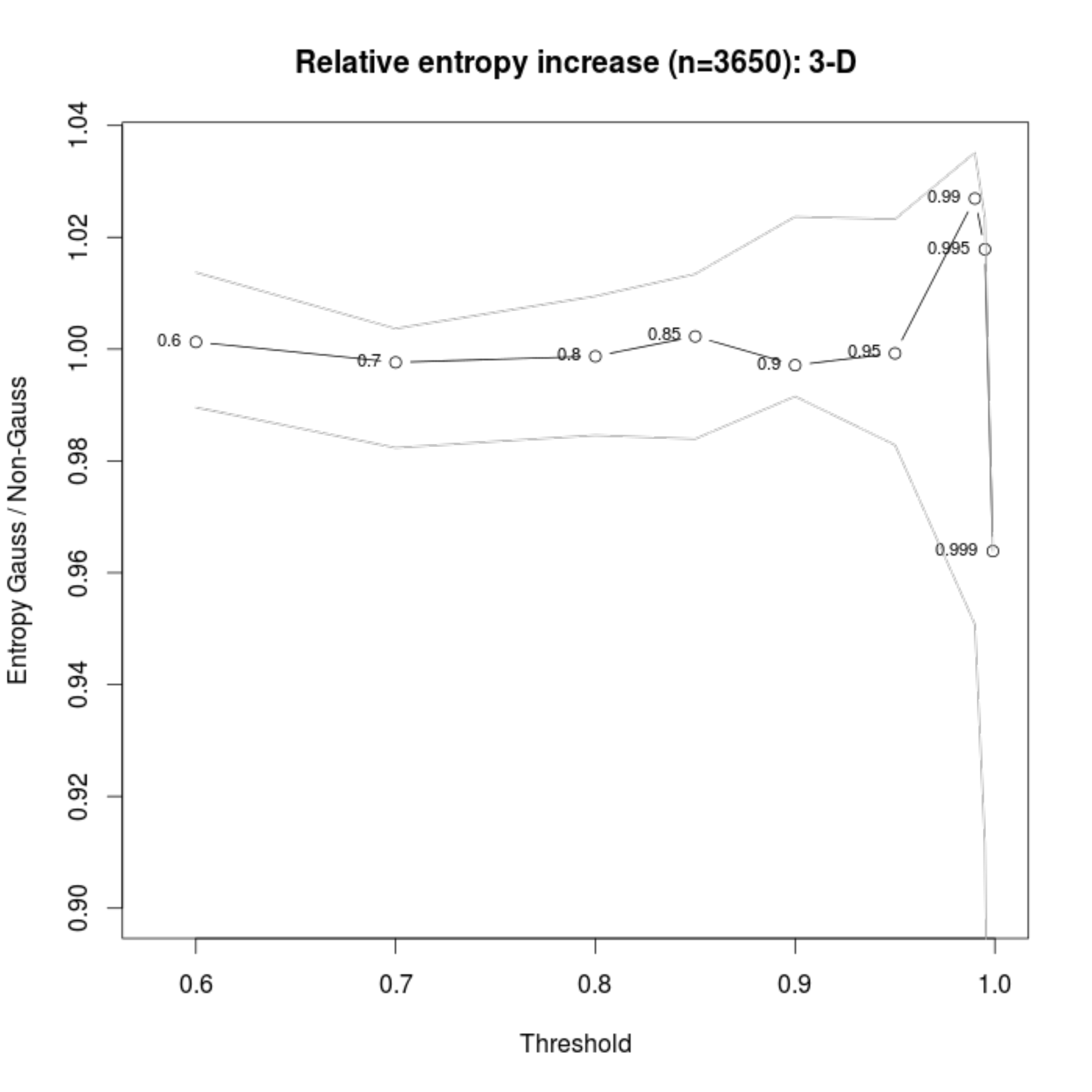}
\par\end{centering}

\caption{\label{fig:Ratio_of_congregation}Ratio, $r_{b}$, of congregation
measures for the percentiles $b\in\left\{ 0.6,0.7,0.8,0.85,0.9,0.95,0.99,0.995,0.999\right\} $.
Association among the three components $\left(Z_{3},Z_{28},Z_{19}\right)$
decreases considerably as compared to that of $\left(X_{3},X_{28},X_{19}\right)$
from the 99\% percentile on. A bootstrap based confidence interval
has been added for significance assessment.}
\end{figure}

The resulting ratio values are shown in figure \ref{fig:Ratio_of_congregation}.
The estimated association of the components $\left(Z_{3},Z_{28},Z_{19}\right)$,
as quantified by this measure, does not seem to decrease considerably
as compared to that of $\left(X_{3},X_{28},X_{19}\right)$. A \textquotedbl{}parametric
bootstrap\textquotedbl{} (see \citet{efron1993introduction}) 90\%
confidence interval has been added for the ratio of the entropies,
computed by simulating a Normal sample of size $n=3650$ with zero
means and correlation matrix the sample correlation matrix of $\left(Z_{3},Z_{28},Z_{19}\right)$.
The procedure is repeated 10000 times to create the confidence interval.

\begin{figure}
\begin{centering}
\includegraphics[width=1\textwidth]{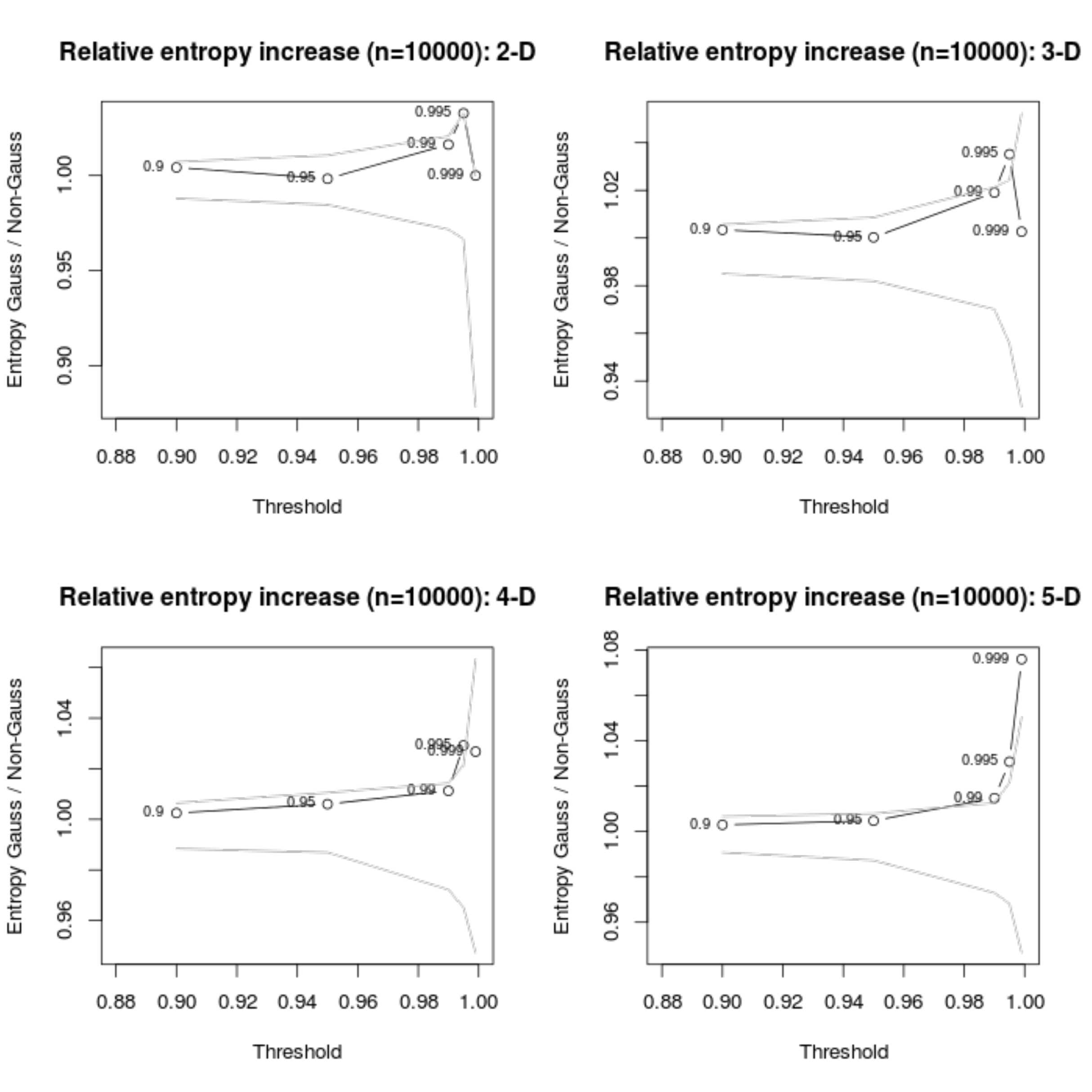}
\par\end{centering}

\caption{\label{fig:Ratio_of_2_3_4_5}Ratio, $r_{b}$, of congregation measures
for the percentiles $b\in\left\{ 0.6,0.7,0.8,0.85,0.9,0.95,0.99,0.995,0.999\right\} $.
Association among two, three, four and five components of $\mathbf{Z}$
decreases considerably as compared to that of subvectors of $\mathbf{X}$
from the 99\% percentile on. A bootstrap based confidence intervals
have been added for significance assessment. Subvectors used have
indexes $\left(3,28\right),\left(3,28,19\right),\left(3,28,19,16\right)$
and $\left(3,28,19,16,25\right)$. }
\end{figure}

However, as shown in figure \ref{fig:Ratio_of_2_3_4_5}, if the sample
size is increased to n=10000, the association among subvectors of
$\mathbf{X}$ can be seen to increase considerably as compared to
that of subvectors of $\mathbf{Z}$. In particular this is the case
as one approaches the uppermost tail of the 2, 3, 4 and 5-dimensional
distributions. Subvectors employed for the evaluation are indicated
in figure \ref{fig:Ratio_of_2_3_4_5}. This was to be expected in
view of the uppermost tail of scaling variable $V$, see the right
panel of figure \ref{fig:Comparison-of-the-scalings}. Hence, on the
basis of the analysis of only three through five components, it is
possible to notice a difference in the dependence structure of the
fields (compare \citet{bardossy_copula_2009}), provided the sample
size is sufficiently large.

A third kind of aggregating statistic comprises the quantiles of the
sum of components above given thresholds. To this end, we took the
n=3650 observations of each vector, $\mathbf{Z}$, $\mathbf{X}$ and
$\mathbf{W}$, and obtained from them $s_{1}^{\mathbf{Z}},\ldots,s_{n}^{\mathbf{Z}},\ldots,s_{n}^{\mathbf{W}}$.
Where, for a given threshold $a$, one has for example,
\begin{equation}
s_{i}^{\mathbf{X}}=\sum_{j=1}^{30}1\left\{ x_{ij}>a\right\} \times x_{ij}\label{eq:suma_sobre_thre}
\end{equation}
with $i=1,\ldots,3650$. 

\begin{figure}
\begin{centering}
\includegraphics[width=1\textwidth]{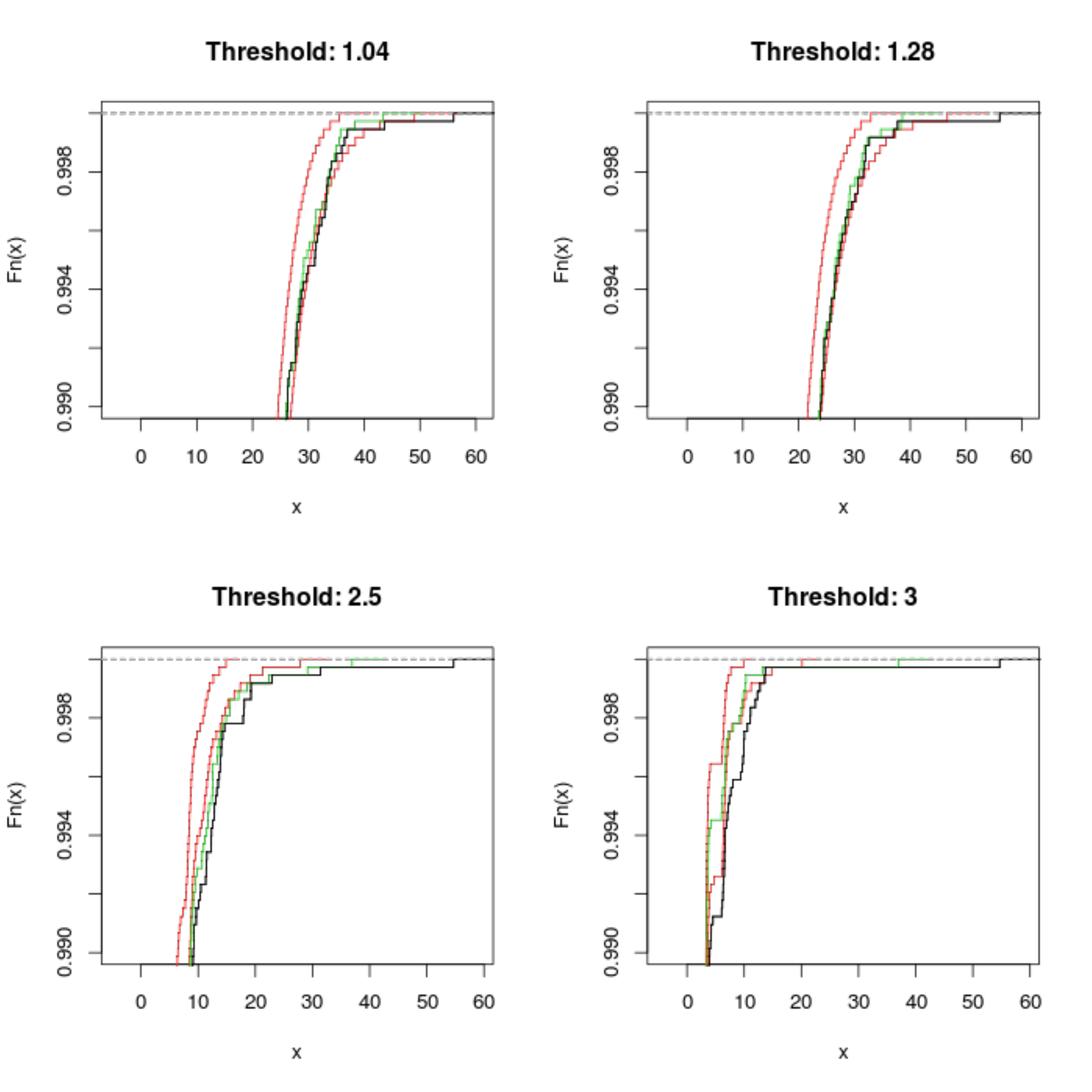}
\par\end{centering}

\caption{\label{fig:Empirical-distributions-of-sumas}Empirical distributions
of sums above various thresholds (1.04, 1.28, 2.5 and 3) for the n=3650
observations of random vectors $\mathbf{X}$ (black) and $\mathbf{W}$
(green). Simulation based 90\% confidence intervals for the empirical
distributions arising from a Gaussian vector with the same means and
covariance matrix as $\mathbf{X}$ appear in red. The tail of the
sums above thresholds 2.5 and 3 is significantly heavier than the
Gaussian model would prescribe. }
\end{figure}

The empirical cumulative distribution functions built from $s_{1}^{\mathbf{X}},\ldots,s_{3650}^{\mathbf{X}}$
and $s_{1}^{\mathbf{W}},\ldots,s_{3650}^{\mathbf{W}}$ are presented
in figure \ref{fig:Empirical-distributions-of-sumas}, for thresholds
$a=\left\{ 1.04,1.28,2.5,3\right\} $. Simulation based 90\% confidence
intervals (appearing in red) for the empirical cumulative distribution
function of $s_{1}^{\mathbf{Z}},\ldots,s_{3650}^{\mathbf{Z}}$ were
also added. These confidence intervals were created for each threshold,
$a$, by repeting 1000 times the following procedure: simulate n=3650
realizations of a Gaussian random vector with mean and covariance
as estimated for $\mathbf{X}$ in section \ref{sub:The-fitted-covariance},
and then apply construction (\ref{eq:suma_sobre_thre}). In this manner
we obtain 1000 empirical cumulative distribution functions; at each
percentile $u\in\left[0,1\right]$, we compute the values of all 1000
e.c.d.f. and take from them the 5\% and 95\% quantile values.

It is clear from figure \ref{fig:Empirical-distributions-of-sumas},
that tails of the distribution functions obtained for $s_{1}^{\mathbf{X}},\ldots,s_{3650}^{\mathbf{X}}$
for thresholds $a\in\left\{ 2.5,3\right\} $ are heavier than expected
from a Gaussian vector having the same means, covariance matrix, and
approximately the same marginal distributions as $\mathbf{X}$. Hence,
the analysis of these statistics is valuable for diagnosis of higher
order interaction. 

\section{\label{sec:MCMC-estimation-of}MCMC estimation of the conditional
scaling variable}

In the following, $\mathbf{x}\in\mathbb{R}^{J}$ represents a partial
observation of a complete field $\left(\mathbf{x},\mathbf{x}^{*}\right)\in\mathbb{R}^{J+M}$.
Also, as indicated in section \ref{sub:Interpolation-to-ungauged},
$p\left(\mathbf{x}-\mu\mid V\right)$ is just $N_{J}\left(\mathbf{0},V\times\Sigma\right)$,
the multivariate Normal distribution on $\mathbb{R}^{J}$ with vector
of means $\mathbf{0}$ and covariance matrix $V\times\Sigma$.

We can build a Markov Chain whose stationary distribution is approximately
that of $V\mid\mathbf{x},\Sigma,\mu$, as follows:
\begin{enumerate}
\item Provide an initial value for the Markov chain, say, $V^{\left(0\right)}=1$.
\item For $b=1,\ldots,B$, do:

\begin{enumerate}
\item Sample a candidate value $C^{\left(b\right)}\sim N\left(V^{\left(b-1\right)},1\right)$.
A variance of 1 for the transition kernel seems to be adequate for
most cases, according to initial exploratory analyses. 
\item If $C^{\left(b\right)}\leq0$, set $V^{\left(b\right)}=V^{\left(b-1\right)}$.
Proceed to iteration $b+1$. 
\item If $C^{\left(b\right)}>0$, compute $w=\frac{p\left(\mathbf{x}-\mu\mid C^{\left(b\right)}\right)p\left(C^{\left(b\right)}\right)}{p\left(\mathbf{x}-\mu\mid V^{\left(b-1\right)}\right)p\left(V^{\left(b.1\right)}\right)}$
and sample $U\sim Unif\left(0,1\right)$. Set $V^{\left(b\right)}=C^{\left(b\right)}$
if $U<\frac{p\left(\mathbf{\mathbf{x}}-\mu\mid C^{\left(b\right)}\right)p\left(C^{\left(b\right)}\right)}{p\left(\mathbf{x}-\mu\mid V^{\left(b-1\right)}\right)p\left(V^{\left(b.1\right)}\right)}$,
else set $V^{\left(b\right)}=V^{\left(b-1\right)}$. Proceed to iteration
$b+1$. 
\end{enumerate}
\end{enumerate}
After sufficiently many iterations, the values $V^{\left(b\right)}$
can be considered as correlated samples from $p\left(V\mid\mathbf{y}-\mu\right)$.
After convergence of the Markov Chain just built, we store additional
$B$ samples $V^{\left(1\right)},\ldots,V^{\left(B\right)}$. We use
these additional samples for conditionally simulating from the total
field, $\left(\mathbf{X},\mathbf{X}^{*}\right)$, in the presence
of observed data, $\mathbf{X}=\mathbf{x}$. 

\section*{References}
\bibliography{Report_QE}

\end{document}